\documentclass[12pt]{article}
\pdfoutput=1

\usepackage[utf8]{inputenc}

\usepackage{draft}
\usepackage{putex}
 
\usepackage[all]{xy}

\usepackage{stackrel,amssymb}

\usepackage{graphicx}
\usepackage{caption}
\usepackage{amsmath,bm}
\usepackage{array}
\usepackage{subcaption}
\usepackage{epstopdf}
\usepackage{enumerate}
\usepackage{cite}
\usepackage{tensor}
\usepackage{slashed}
\usepackage[utf8]{inputenc}
\usepackage{rotating}
\usepackage{bigfoot}
\usepackage[
      colorlinks=true,
      linkcolor=blue,
      urlcolor=blue,
      filecolor=black,
      citecolor=red,
      ]{hyperref}

\usepackage{tikz-cd}

\def\Xint#1{\mathchoice
	{\XXint\displaystyle\textstyle{#1}}%
	{\XXint\textstyle\scriptstyle{#1}}%
	{\XXint\scriptstyle\scriptscriptstyle{#1}}%
	{\XXint\scriptscriptstyle\scriptscriptstyle{#1}}%
	\!\int}
\def\XXint#1#2#3{{\setbox0=\hbox{$#1{#2#3}{\int}$ }
		\vcenter{\hbox{$#2#3$ }}\kern-.6\wd0}}

\def\dashint{\Xint-}

\numberwithin{equation}{section}

\newcommand{\CC}{\Gamma}

\def\<{\langle}
\def\>{\rangle}
\def\pa{\partial}
\def\ve{\varepsilon}
\def\ep{\epsilon}

\def\sP{\slashed{\nabla}}
\def\sf{\mathsf}

\newcommand{\leftrarrows}{\mathrel{\raise.75ex\hbox{\oalign{%
				$\scriptstyle\leftarrow$\cr
				\vrule width0pt height.5ex$\hfil\scriptstyle\relbar$\cr}}}}
\newcommand{\lrightarrows}{\mathrel{\raise.75ex\hbox{\oalign{%
				$\scriptstyle\relbar$\hfil\cr
				$\scriptstyle\vrule width0pt height.5ex\smash\rightarrow$\cr}}}}
\newcommand{\Rrelbar}{\mathrel{\raise.75ex\hbox{\oalign{%
				$\scriptstyle\relbar$\cr
				\vrule width0pt height.5ex$\scriptstyle\relbar$}}}}

\makeatletter
\def\leftrightarrowsfill@{\arrowfill@\leftrarrows\Rrelbar\lrightarrows}
\newcommand{\xleftrightarrows}[2][]{\ext@arrow 3399\leftrightarrowsfill@{#1}{#2}}
\makeatother

\begin{document}

\preprint{PUPT-2608}

	\institution{PU}{Joseph Henry Laboratories, Princeton University, Princeton, NJ 08544, USA}
	\institution{CMSA}{Center of Mathematical Sciences and Applications, Harvard University, Cambridge, MA 02138, USA}
	\institution{HU}{Jefferson Physical Laboratory, Harvard University,
		Cambridge, MA 02138, USA}

\title{
Taming Defects in $\mathcal{N}=4$ Super-Yang-Mills  
}

\authors{Yifan Wang\worksat{\PU,\CMSA,\HU}}

\abstract{
We study correlation functions involving extended defect operators in the four-dimensional $\cN=4$ super-Yang-Mills (SYM). The main tool is supersymmetric localization with respect to the supercharge $\cQ$ introduced in \cite{Pestun:2009nn} which computes observables in the $\cQ$-cohomology. We classify general defects of different codimensions in the $\cN=4$ SYM that belong to the $\cQ$-cohomology, which form ${1\over 16}$-BPS defect networks.
By performing the $\cQ$-localization of the $\cN=4$ SYM on the four-dimensional hemisphere, we discover a novel defect-Yang-Mills (dYM) theory on a submanifold given by the two-dimensional hemisphere and described by (constrained) two-dimensional Yang-Mills coupled to topological quantum mechanics on the boundary circle. This also generalizes to interface defects in $\cN=4$ SYM by the folding trick. We provide explicit dictionary between defect observables in the SYM and those in the dYM, 
  which enables extraction of general  ${1\over 16}$-BPS defect network observables of the SYM from two-dimensional gauge theory and matrix model techniques.
  Applied to the D5 brane interface in the $SU(N)$ SYM, we explicitly determine a set of defect correlation functions in the large $N$ limit and obtain precise matching with
   strong coupling results from IIB supergravity on $AdS_5\times S^5$.
}
\date{}

\maketitle

\tableofcontents

\pagebreak

\section{Introduction}
The $\cN=4$ super-Yang-Mills (SYM)  theory in four spacetime dimensions is one of the most well-studied quantum field theories in recent decades. On one hand, formulated as a Lagrangian theory, it has been an active arena to understand general features of gauge theories,   such  instanton effects,  resurgence in perturbation series, and strong-weak dualities. On the other hand, via the conjectured AdS/CFT correspondence \cite{Maldacena:1997re,Witten:1998qj,Gubser:1998bc}, the $\cN=4$ SYM provides a non-perturbative definition of the type IIB string theory on  $AdS_5\times S^5$ from which one can draw important lessons about quantum gravity.  These kinds of investigations in the $\cN=4$ SYM are made possible by an array of methods to explore the rich dynamics of the theory, including supersymmetric localization, integrability and conformal bootstrap.\footnote{There is a vast amount of literature on each of the three subjects. For a review, see \cite{Pestun:2016zxk},\cite{Beisert:2010jr} and \cite{Poland:2018epd}  respectively.} In particular, the conformal invariance of the theory allows for a non-perturbative exact formulation of the $\cN=4$ SYM in terms of fundamental building blocks: the two-point and three-point functions of local operators. Thanks to the maximal supersymmetry, a large number of such structure constants can be extracted efficiently and analytically via integrability and localization methods, even in the strong coupling regime. Combined with the conformal bootstrap technique, they provide a powerful way to potentially solve the $\cN=4$ SYM at the level of local operator algebra. For recent fruitful attempts in this direction, see for example \cite{Binder:2019jwn,Chester:2019pvm,Chester:2019jas,Chester:2020dja}.

However the richness of the $\cN=4$ SYM extends well beyond the local operator algebra. The theory is known to admit extended  defect operators of various codimensions that exhibit nontrivial interactions with local operators and among themselves. The most familiar examples are perhaps the Wilson and 't Hooft loop operators. These defect operators play an important role in elucidating the phase  diagram of the gauge theory (SYM and its closely-related cousins) \cite{Witten:1998zw,Gukov:2013zka,Gaiotto:2014kfa}, as well as refining the notion of dualities \cite{Aharony:2013hda}. In the context of AdS/CFT, the defect operators correspond to branes or solitons in the type IIB string theory on $AdS_5\times S^5$, which are  crucial in non-perturbative aspects of quantum gravity. The defects themselves may also harbor local operators restricted to their worldvolume, which map to open string excitations of branes in IIB. Moreover they may split or join with other defect operators of different codimensions, coming from brane intersections. Altogether they give rise to complicated networks of observables in the SYM.
For defects that preserve a conformal subalgebra, a natural generalization of the conformal bootstrap program for local operators applies and constrains the spectrum and operator-product-expansion (OPE) data on the defect in relations to those of the conventional bulk local operators  \cite{Liendo:2012hy,Billo:2016cpy}.\footnote{An equally interesting problem is to constrain the spectrum of defect operators. But we will not address that in this paper.} However to solve such defect bootstrap problems for the SYM requires additional dynamical inputs, namely intrinsic defect structure constants (e.g. one-point-functions of bulk local operators and defect-bulk two-point functions) in the SYM. This calls for extensions of the localization and integrability methods to incorporate defect observables.

In \cite{Drukker:2007yx,Drukker:2007qr}, Drukker-Giombi-Ricci-Trancanelli identified an interesting 2d sector of the   $\cN=4$ SYM. By studying ${1\over 8}$-BPS Wilson loops restricted to a two-sphere in the SYM, they conjectured that this 2d sector is described by a  bosonic Yang-Mills (YM) theory.\footnote{To be more precise,  the 2d Yang-Mills theory here is constrained to the zero instanton sector, also known as the constrained 2d Yang-Mills in \cite{Pestun:2009nn}.  } For this reason we will refer to this two-sphere as $S^2_{\rm YM}$. This conjecture was later derived from a localization computation in \cite{Pestun:2009nn}: by choosing a particular supercharge $\cQ$ of the 4d SYM which is nilpotent when restricted to  $S^2_{\rm YM}$ in the 4d spacetime $\mR^4$ (or $S^4$ by a Weyl transformation), the 2d Yang-Mills emerges as an effective description of the $\cQ$-cohomology in the space of all field configurations of the original SYM. A dictionary was provided between certain observables in the 4d  SYM and the 2d YM. In particular, the ${1\over 8}$-BPS Wilson loops and ${1\over 8}$-BPS local operators are mapped to insertions of ordinary Wilson loops and field strength in the 2d YM on $S^2_{\rm YM}$ \cite{Giombi:2009ds}. This dictionary was later extended to include ${1\over 2}$-BPS {}'t Hooft loop on a great $S^1$ that links with the  $S^2_{\rm YM}$ \cite{Giombi:2009ek}. Unlike the chiral algebra sector of general 4d $\cN=2$ SCFTs \cite{Beem:2013sza}, the 2d YM sector of 4d $\cN=4$ SYM carries nontrivial dependence on the gauge coupling $g_4$. This has lead to substantial progress in understanding perturbation series and non-perturbative effects in gauge theories, as well as many sophisticated precision checks of AdS/CFT \cite{Giombi:2009ms,Bassetto:2009rt,Gerchkovitz:2016gxx,Giombi:2017cqn,Giombi:2018qox,Giombi:2018hsx}.

\begin{figure}[!htb]
	\centering
	\includegraphics[scale=.25]{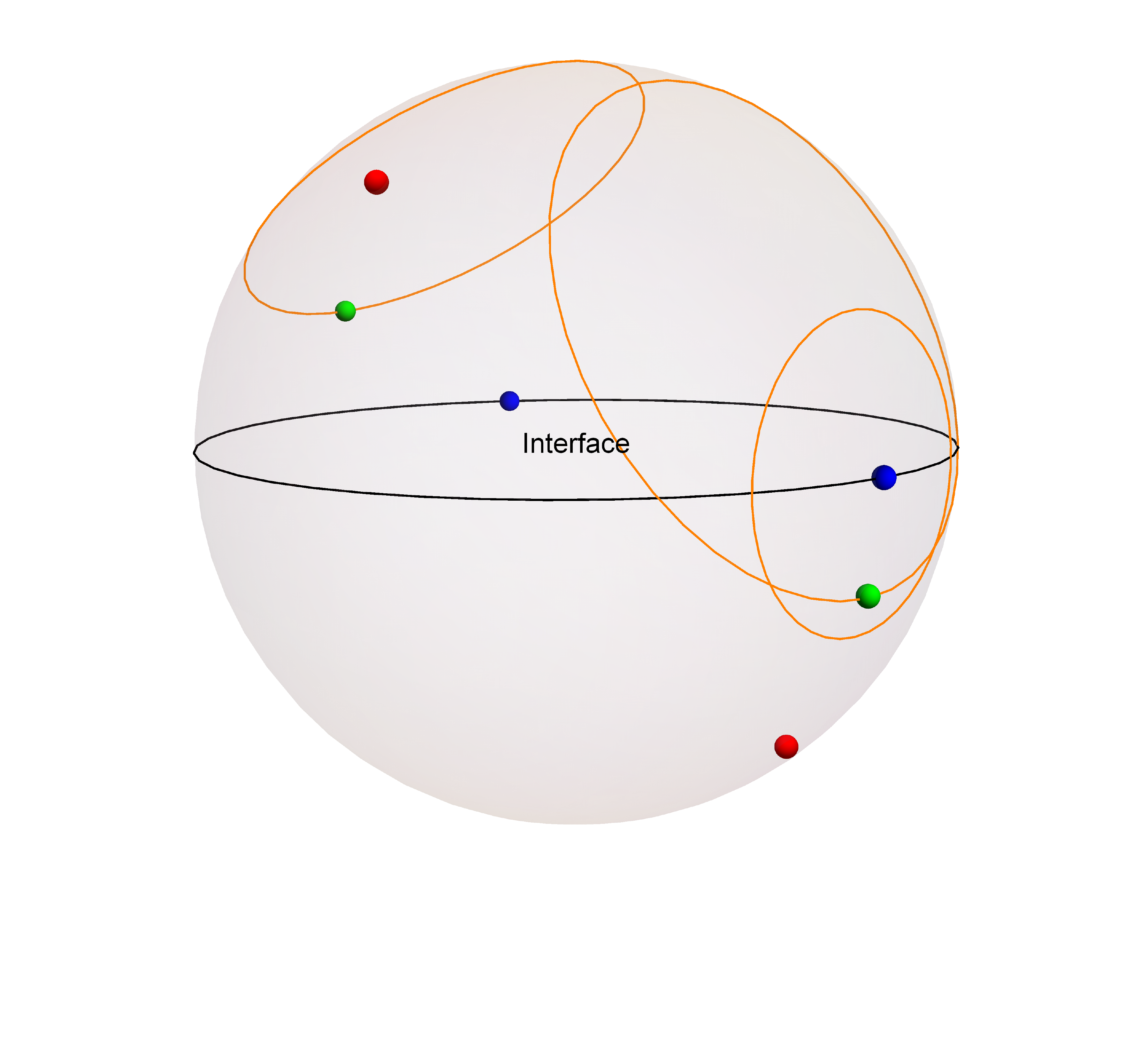}
	\caption{An example of a defect network in the $\cN=4$ SYM on $S^4$ that is captured by the 2d dYM on $S^2_{\rm YM}$. In the figure above we have suppressed the two transverse directions to the $S^2_{\rm YM}$ in $S^4$. The black circle at the equator $S^1_{\rm TQM}$ denotes an interface defect. The orange loops are intersecting Wilson loops. Red dots correspond to bulk local operator insertions, blue dots are local operators on the interface, and green dots are local operators on the Wilson loops. }
	\label{fig:network}
\end{figure}

In this paper, we extend the  4d/2d setup of \cite{Drukker:2007yx,Drukker:2007qr,Pestun:2009nn}  by classifying general conformal defects of the 4d $\cN=4$ SYM in the $\cQ$-cohomology, which include, in addition to the Wilson loops and 't Hooft loops,\footnote{From the classification, we also discover new line operators in the $\cQ$-cohomology beyond those ones considered in \cite{Drukker:2007qr} (see type $\cD_1^{\rm II}$ Wilson line defects in Section~\ref{sec:linedefects}).}  interfaces  (or boundaries) and surface operators. Carrying out the $\cQ$-localization in the presence of these defects leads to interesting refinement of the 2d YM sector, which we will refer to as the 2d defect-Yang-Mills (dYM). In particular, the BPS interface (boundary) intersects with the  $S^2_{\rm YM}$ at an equator $S^1$  (boundary of hemisphere $HS^2_{\rm YM}$), thus inducing a codimension-one defect in the 2d YM. The $\cQ$-cohomology and thus the dYM are naturally extended by local operator insertions on the interface restricted to this $S^1$. When the interface hosts a local 3d $\cN=4$ SCFT, this includes a 1d protected subsector of the full 3d operator algebra, known as the   1d topological quantum mechanics (TQM) on this $S^1$ \cite{Chester:2014mea,Beem:2016cbd,Dedushenko:2016jxl}. For this reason, we refer to the equator (boundary) $S^1$ as $S^1_{\rm TQM}$. For a large number of cases where the original 3d $\cN=4$ SCFT admits a UV Lagrangian, such TQM sector is described by a gauged quantum mechanics of anti-periodic scalars on $S^1_{\rm TQM}$ with topological kinetic terms \cite{Dedushenko:2016jxl}. In general, the 1d TQM couples non-trivially to the 2d YM fields through its flavor symmetries.  In a sense, our setup generalizes that of \cite{Pestun:2009nn} and \cite{Dedushenko:2016jxl} by identifying 1d TQM coupled to 2d YM as a consistent sector of 3d $\cN=4$ SCFT coupled to bulk 4d $\cN=4$ SYM. Combined with insertions of defect observables of other codimensions as well as local operators in the $\cQ$-cohomology, our setup provides a systematic framework to extract exact correlation functions of defect networks (see Figure~\ref{fig:network}) in the SYM that preserve a common single supercharge $\cQ$ (i.e. ${1\over 16}$-BPS).

There have been steady progress on the integrability side in computing observables of the SYM in the planar large $N$ limit with interface (boundary) defects (see \cite{Andrei:2018die,deLeeuw:2019usb} for an overview). One-point functions of both BPS and non-BPS operators have been obtained using the spin chain method. Here the interface defect is represented by a matrix product state (MPS) of the spin chain and the local operators in the bulk correspond to Bethe eigenstates of the spin chain Hamiltonian (and excitations). Consequently the one-point functions simply follows from overlaps of the MPS with the Bethe states.  However, the computation relies on the expansion in small 't Hooft coupling $\lambda \ll 1$ and is highly loop dependent, thus little is known beyond one-loop \cite{deLeeuw:2015hxa,Buhl-Mortensen:2016jqo,Buhl-Mortensen:2017ind}. As a result, up to now there were no direct comparisons with the strong coupling $\lambda \gg 1$ results in $AdS_5\times S^5$ predicted in \cite{DeWolfe:2001pq}.

In this paper, as a simple application of the dYM setup, we show such defect correlators are computed by standard matrix model techniques in the leading strong coupling limit, and in perfect match with results from IIB string theory on $AdS_5\times S^5$ \cite{Nagasaki:2012re}. In a subsequent publication \cite{nahmp}, we consider more general interface defects as in \cite{deLeeuw:2015hxa, Buhl-Mortensen:2016jqo,Buhl-Mortensen:2017ind} and obtain exact expressions in the 't Hooft coupling $\lambda$. We emphasize that for simple defect observables considered in this paper such as one-point functions and bulk-defect two-point-functions, the correlators can be related to the familiar single hermitian matrix model albeit with a non-polynomial potential. For more general defect network observables, we obtain novel multi-matrix models from the dYM as an extension of those in \cite{Giombi:2009ds,Giombi:2012ep}. The study of such matrix models are deferred to a future publication.
 
The rest of the paper is organized as follows. We will begin in Section~\ref{sec:condef} by reviewing general conformal defects in the $\cN=4$ SYM, in relation to the subalgebras they preserve in the full $\cN=4$ superconformal algebra and corresponding brane constructions in IIB string theory. In Section~\ref{sec:Qcoh}, we classify the conformal defects that preserve the supercharge $\cQ$ of \cite{Pestun:2009nn}. Focusing on the interface defects, we  perform the supersymmetric localization of $\cN=4$ SYM in the presence such defects and identify the two-dimensional defect-Yang-Mills theory in Section \ref{sec:loctodym}. We explain how to compute general defect observables in the $\cQ$-cohomology using the dYM in Section~\ref{sec:defectindym} and comment on comparisons to known results in the literature. In Section~\ref{sec:applications}, we apply the methods developed in the previous sections to compute simple defect correlation functions in the $\cN=4$ SYM with interface defects and compare to holographic computations in the large $N$ limit. We end by a brief summary and discuss future directions in Section~\ref{sec:conclusion}.
  
\section{Conformal Defects in $\cN=4$ Super-Yang-Mills} 
\label{sec:condef}

\subsection{Review of $\cN=4$ superconformal symmetry }

The $\cN=4$ SYM is symmetric under the superconformal group $PSU(2,2|4)$ which includes the bosonic conformal group $SO(4,2)$, the  R-symmetry group $SO(6)_R$, as well as 16 Poincar\'e supercharges $Q$ and 16 conformal supercharges $S$.

It is convenient to parametrize the supercharges by a 16-component conformal Killing spinor $\ve$
subjected to the conformal killing spinor equation
\ie
\nabla_\m \ve ={1\over 4}\C_\m \sP \ve \,.
\label{CKV}
\fe
In flat space, the solutions are parametrized by
\ie
\ve=\ep_s+x^\m \C_\m \ep_c\,,
\label{SYMCKS}
\fe
where $\epsilon_{s,c}$ are 16-component complex Weyl spinors of $Spin(10,\mR)$ with  positive and negative chiralities respectively.\footnote{We will be mainly working with the Euclidean signature by a Wick rotation.} The $Spin(10,\mR)$ arises naturally when viewing the 4d  $\cN=4$ SYM as coming from Kaluza-Klein (KK) reduction of the 10d $\cN=1$ SYM.

The  general superconformal transformations denoted by
\ie
\D_\ve=\ep^\A_s Q_\A +\ep^c_\B S^\B
\fe
generate the full $\cN=4$ superconformal algebra by anti-commutators
\ie
\{\D_{\ve_1},\D_{\ve_2}\}
=-2(\cL_v+R_w+\Omega_\lambda)\,.
\label{sca}
\fe
Here $\cL_v$ denotes the Lie derivative with respect to the vector field (in general a conformal killing vector)
\ie
v^\m=\ve_{(1} \Gamma^\m \ve_{2)}\,.
\label{vecpara}
\fe
$R_w$ is the $SO(6)_R$ rotation which acts by
\ie
R_w={1\over 2}w_{IJ} R^{IJ}\,,
\fe
with parameter
\ie
w_{IJ}=2\ve_{(1} \tilde \Gamma_{IJ} \tilde \ve_{2)}\,,
\label{Rpara}
\fe
where
\ie
\tilde \ve\equiv {1\over 4}\Gamma^\m \p_\m \ve \,.
\fe
The generators $R^{IJ}$ with $I,J=5,6,7,8,9,0$ satisfy the $SO(6)_R$ commutation rules
\ie
&[R_{IJ},R_{KL}]=2\D_{L[J}R_{I]K}-2\D_{K[I} R_{J]L}\,,
\fe
and have the following matrix representations in the vector and spinor basis \ie
(R^{IJ})_K{}^L=2\D^{[I}_K \D^{J]L},\quad 
(R^{IJ})_\A{}^\B={1\over 2}(\tilde \Gamma^{IJ})_\A{}^\B\,.
\label{RMatrix}
\fe
Lastly $\Omega_\lambda$ is the dilation that acts with scaling factor $\lambda=2\ve_{(1} \tilde \ve_{2)}$.

 \subsection{The supersymmetric action for $\cN=4$ SYM}

 The $\cN=4$ SYM theory in four dimensions can be obtained from dimensional reduction of the 10d $\cN=1$ SYM.  We follow \cite{Pestun:2009nn} in using the notation from the 10d SYM and split the 10d Gamma matrices as $\Gamma_M=\{\Gamma_\m,\Gamma_I\}$ with $\m=1,\dots,4$ and $I=5,\dots,9,0$. The action for 4d $\cN=4$ SYM with gauge group $G$ on a general compact four manifold $\cM$ is \cite{Berkovits:1993hx,Pestun:2007rz}
 \ie
 S = -{1\over 2 g_4^2}\int_{\cM} d^4 x\, \sqrt{g} \tr \Bigg(
 {1\over 2}F_{MN}F^{MN}-\Psi \Gamma^M D_M\Psi  +{\cR\over 6} \Phi^I \Phi_I -K^m K_m
 \Bigg)\,,
 \label{SYMos}
 \fe
 where $\cR$ denotes the scalar curvature of $\cM$, and $K_m$ with $m=1,\dots 7$ are auxiliary fields which serve to give an off-shell realization of the supercharge that we will use to localize the theory. We adopt the convention of \cite{Pestun:2009nn} for the covariant derivative $D\equiv d+A$ and curvature $F_{MN}\equiv [D_M,D_N]$. The SYM fields expand as $A_M\equiv A_M^a T_a$ with real coefficients $A_M^a$ and \textit{anti-hermitian} generators $T^a$ of the Lie algebra $\mf{g}$ of the gauge group. The trace $\tr(\cdot,\cdot)$ corresponds to the Killing form of $\mf{g}$ and is related to the usual trace  in a particular representation $R$ by $\tr={1\over 2T_R}\tr_R$, where $T_R$ denotes the Dynkin index of $R$. For example for $\mf{g}=\mf{su}(N)$, the Killing form is identical to the trace in the fundamental representation $\tr=\tr_F$. Finally the generators $T^a$ are normalized by $\tr(T_a T_b)=-{1\over 2}\D_{ab}$. 
 
  We discuss the $\cN=4$ superconformal symmetries of \eqref{SYMos} below.
The SUSY transformations are
 \ie
 &\D_\ve A_M= \varepsilon \Gamma_M \Psi  \,,
 \\
 &\D_\ve\Psi ={1\over 2}F_{MN}\Gamma^{MN}\varepsilon+{1\over 2}\Gamma_{\m I}\Phi^I \nabla^\m \varepsilon+K^m \n_m \,,
 \\
 &\D_\ve K^m=-\n^m\Gamma^M D_M\Psi\,,
 \label{SUSYos}
 \fe
 where the conformal Killing spinor $\varepsilon$ is a 10d chiral spinor introduced in the previous section satisfying the killing spinor equation \eqref{CKV} which implies
 \ie
 \nabla_\m \varepsilon=\Gamma_\m \tilde{\varepsilon}\,,
 \qquad  \tilde\CC^\m \nabla_\m \tilde\varepsilon= -{1\over 12}\cR {\varepsilon} \,.
 \label{CKSeqn}
 \fe
 Here $\Gamma_\m=e_{\hat\m \m}\Gamma^{\hat \m}$, where $e_{\hat\m}^\m$ is the vielbein and  $\Gamma^{\hat M}$ denotes flat space 10d Gamma matrices in the chiral basis (we will not distinguish between $\Gamma^{\hat I}$ and $\Gamma^{ I}$ for Gamma matrices in the internal directions). The auxiliary 10d chiral spinors $\n^m$ with $m=1,\dots,7$ in \eqref{SUSYos} are chosen to satisfy
 \ie
\varepsilon \Gamma^M \n_m=0\,,~~ \n_m\Gamma^M\n_n=\D_{mn}\varepsilon \Gamma^M \varepsilon\,,~~\n^m_\A n^m_\B +\ep_\A \ve_\B ={1\over 2}\ve \CC_M \ve \tilde \CC^M_{\A\B}\,.
 \label{PSpinor}
 \fe
Furthermore the SYM action \eqref{SYMos} is invariant under the  Weyl transformation with parameter $\lambda$,
 \ie
 g_{\m\n}\to g_{\m\n}e^{2\lambda}\,,\quad A_\m \to A_\m  \,,\quad \Phi_A\to \Phi_A e^{-\lambda}\,,\quad \Psi\to \Psi e^{-{3\over 2}\lambda}\,,\quad K_m \to  K_m e^{-2\lambda}\,.
 \fe
 The conformal Killing spinors also transform as\footnote{The auxiliary pure spinors  transform as $\n_m \to e^{{1\over 2}\lambda} \n_m$.}
 \ie
 \varepsilon \to e^{{1\over 2}\lambda} \varepsilon\,,\qquad \tilde\varepsilon \to  e^{-{1\over 2}\lambda} \left( \tilde\varepsilon +{1\over 2} \CC^\m \pa_\m \lambda \varepsilon \right)\,,
 \label{ksWeyl}
 \fe
 such that \eqref{CKSeqn} is invariant and the SUSY transformations \eqref{SUSYos} are also preserved.
 
 Finally the action \eqref{SYMos} has $SO(6)_R$ R-symmetry which is generated by $R_{IJ}$ which act on the fields depending on their $SO(6)_R$ representations as in \eqref{RMatrix}.
 
 It is easy to compute   the anti-commutators of the supersymmetry transformations acting on the SYM fields. They take the following form\footnote{In writing this equation we take $\D_{\varepsilon_{1,2}}$ to be the on-shell supersymmetry transformation generators (turning off the auxiliary fields $K_m$ in \eqref{SUSYos}).}
 \ie
 \{\D_{\varepsilon_1},\D_{\varepsilon_2}\}=-2(\cL_{v}+R_\omega+\Omega_\lambda)-2\cG_\zeta+({\rm e.o.m})\,,
 \label{SUSYac}
 \fe
 in agreement with \eqref{sca} up to equation of motion and gauge transformation $\cG_\zeta$ with gauge parameter $\zeta=v^M A_M$, where  $v^M\equiv \varepsilon_{(1}\CC^M \varepsilon_{2)}$.

\subsection{Half-BPS superconformal defects and branes}
\label{sec:halfbpsalg}
Conformal defects of codimension $p$ in a $d$-dimensional CFT breaks the conformal group $SO(d,2)$ to (subgroups of) its maximal subgroup $SO(d-p,2)\times SO(p)$. In flat spacetime, the maximally symmetric conformal defect takes the shape of a $d-p$-dimensional plane or sphere related by conformal transformations. In supersymmetric theories, the defect conformal algebra can be further extended to  BPS subalgebras of the full superconformal algebra. Among them the maximally supersymmetric ones are half-BPS. We refer to such conformal defects preserving  half-BPS subalgebras as half-BPS superconformal defects. We'll comment on more general (conformal) defects in Section~\ref{sec:gendef}.

Half-BPS subalgebras of $\mf{psu}(2,2|4)$ are classified as centralizers of involutions in the $\mf{psu}(2,2|4)$ algebra \cite{DHoker:2008wvd}. We take the spacetime to be $\mR^4$ and consider involutions that fix a hyperplane (the other cases are related by conformal transformations).  An involution $\iota$ induces a reflection $\cP_\iota$ on the conformal killing spinors and the invariant supercharges satisfy
\ie
\ve =\cP_\iota \ve \,.
\label{ksproj}
\fe
Therefore classifying half-BPS subalgebras is equivalent to looking for $\cP_\iota$ that preserves the anti-commutation relation \eqref{sca} when restricted to the plane fixed by $\iota$.

It is easy to see that up to conjugation by the bosonic conformal group, such reflection matrices $\cP_{\iota}$ are simply given by one of the following 8 types
\ie
\cP_{\iota}= i\Gamma^{IJ },~\Gamma^{IJ KL},~\Gamma^{\m IJK},~\Gamma^{\m \n IJ},~
i\Gamma^{\m \n},~\Gamma^{\m \n\rho I},~i\Gamma^{\m I},~\Gamma^{\m \n\rho\sigma}.
\label{Projs}
\fe
We are interested in defects that are half-BPS in the original Lorentzian theory. This requires $\cP_\iota$ to be real after a Wick rotation of the $x^1$ direction. Since the gamma matrices are manifestly real in the 10d Majorana-Weyl basis, this condition exclusions two cases from \eqref{Projs} given by $i\Gamma^{IJ}$ and $\Gamma^{\m\n\rho\sigma}$.  

Below we will elaborate on each of the remaining 6 cases by identifying the corresponding defects in the 4d  $\cN=4$ SYM.~In the large $N$ limit, via AdS/CFT, such defects are realized by probe branes in $AdS_5\times S^5$ with metric,\footnote{Before the near horizon limit, these defects are realized by branes intersecting the stack of D3 branes that engineer the $\cN=4$ SYM.} 
\ie
ds^2={dx_\m dx^\m  +dy^I dy^I \over |y|^2} \,.
\label{AdS5S5m}
\fe
The $\cN=4$ superconformal symmetry is realized in the bulk by killing spinors on $AdS_5\times S^5$ 
\ie
\ve_{\rm AdS}(x,y)={1\over \sqrt{|y|}} (\ep_s+ (x^\m \Gamma_\m+ y^I \Gamma_I)\ep_c)\,,
\fe
and they are related to the conformal killing spinor \eqref{SYMCKS} on the boundary by taking the asymptotic limit
\ie
\lim_{|y| \to 0} \sqrt{|y|}\ve_{\rm AdS} = \ve\,.
\fe
In the IIB string theory, the probe D-brane   preserves a subset of the supersymmetries that satisfies the $\kappa$-symmetry constraint
\ie
\ve_{\rm AdS} = \Gamma_{\rm vol}\ve_{\rm AdS}\,,
\label{AdSqsproj}
\fe
with 
\ie
\Gamma_{\rm vol} \equiv {\ep^{\m_0\m_1\dots \m_p}\over (p+1)!\sqrt{G}} \Gamma_{M_0 M_1\dots M_p}\pa_{\m_0}X^{M_0}\pa_{\m_1}X^{M_1}\dots \pa_{\m_p}X^{M_p}\,.
\fe
in the absence of world-volume flux, where $G_{\m\n}$ is the induced metric on the brane and $X^M$ are the embedding coordinates.  The $\kappa$-symmetry constraint for BPS branes is naturally related to the boundary BPS defect condition \eqref{ksproj} by \cite{Park:2017ttx}
\ie
\lim_{|y| \to 0}\Gamma_{\rm vol}=\cP_\iota \Gamma_{1234}\,.
\fe
In Table~\ref{Table:bpssub}, we summarize the half-BPS defects in the $\cN=4$ SYM and the corresponding extended objects in IIB string theory. Below we give more details about each cases.

We start with the simpler and more familiar cases. The line defects arises at the fixed locus of $\iota$ at $x_1=x_2=x_3=0$ when 
 $\cP_\iota=i\Gamma^{4 I }$ or $\cP_\iota=\Gamma^{123 I }$. The corresponding half-BPS subalgebras are isomorphic in these cases. The former is realized by D5/NS5 branes while the latter is realized by D1/F1 branes.

The case $\cP_\iota= \Gamma^{IJ KL}$ corresponds a spacetime-filling defect or flavor brane and has two kinds of realizations. One is realized by an ALE instanton in IIB string theory longitudinal to the spacetime (T-dual to the NS5 brane).\footnote{More explicitly, for an ALE instanton with transverse directions $6,7,8,9$, we have $\cP_\iota=\CC_{6789}$. Note that \eqref{AdSqsproj} does not apply to this defect. The BPS condition for the ALE instanton is simply $\ve_{\rm AdS} = \cP_\iota\ve_{\rm AdS}$.
 } It introduces flavor symmetry and matter carrying the flavor symmetry to the 4d theory while breaking the $\cN=4$ supersymmetry to an $\cN=2$ subalgebra. Another (perhaps more familiar) flavor brane corresponds to D7 branes parallel to the spacetime. 

Next, we have codimension-one interfaces at $x_{1}=0$ from involution $\cP_\iota=\Gamma^{1 IJK }$ . They can be realized by D5 branes intersecting with the D3 branes along 3 longitudinal directions in the spacetime. The interface has 3d $\cN=4$ superconformal symmetry on its worldvolume which will be important for the boundary topological quantum mechanics (TQM) sector we identify in Section~\ref{sec:Qcoh}.

The codimension-two surface defects arise for the  case $\cP_\iota=i\Gamma^{12 }$ at $x_3=x_4=0$ and for $\cP_\iota=\Gamma^{12 IJ }$ at $x_1=x_2=0$. The former  gives rise to chiral surface defects with $\cN=(8,0)$ world-volume supersymmetry and can be realized by probe D7-branes intersecting the D3s \cite{Harvey:2008zz}. The latter gives non-chiral surface defects with $\cN=(4,4)$ supersymmetry and comes from probe D3-branes \cite{Gukov:2006jk}.

\begin{table}[htb]
	\centering
	\begin{tabular}{|c|c|c|c|c|c|}
		\hline
		Dimension &Involution & Symmetry & Branes & $n_{ND}$  
		\\	\hline
		4 &  $ \Gamma^{IJ KL }$  & $\mf{su}(2,2|2)\oplus \mf{su}(2)$  & D3/ALE & 4
		\\	\hline
		4 & $\Gamma^{IJ KL }$ & $\mf{su}(2,2|2)\oplus \mf{su}(2)$ &  D3/D7 & 4
		\\	\hline
		3 & $\Gamma^{\m IJK }$  &  $\mf{osp}(4|4,\mR)$ & D3/D5 & 4
		\\	\hline
		2 & $\Gamma^{\m\n IJ }$& $(\mf{psu}(1,1|2)\times\mf{psu}(1,1|2))\rtimes \mf{so}(2)  $ & D3/D3 & 4
		\\	\hline
		2 & $i\Gamma^{\m\n }$& $\mf{su}(1,1|4)\times\mf{su}(1,1)  $ & D3/D7 & 8
		\\	\hline
		1 & $\Gamma^{\m\n\rho I}$& $\mf{osp}(4^*|4)$ & D3/D1 & 4
		\\	\hline
		1 &$i\Gamma^{\m I }$& $\mf{osp}(4^*|4)$ & D3/D5 & 8
		\\	\hline
	\end{tabular}
	\caption{Half-BPS subalgebras of $\mf{psu}(2,2|4)$ fixed by involutions, and realizations by IIB branes. }
	\label{Table:bpssub}
\end{table}
In each case above, the $SL(2,\mZ)$ orbits share the same supersymmetry subalgebra (up to isomorphisms). For example, the D1-brane gets mapped to $(p,q)$-strings and similarly for 5-branes and 7-branes.

\subsection{More general defects}
\label{sec:gendef}

In the previous section we have focused on conformal defects in the $\cN=4$ SYM that preserve  the maximal symmetry at each particular codimension. Upon worldvolume deformations, they give rise to large classes of general defects in the SYM, preserving a subalgebra of the corresponding half-BPS algebra.

Such deformations may come from putting the half-BPS conformal defect on a less symmetric submanifold, or turning on symmetry breaking interactions on the defect (typically one needs to combine these deformations to preserve a subset of the supercharges). For example, they include the generalized (not necessarily supersymmetric) Wilson line operators \cite{Maldacena:1998im} in $\cN=4$ along a general loop (or infinite line) $\cC$ in the spacetime,
\ie
W_R(\cC)\equiv \tr_R {\text{Pexp}\,}\oint_\cC ds \left( A_\m(x) {d x^\m\over ds}+ \theta^I(s) \Phi^I(x) \sqrt{g_{\m\n}{d x^\m\over ds}{d x^\n\over ds}}\right)
\fe
with $\theta^I(s)$ specifying the couplings of the Wilson line operator to the SYM scalar fields. In particular the half-BPS Wilson lines preserving $\mf{osp}(4^*|4)$ are for example given by $\cC$ equal to a straight line in $\mR^4$ and $\theta^I=\D^I_5$. The general Wilson lines can be obtained by marginal deformations on the worldvolume of the half-BPS line corresponding to $\D \theta^I$ as well as deformation of the curve $\D \cC$ by the displacement operator. For special cases of $(\theta^I(s),\cC)$, one can preserve a nonempty subset of the supercharges in the full half-BPS superalgebra $\mf{osp}(4^*|4)$, corresponding to ${1\over 16},~{1\over 8}$ and ${1\over 4}$-BPS Wilson loops. This was analyzed in detail in \cite{Drukker:2007qr}. Under $SL(2,\mZ)$ duality, the Wilson loop operators are mapped to disorder type line operators in the SYM, known as 't Hooft or more generally dyonic line operators. They are specified by codimension-three singularities (boundary conditions) of the SYM fields \cite{Kapustin:2005py}. Once again, for specific forms of the singularity, we obtain half-BPS 't Hooft (dyonic) lines whereas the more general ones are obtained by deforming the locus of the singularity as well as introducing boundary condition changing couplings along the singularity. The same discussion applies to lower-codimension defects, i.e. surfaces and interfaces. In the rest of the paper, we will focus on   defects obtained from deforming the half-BPS ones that are still invariant under a  certain supercharge $\cQ$ (and its conjugate). As we will explain, such defect observables can be analyzed using the localization technique.
 
Another interesting generalization is to consider conformal defects that transform nontrivially under the half-BPS subalgebras of the relevant codimension. They include the spinning conformal defects of \cite{Kobayashi:2018okw,Lauria:2018klo} and supersymmetric generalizations. We will not discuss such defects in this paper.

\section{Conformal Defects in the $\cQ$ Cohomology}
\label{sec:Qcoh}
\subsection{Review of the 2d sector}
\label{sec:review2d}
In the $\cN=4$ superconformal algebra $\mf{psu}(2,2|4)$, we will now denote the Poincar\'e supercharges as $Q_{\A }^A,\tilde Q_{\dot\A A}$ and superconformal charges as $S_{ A}^\A, \tilde S^{\dot \A A }$. Here $(\A,\dot \A)$ are the $\mf{su}(2)_L\times \mf{su}(2)_R$ spinor indices and $A=1,2,3,4$ is the fundamental (anti-fundamental) indices for $\mf{su}(4)_R$ symmetry. For convenience, we will split the 10d Gamma matrices as $\C_\m=\Gamma_\m$ for $\m=1,2,3,4$ and $(\rho_1,\rho_2,\rho_3,\rho_4,\rho_5,\rho_6)= (\CC_7,\CC_9,\CC_0,\CC_5,\CC_6,\CC_8)$. We refer the readers to Appendix~\ref{app:4dsca} for our conventions for the $\cN=4$ superconformal algebra.

The $\cN=4$ SYM with gauge group $G$ on $\mR^4$  contains a nontrivial 2d sector \cite{Drukker:2007yx,Drukker:2007qr} on an $S^2$ of radius $2R$ at\footnote{This is chosen such that, after the Weyl transformation \eqref{Weyltransform}, it corresponds to an $S^2$ of radius $R$ inside an $S^4$ of the same radius.}
\ie
x_4=0,\quad \sum_{i=1}^3 x_i^2=4R^2\,,
\fe
which is invariant under the $SO(3)$ isometry of the $S^2$ as well as a transverse rotation generated by a combination of translation and special conformal transformations in the $x_4$ direction
\ie
M_\perp \equiv {1\over 2}(K_4-P_4)\,.
\fe
For convenience we will set $R={1\over 2}$ for most of the analysis below and only restore the units when necessary.

By studying the supersymmetric transformations of the SYM fields, one find that for certain observables on the $S^2$, the bosonic symmetries (twisted by certain generators in the $SO(6)_R$ R-symmetry group) extend to invariance under an $\mf{su}(2|1) \times \mf{su}(2)_{\rm diag} $ subalgebra of $\mf{psu}(2,2|4)$ (whose bosonic subalgebra contains the R-symmetry twisted versions of $\mf{so}(3)\times \mf{u}(1)$ isometry in addition to the $\mf{so}(3)_{568}$ R-symmetry). In terms of the supersymmetry parameters  $\ep_s$ and $\ep_c$ in \eqref{SYMCKS}, this subalgebra is specified by the following projectors on $\ep_s$ \cite{Giombi:2009ds}
\ie
&(\C_{ij}+\rho_{ij})\ep_s=0 \,,
\fe
with $i,j=1,2,3$ (note that only two of the three equations are independent) and the relation
\ie
\ep_c=-i   \rho_{123}   \ep_s\,.
\fe
These constraints ensure that the twisted YM connection restricted to the $S^2$
\ie
\cA\equiv  A +i\ve_{ijk} \phi_i {x^k }  dx^j\,,
\label{tA}
\fe
where for convenience we define
\ie
(\phi_1,\phi_2,\phi_3)\equiv (\Phi_7,\Phi_9,\Phi_0)\,,
\label{phiPhi}
\fe
is invariant under the corresponding supersymmetry transformation $\D_\ve$, so are the ${1\over 8}$-BPS Wilson loops in \cite{Drukker:2007qr}
\ie
W_R(C)={1\over d_R}\tr_R {\rm Pexp\,}\oint_C  \cA\,.
\label{18WL}
\fe
Solving these constraints, we find explicit generators of the $\mf{su}(2|1)$ algebra\footnote{Here the Pauli matrices are defined as the usual ones regardless of the position of the indices.}
\ie
{\bm Q}_1 \equiv  & 
i\D^{\A \dot a}Q_{\A 1 \dot a}+\sigma_1^{\A \dot a}Q_{\A 1 \dot a}
+
 i\D^{\A \dot a}S_{\A 1 \dot a}- \sigma_1^{\A \dot a}S_{\A 1 \dot a} \,,
\\
  {\bm Q}_{ 2}\equiv & 
  -\D^{\A \dot a}Q_{\A 2 \dot a}-i\sigma_1^{\A \dot a}Q_{\A 2 \dot a}
    -\D^{\A \dot a}S_{\A 2 \dot a}+i\sigma_1^{\A \dot a}S_{\A 2 \dot a} \,,
\\
\tilde {\bm Q}_1\equiv & 
-(\sigma_3)^{\A \dot a}\tilde Q_{\A 1 \dot a}-\sigma_2^{\A \dot a}\tilde Q_{\A 1 \dot a}
-
(\sigma_3)^{\A \dot a}\tilde S_{\A 1 \dot a}+ \sigma_2^{\A \dot a}\tilde S_{\A 1 \dot a} \,,
\\
\tilde{\bm Q}_{  2} \equiv & 
i(\sigma_3)^{\A \dot a}\tilde Q_{\A 2 \dot a}-i\sigma_2^{\A \dot a}\tilde Q_{\A 2 \dot a}
-
i(\sigma_3)^{\A \dot a}\tilde S_{\A 2 \dot a}+i\sigma_2^{\A \dot a}\tilde S_{\A 2 \dot a} \,,
\label{su21scs}
\fe
where ${\bm Q}_a$ and $\tilde {\bm Q}_{ a}$ transform as doublets under the  $SO(3)_{568}$ R-symmetry generated by $(R_1,R_2,R_3)=-i(R_{58},R_{68},R_{56})$, and carry $\pm {1\over 2}$ charges under $R_3=R_{56}$. 

They satisfy the following (anti)commutation relations
\ie 
&\{{\bm Q}_a,{\bm Q}_b\}=-\{\tilde{\bm Q}_{  a},\tilde{\bm Q}_{  b}\}=4 i (\sigma^i)_{ab}  R_i,\quad 
\{{\bm Q}_a, \tilde{\bm Q}_{ b}\}=-4\ep_{ab}  M_\perp,\\
&[M_\perp, {\bm Q}_a]= {i\over 2}\tilde {\bm Q}_a,\quad [M_\perp,\tilde  {\bm Q}_a]= {i\over 2}{\bm Q}_a\,.
 \fe
 In addition, the twisted connection \eqref{tA} and the associated Wilson loops \eqref{18WL} are also invariant under the diagonal subalgebra $\mf{su}(2)_{\rm diag}\subset\mf{su}(2)_{790}\oplus \mf{su}(2)_{S^2}  $ generated by
  \ie
  \mf{su}(2)_{\rm diag}:\quad  (\hat M_{12},\hat M_{13}, \hat M_{23})=(M_{12}- R_{79},M_{13}-R_{70},M_{23}-R_{90}) \,,
  \label{twistsu2}
  \fe
which commute with the generators of the $\mf{su}(2|1)$ subalgebra, and come from anti-commutators that involve any of the four supercharges $({\bm Q}_a,\tilde {\bm Q}_{ a})$  and other $\mf {psu}(2,2|4)$ fermionic generators outside the $\mf {su}(2|1)$ subalgebra.   

The fact that the $\mf{su}(2)_{\rm diag}$ symmetry is $\bm Q$-exact  has the following implication. The correlation function of $1\over 8$-BPS Wilson loops \eqref{18WL} on the $S^2$, which are individually $\bm Q$-closed,
\ie
\la W_{R_1}(\cC_1)W_{R_2}(\cC_2)\cdots W_{R_n}(\cC_n)\ra
\fe
remains unchanged if we act by $\mf{su}(2)_{\rm diag}$ on any collection of the Wilson loop insertions $W_{R_i}(\cC_i)$, as long as it does not change the topology of the intersecting graph of the loops $(\cC_1,\cC_2,\dots,\cC_n)$.
  
It is natural to expect that for SYM observables on $S^2$ that are built out of the twisted connection $\cA$ in \eqref{tA}, there is an emergent 2d quantum field theory that computes their expectation values. Indeed, by studying the perturbative expansion of the expectation value of the $1\over 8$-BPS Wilson loops \eqref{18WL} in the SYM, it was argued in \cite{Drukker:2007qr} that the 2d theory is a bosonic Yang-Mills theory on $S^2$ 
\ie
S_{\rm YM}={1\over 4g_{\rm YM}^2} \int_{S^2} d\sigma \sqrt{g} \tr \cF^2\,,
\label{S2d}
\fe
with gauge group $G$ and field strength $\cF\equiv d\cA+\cA \wedge \cA$. The 2d YM coupling is given by
\ie
g_{\rm YM}^2=-{g_{4}^2\over 2 \pi R^2 } 
\fe
which implies $g_{\rm YM}$ is imaginary.
This was later made more precise by a localization computation in \cite{Pestun:2009nn} on $S^4$ (related to the $\mR^4$ by the simple stereographic map), where the  the 2d YM lives on a great $S^2$ in $S^4$. The localization computation of \cite{Pestun:2009nn} requires a supercharge $\cQ$ in $\mf{su}(2|1)$ with the property that
\ie
\cQ^2=- 2\hat M_\perp,\quad 
\hat M_\perp
\equiv 
M_\perp - R_{56}
\fe
and thus nilpotent on $S^2$ (when acting on fields uncharged under $R_{56}$). Up to conjugation, we can take
\ie
\cQ={i\over 2\sqrt{2}}\left( {\bm Q}_1+i\tilde {\bm Q}_{1} +i{\bm Q}_{2}-\tilde  {\bm Q}_{2}\right)
\label{LocQdef}
\fe 
The 2d YM connection $\cA$ in \eqref{tA} arises naturally from studying $\cQ$-cohomology at the level of the (gauge-variant) SYM fields: the smooth solutions to the BPS equation $\cQ \Psi=0$ (with $\Psi=0$)  are parametrized by $\cA$ on the $S^2$ (and determined elsewhere by certain elliptic differential equations as well as covariance along the vector field corresponding to $\hat M_\perp$). The 4d SYM action \eqref{SYMos} on $S^4$ reduces on the BPS locus to the 2d YM action \eqref{S2d} on $S^2$. From now on we will naturally refer to this $S^2$ as $S^2_{\rm YM}$.
 
This localization setup allows extraction of observables of the SYM in the $\cQ$-cohomology. In general such objects may not be of order type (i.e. written in terms of $\cA$ such as the ${1\over 8}$-BPS Wilson loops). Instead they may be of disorder type, and give rise to singularities (or boundary conditions) for $\cA$ on certain submanifolds of the  $S^2_{\rm YM}$.

The previously known SYM observables in the 2d sector consists of the  ${1\over 8}$-BPS Wilson loops \eqref{18WL} and ${1\over 8}$-BPS local operators on  $S^2_{\rm YM}$ \cite{Giombi:2009ds}, as well as ${1\over 2}$-BPS 't Hooft loops on a great circle that links with the  $S^2_{\rm YM}$ in $S^4$ (or along the $x_4$ axis on $\mR^4$) \cite{Giombi:2009ms}. In the next section, we discuss general defect observables in the $\cQ$-cohomology.

We will find useful the following constraints satisfied by the constant spinors parametrizing $\cQ$,
\ie
\boxed{
\Gamma_{1890}\ep_s=\Gamma_{1279}\ep_s=\Gamma_{1370}\ep_s=\Gamma_{2390}\ep_s=-\Gamma_{1238}\ep_s=\ep_s,\quad \ep_c=-i   \Gamma_{790}   \ep_s
}\,,
\label{Qspinor}
\fe
which amounts to four independent commuting projectors that determines the $\mf{su}(1|1)$ subalgebra generated by $\cQ$.   

\subsubsection*{Comparison to the localization supercharge in \cite{Pestun:2007rz}}
Recall that the localization in \cite{Pestun:2007rz} relies only on the massive $\cN=2$ subalgebra 
\ie
\mf{osp}(4|2)\subset \mf{su}(2,2|2)\subset \mf{psu}(2,2|4)\,.
\fe 
Here the relevant $\mf{osp}(4|2)$ sub-algebras are parametrized in terms the constant spinors satisfying the constrains
\ie
\CC_{5690}\ep_s=\pm\ep_s,\quad \ep_c=-i\CC_{178} \ep_s\,.
\fe
and labelled as $\mf{osp}(4|2)_\pm$.

The supercharge $\cQ$ in $\mf{psu}(2,2|4)$ is not contained in either of the  $\mf{osp}(4|2)_\pm$ subalgebras.
Rather, by projecting to the $\pm1 $ eigenspace of $\CC_{5690}$, $\cQ$ decomposes into two supercharges in $\mf{osp}(4|2)_+$ and  $\mf{osp}(4|2)_-$ respectively,
\ie
\cQ =\cQ_{\cN=2}^+ +\cQ_{\cN=2}^-\,,
\fe
which satisfy
\ie 
&\{\cQ^+_{\cN=2},\cQ^+_{\cN=2}\}=-M_\perp-M_{23}+R_{56}+R_{89}\,,
\\
&\{\cQ^-_{\cN=2},\cQ^-_{\cN=2}\}=-M_\perp+M_{23}+R_{56}-R_{89}\,,
\\
&\{ \cQ^+_{\cN=2},\cQ^-_{\cN=2}\}=0\,.
\fe
These $\cN=2$ supercharges $\cQ^\pm_{\cN=2}$ are precisely the ones (chiral and anti-chiral versions thereof) used in \cite{Pestun:2007rz} to localize $\cN=4$ SYM on $S^4$ to a zero-dimensional Gaussian matrix model.

\subsection{General defects in the 2d sector}
In this section, we study general defect observables of the $\cN=4$ SYM in the $\cQ$-cohomology of the \textit{scalar} type (in the sense of \cite{Kobayashi:2018okw}). We take the spacetime to be $\mR^4$ for simplicity. Since   $\cQ$ squares to the sum of the vector field $M_\perp=P_4-K_4$ and R-symmetry rotation $R_{56}$. The defect observable must be defined on a submanifold $\cD_d\subset \mR^4$ of dimension $0\leq d\leq 3$ that is preserved by $M_\perp $ and the bulk-defect coupling must be invariant under $R_{56}$. Since the vector field $M_\perp$ is complete and non-vanishing everywhere except for the  $S^2_{\rm YM}$ submanifold, an $M_\perp$-preserving submanifold  $\cD_d$ can be described by its cross-section $\pi(\cD_d)$ at $x_4=0$, which is one-dimension lower if $\pi(\cD_d) \not\subset S^2$ and of the same dimension as $\cD_d$ if $\pi(\cD_d)   \subset S^2$. Furthermore, on the $x_4=0$ slice, the exterior of the  $S^2_{\rm YM}$ is connected to the interior three-ball $B^3$ by flow lines of $M_\perp$, consequently  it suffices to specify the intersection $\pi^{\rm in}(\cD_d)\equiv \cD_d \cap \bar B^3$ between $\cD_d$ and the closure of the three-ball.

\subsubsection{Point-like defects}
For $d=0$, namely a point-like defect, to preserve $M_\perp$, the defect insertion has to lie on the $S^2_{\rm YM}$ where the vector field $M_\perp$ vanishes,
\ie
\cD_0 \subset S^2_{\rm YM}\,.
\fe
They correspond to local operator insertions of the ${1\over 8}$-BPS type on the $S^2_{\rm YM}$ \cite{Giombi:2009ds} (recall \eqref{phiPhi}),
\ie
\cO_p(x^i) \equiv\tr \left({x^i} \phi_i+i \Phi_8\right)^p\,,
\label{pointd}
\fe
which preserves the following subalgebra of $\mf{psu}(2,2|4)$,
\ie
(\mf{psu}(1|1)\oplus \mf{psu}(1|1)) \oplus    \widehat{\mf{so}(2)}_{\perp} 
\fe
where the last $ \widehat{\mf{so}(2)}_{\perp} $ is generated by $\hat M_{\perp}$, and gives the central extension of the two $\mf{psu}(1|1))$ factors. It contains $\mf{psu}(1|1)$  algebra generated by $\cQ$ as a diagonal subalgebra of $ \mf{psu}(1|1)\oplus \mf{psu}(1|1) $.
 
 In terms of the 16-component spinors $\ve_s,\ve_c$ parametrizing the conformal killing spinor, the supercharges in this subalgebra are given by 
 \ie
  &(\C_{ij}+\rho_{ij})\ep_s=0,\quad \ep_c=-i \C_1  \rho_{14} \ep_s=-i \Gamma_{178}\ep_s\,,
 \fe 
which amounts to two independent projectors on $\ep_s$ while $\ep_c$ is completely determined by $\ep_s$.
 
As a consequence of the $\cQ$-exact twisted $SU(2)$ rotations \eqref{twistsu2}, correlations functions in the $\cQ$-cohomology involving the local operator \eqref{pointd} are independent of their locations on $S^2_{\rm YM}$ as long as they don't move across other insertions.
 
The symmetry is enhanced for special values of $x_i$ on $S^2_{\rm YM}$. In particular at $x=(1,0,0,0)$
\ie
\cO_p= \tr (\Phi_7+i \Phi_8)^p
\fe
which preserves the following half-BPS subalgebra\footnote{Precisely this is the subalgebra preserved by $\cO_p$ at $x=(1,0,0,0)$ and its conjugate $\bar \cO_p$ at $x=(-1,0,0,0)$.} 
\ie
(\mf{psu}(2|2)\oplus \mf{psu}(2|2)) \oplus \widehat{\mf{so}(2)}_{78}\supset \mf{so}(4)_{\rm rot} \oplus  \mf{so}(4)_{5690}  
\oplus \widehat{\mf{so}(2)}_{78}
\fe
where $\widehat{\mf{so}(2)}_{78}$ is generated by $D-R_{78}$ which gives the central extension of the two $\mf{psu}(2|2)$ factors. The $\mf{so}_{\rm rot}(4)$ factor is generated by rotations
\ie
M_{23},~M_{24},~M_{34},~P_2-K_2,~P_3-K_3,~P_4-K_4\,,
\fe
which split into two commuting $\mf{su}(2)$ algebras $\{L_1,L_2,L_3\}$ and $\{\bar L_1,\bar L_2,\bar L_3\}$
\ie
&[L_i,L_j]=2\epsilon_{ijk} L_k,\quad [\bar L_i, \bar L_j]=2\epsilon_{ijk} \bar L_k\,.
\fe
The supercharges are
\ie
Q_{\A 1 \dot a}+  i(\sigma_2)^{\dot \A}_\A\bar S_{\dot \A 1 \dot a}\,,
\quad
\bar  Q_{\A 2 \dot a}+i(\sigma_2)^{\dot \A}_\A  S_{\dot \A 2 \dot a}\,,
\fe
and
\ie
Q_{\A2 \dot a}-i  (\sigma_2)^{\dot \A}_\A\bar S_{\dot \A 2 \dot a}\,,
\quad
\bar  Q_{\A 1 \dot a} -i  (\sigma_2)^{\dot \A}_\A S_{\dot \A 1 \dot a}\,,
\fe
or equivalently
 in terms of the 16-component spinors $\ep_s,\ep_c$  
\ie
& \ep_c=-i \C_1  \rho_{14} \ep_s\,.
\fe

\subsubsection{Line defects}
\label{sec:linedefects}
At $d=1$, we have two possibilities. Either (I)~$\cD_1 \subset S^2_{\rm YM}$, or (II)~$\cD_1$ is the orbit of $M_\perp$ through a point   $\pi^{\rm in}(\cD_1) \in B^3$ (namely $(\vec x,x_4=0)$ with $|\vec x|<1$), which is a circle of radius
\ie
{1-|\vec x|^2\over 2|\vec x| }\,,
\fe
that links with the $S^2_{\rm YM}$. We will denote the two types of line defects by $\cD_1^{\rm I}$ and $\cD_1^{\rm II}$ respectively.

 The type $\cD_1^{\rm I}$ defects are given by $1\over 8$-BPS Wilson loops \eqref{18WL} on $S^2_{\rm YM}$ at $\sum_{i=1}^3 (x^i)^2=1$
\ie
W_R(\cD_1^{\rm I})\equiv \tr_R {\text{Pexp}\,}\oint_{\cD_1^{\rm I}} \left( A+ {i }\phi^k  \epsilon_{ijk} {x^j} dx^i \right)
\fe
where $\phi_k\equiv (\Phi_7,\Phi_9,\Phi_{0})$ as reviewed in the Section~\ref{sec:review2d}. These Wilson loops preserves the following subalgebra of $\mf{psu}(2,2|4)$
\ie
 \mf{su}(2|1)  \oplus \mf{su}(2)_{\rm diag}\supset (\mf{so}(3)_{568} \oplus  \widehat{\mf{so}(2)}_{\perp} ) \oplus \mf{su}(2)_{\rm diag}\,.
\fe
The $\mf{su}(2|1)$ algebra is generated by the supercharges $({\bm Q}_a,\tilde{\bm Q}_a)$ in \eqref{su21scs},  
and $\mf{su}(2)_{\rm diag}$ is generated by $\cQ$-exact twisted rotations on the $S^2_{\rm YM}$. By the $SL(2,\mZ)$ duality of the theory, we also expect there to be 't Hooft (and dyonic) loop operators on the $S^2_{\rm YM}$ but we will discuss the details elsewhere.\footnote{Note that the modular S-transform generally does not preserve the observables in the $\cQ$-cohomology  (since $\cQ$ transforms).}

The special Wilson loop defect of type $\cD_1^{\rm I}$ along a great circle, e.g. at $x_1=0$,
\ie
W_R(\cD_1^{\rm I})\equiv \tr_R {\text{Pexp}\,}\oint_{x_1=x_4=0,\,x_2^2+x_2^2=1} \left( A+ {i }\Phi_7  (x_2 dx_3-x_3 dx_2) \right)\,,
\label{hBWL}
\fe
 enjoys the enhanced symmetry under 
\ie
\mf{osp}(4|4)  \supset  \mf{so}(5)_{56890} \oplus  \mf{sl}(2)_{\perp } \oplus \mf{sl}(2)_{\parallel} \,,
\fe
where $\mf{so}(5)_{56890}$ denotes the R-symmetry subgroup that preserves the Wilson loop, and $\mf{sl}(2)_{\perp }$ and  $\mf{sl}(2)_{\parallel }$ are transverse and longitudinal conformal symmetries generated by $\{M_{14},K_1+P_1,K_4+P_4\}$ and  $\{M_{23},K_2+P_2,K_3+P_3\}$ respectively.
The supercharges generating $\mf{osp}(4|4)$ are parametrized by
 \ie
\ep_c=-i   \CC_{237}   \ep_s\,.
\fe

For type $\cD_1^{\rm II}$  defects, they are given by Wilson loops of the form\footnote{This is similar to the ${1\over 4}$-BPS Wilson loops in 5d $\cN=2$ SYM discussed in Appendix C of  \cite{Mezei:2017kmw}.} 
\ie
W_R(\cD_1^{\rm II})\equiv \tr_R {\text{Pexp}\,}\oint_{\cD_1^{\rm II}} \left( A+      {\dot x_\mu   \over \dot x \cdot v} v^I \Phi^I dx^\mu  \right)\,,
\fe
where $v^M=(v^\m,v^I)=\ve \Gamma^M \ve$. Here $\D_\ve W_R(\cD_1^{\rm II})=0$ follows from the identity
\ie
v^M \Gamma_M \ve=v^\m \Gamma_\m \ve+ v^I \Gamma_I \ve =0\,.
\fe
These Wilson loops are ${1\over 4}$-BPS and invariant under
\ie
  \mf{psu}(1|1)^4_{\rm c.e.}    \rtimes  \mf{so}(4)_{7890}  \supset  \widehat{\mf{so}(2)}_{\perp}   \oplus  \mf{so}(4)_{7890}  \,,
\fe
where $\mf{psu}(1|1)^4_{\rm c.e.} $ denotes four copies of $\mf{psu}(1|1)$ centrally extended by a common $\widehat{\mf{so}(2)}_\perp$. The $\mf{so}(4)_{7890}$ R-symmetry subgroup acts on these four copies of $\mf{psu}(1|1)$ as $   (\bm 2,\bm1)\oplus (\bm 1,\bm 2)$. In terms of the constant spinors, the preserved supercharges are determined by
 \ie
(\Gamma_{4}-i\Gamma_5)\ep_s=0,\quad \ep_c=i\CC_6\ep_s\,.
 \fe

Now there are also type $\cD_1^{\rm II}$  defects of the disorder type, given by 't Hooft loops. Here we focus on 
the special case with $\pi^{\rm in}(\cD_1^{\rm II})$ located at the center of the $B^3$ (so that $\cD_1^{\rm II}$ corresponds to a straight infinite line in $\mR^4$)  \cite{Giombi:2009ek}, leaving the general analysis to a future publication. The 't Hooft loop of \cite{Giombi:2009ds} is half-BPS with the symmetry algebra
\ie
\mf{osp}(4|4)  \supset \mf{so}(5)_{56790} \oplus \mf{so}(3)_{123}\oplus \mf{sl}(2)\,,
\fe
where $\mf{so}(5)_{56790}$ is the R-symmetry subgroup that preserves the half-BPS 't Hooft loop (which only couples to one of the six scalars $\Phi^8$), $\mf{so}(3)_{123}$ denotes the transverse spacetime rotation group, and $\mf{sl}(2)$ the conformal group  longitudinal to the defect. The 't Hooft loop is defined by a singularity of the SYM fields along the contour $\cD_1^{\rm II}$, 
 \ie
 F_{ij}(y)=&{1\over 2} \ep_{ijk} {(y_k-x_k)\over |y-x|^3} T_{\vec m} + {\rm regular} \,,
 \\
 \Phi_8(y)=&-{1\over 2|y-x| } T_{\vec m} + {\rm regular} \,,
 \fe
 and $T_{\vec m}$ denotes a Cartan element of the Lie algebra $\mf{g}$ for the gauge group $G$. For $G=U(N)$, we write
 \ie
 T_{\vec m}=-i{\rm diag}(m_1,\dots, m_N)\,.
 \fe
 One can easily check that the above configuration solves the BPS condition,
 \ie
 \D_\ve\Psi =&{1\over 2}F_{MN}\Gamma^{MN}\varepsilon+{1\over 2}\Gamma_{\m A}\Phi^A \nabla^\m \varepsilon 
\\
=&  \left[{1\over 2 |\vec x|^3} x^k \Gamma^k \left(     \Gamma^{123} +
\Gamma^{ 8}
\right) \ep_s
+
{1\over 2 |\vec x|^3}(|\vec x|^2- x^k  x^4\Gamma^k  \CC^4) \left(     \Gamma^{123} +
\Gamma^{ 8}
\right)  \ep_c \right]  T_{\vec m}
=0\,,
\fe 
as long as the constant 16-component spinors satisfy
\ie
\Gamma^{1238}\ep_s=-\ep_s,\quad \Gamma^{1238}\ep_c=-\ep_c\,,
\fe
which is indeed the case for the spinors parametrizing the supercharge $\cQ$ satisfying \eqref{Qspinor}.

Let us briefly comment on the relation between the half-BPS 't Hooft loop and the familiar half-BPS Wilson loop in light of the S-duality of $\cN=4$ SYM.
Recall  $SL(2,\mZ)$ acts on the $\cN=4$ superconformal algebra as an outer-automorphism 
 \ie
 (Q,\tilde S) \to  \left( c\tau +d\over c\bar\tau +d \right)^{1\over 4} (Q,\tilde S),\quad 
  (\tilde Q, S) \to  \left( c\tau +d\over c\bar\tau +d \right)^{-{1\over 4}} (\tilde Q, S)\,.
  \label{susysl2}
 \fe
 In particular, for $\tau={4\pi i \over g_{\rm YM}^2}$ and under the S-transform (which is a chiral rotation by $\pm \pi/2$ and does not preserve $\ve$) we have 
 \ie
 \ep_s\to e^{{\pi i\over 4}\CC_{1234}} \ep_s,\quad  \ep_c\to e^{-{\pi i\over 4}\CC_{1234}} \ep_c\,.
 \fe
The  supercharges preserved by the 't Hooft are thus mapped to
\ie
\Gamma^{48}\ep_s=i\ep_s,\quad \Gamma^{48}\ep_c=-i\ep_c\,,
\fe
which are precisely the BPS conditions for supercharges preserved for the half-BPS Wilson line along the $x_4$ axis
\ie
W_R \equiv \tr_R {\text{Pexp}\,}\oint_{x_i=0} \left( A+      {i }
  \Phi^8
dx^4  \right)\,.
\fe
Note that while this half-BPS Wilson loop is not in the $\cQ$-cohomology (for our chosen $\cQ$ \eqref{LocQdef}), it is  related by conformal and $SO(6)_R$ symmetry transformation to the half-BPS Wilson loop  \eqref{hBWL} which is $\cQ$-closed.

\subsubsection{Surface defects}
At $d=2$, we have three possibilities for the worldvolume submanifold $\cD_2$. Either (I)~$\cD_2  =S^2_{\rm YM}$ , or (II)~$\cD_2$ is generated by flowlines of $M_\perp$ through a curve  $\pi^{\rm in}(\cD_2)\subset \bar B^3$ which intersects with the $S^2_{\rm YM}$ boundary at isolated points, or (III)~$\cD_2$ is generated by $M_\perp$ from a curve $\pi^{\rm in}(\cD_2)\subset B^3$. We will denote these defects by $\cD_2^{\rm I}$, $\cD_2^{\rm II}$ and $\cD_2^{\rm III}$ respectively. Embedded in $\mR^4$, these defects have the topology of $S^2$, $D^2_\infty$ (disk of infinite size) and $T^2$ respectively.

Supersymmetric surface defects in $\cN=4$ SYM are specified by codimension two singularities of the gauge fields and adjoint scalars \cite{Gukov:2006jk,Gukov:2008sn}.\footnote{Here we have focused on the half-BPS surface defects in the $\cN=4$ SYM. See \cite{Alday:2009fs,Gaiotto:2009fs,Alday:2010vg,Gadde:2013dda,Gaiotto:2013sma,Gomis:2014eya,Frenkel:2015rda,Pan:2016fbl,Gomis:2016ljm,Ashok:2017odt,Gorsky:2017hro,Ashok:2017lko,Nekrasov:2017rqy,Jeong:2018qpc,Ashok:2018zxp,Ashok:2019rwa} for surface defects in general $\cN=2$ supersymmetric gauge theories.} We review the description of the half-BPS surface defect along a two-dimensional loci $\Sigma$ in the $U(N)$ SYM below. We first define a complex scalar field ${\bf \Phi}$ (a combination of two of the six  scalars $\Phi_I$). In the local normal bundle to $\Sigma$, we take the transverse distance to be $r$ and polar angle  to be $\psi$. The scalar field ${\bf \Phi}$ acquires the following singularity
\ie
{\bf \Phi}={{\rm Diag}[(\B_1+\C_1)\otimes 1_{N_1},(\B_2+\C_2)\otimes 1_{N_1},\dots, (\B_k+\C_k)\otimes 1_{N_k } ] \over z}\,,
\fe
with $z=r e^{i\psi}$ the complex coordinate in the local normal bundle fiber direction. This breaks the full gauge symmetry to the Levi subgroup $L=\prod_{i=1}^k U(N_i) \subset U(N) $. The gauge field  in the vicinity of the defect takes the form
\ie
A={{\rm Diag}[ \A_1 \otimes 1_{N_1},\A_2\otimes 1_{N_1},\dots,  \A_k \otimes 1_{N_k} ]}d\psi 
\fe
with $\{\A_i \otimes 1_{N_i}\}$ taking values in the maximal torus ${\mathbb T}^N={\mR^N/\mZ^N}$ of $U(N)$. Furthermore we can decorate the defect with 2d theta terms
\ie
\exp \left(
i\sum_{i=1}^k \eta_i \int_{\Sigma} \tr F^{(i)}
\right)\,,
\fe
where 
\ie
\eta={{\rm Diag}[ \eta_1 \otimes 1_{N_1},\eta_2\otimes 1_{N_1},\dots,  \eta_k \otimes 1_{N_k} ]}\,,
\fe
so $\eta$ naturally takes value in the maximal torus of the Langlands dual of $G=U(N)$ which is $G^\vee=U(N)$. Together the quadrupole $(\A_i, \B_i,\C_i,\eta_i)$ furnish the parameters that specify the half-BPS surface defect. In the conformal setting (e.g. when $\Sigma$ is a plane or a sphere), the defect SCFT on $\Sigma$ for the half-BPS surface defect has $\cN=(4,4)$ supersymmetry. The parameters $(\A_i, \B_i,\C_i,\eta_i)$ correspond to the conformal moduli of the 2d SCFT.

Let's now identify these surface defects in the $\cQ$-cohomology of the $\cN=4$ SYM. We will focus on the type $\cD_2^{\rm I}$ and $\cD_2^{\rm II}$ defects here. It would be interesting to see whether there are realizations of surface defects of type $\cD_2^{\rm III}$ in the $\cN=4$ SYM.

 The type $\cD_2^{\rm I}$ defects are given by ${1\over 2}$-BPS surface defects on the  $S^2_{\rm YM}$. It preserves the subalgebra 
\ie
(\mf{psu}(2|2)\oplus \mf{psu}(2|2)) \oplus \widehat{\mf{so}(2)}_{\perp }\supset \mf{so}(3,1)_{\rm conf} \oplus  \mf{so}(4)_{7890}  
\oplus \widehat{\mf{so}(2)}_{\perp}
\label{sda1}
\fe
where  $ \widehat{\mf{so}(2)}_{\perp} $ is generated by $\hat M_{\perp}$ which gives the central extension of the two $\mf{psu}(2|2)$ factors. The $\mf{so}(3,1)_{\rm conf}$ factor is generated by rotations
\ie
M_{12},M_{13},M_{23}\,,
\fe
and conformal transformations
\ie
P_1+K_1,~P_2+K_2,~P_3+K_3\,,
\fe
which split into two commuting $\mf{su}(2)$ algebras
\ie
&[L_i,L_j]=2\epsilon_{ijk} L_k,\quad [\bar L_i, \bar L_j]=2\epsilon_{ijk} \bar L_k\,.
\fe
The $16$ preserved  supercharges are
\ie
Q_{\A 1 \dot a}- i(\sigma_2)^{\dot \A}{}_\A\bar S_{\dot \A 1 \dot a},
\quad
\bar  Q_{\A 2 \dot a}+i(\sigma_2)^{\dot \A}{}_\A  S_{\dot \A 2 \dot a}\,,
\fe
and
\ie
Q_{\A2 \dot a}+i  (\sigma_2)^{\dot \A}{}_\A\bar S_{\dot \A 2 \dot a},
\quad
\bar  Q_{\A 1 \dot a} -i  (\sigma_2)^{\dot \A}{}_\A S_{\dot \A 1 \dot a}
\fe
respectively. Equivalently, these supercharges are specified in terms of the constant spinors by\footnote{This is so that the surface defect supersymmetry are determined by the projector 
	\ie
	{x^i } \Gamma_i \Gamma_{456} \ve=-\ve
	\fe
	at  $x_4=0$ and $\sum_{i=1}^3 x_i^2=1$.
}
\ie
\ep_c= - \CC_{456} \ep_s\,.
\fe

On the other hand, half-BPS surface defects of the type $\cD_2^{\rm II}$ arises when $\pi^{\rm in}(\cD_2)$ is a straight segment passing through the center of $B^3$ and intersecting the $S^2_{\rm YM}$ at two antipodal  points. Up to an $\mf{so}(3)$ rotation, we can take $\pi^{\rm in}(\cD_2)$ to be given by $x_1=x_2=0$ and $x_3\in [-1,1]$. Embedded in $\mR^4$, $\cD_2^{\rm II}$ is a simply the plane at $x_1=x_2=0$.\footnote{Upon Weyl transformation (via stereographic mapping), this becomes another great $S^2\subset S^4$.} 
 
This surface defect preserves a subalgebra isomorphic to the one in \eqref{sda1}  
\ie
\mf{psu}(2|2)\oplus \mf{psu}(2|2) \oplus \widehat{\mf{so}(2)}_{79}\supset \mf{so}(3,1)_{\rm conf} \oplus  \mf{so}(4)_{5680}  
\oplus \widehat{\mf{so}(2)}_{79}
\fe
where $\widehat{\mf{so}(2)}_{79}$ is generated by $M_{12}-R_{79}$ (or $i(J_{12}-J_{\dot 1\dot 2})-R_{79}$)  which gives the central extension of the two $\mf{psu}(2|2)$ factors. The $\mf{so}(3,1)_{\rm conf}$ factor is generated by 
\ie
D,M_{34},P_3,P_4,K_3,K_4
\fe
which split into two commuting $\mf{su}(2)$ algebras
\ie
\mf{su}(2)_L:~&\{P_{1\dot1},~K_{2\dot 2},~D+J_{12}+J_{\dot 1\dot 2}\}\,,
\\
\mf{su}(2)_R:~&\{P_{2\dot2},~K_{1\dot 1},~D-J_{12}-J_{\dot 1\dot 2}\}\,.
\fe
The $16$ preserved  supercharges for the two $\mf{psu}(2|2)$ factors are
\ie
&Q_{1 a \dot a}-(\sigma^3)^{b}{}_a (\sigma^2)^{\dot b}{}_{\dot a}Q_{1 b \dot b},\quad \tilde Q_{1 a \dot a}- (\sigma^3)^{b}{}_a (\sigma^2)^{\dot b}{}_{\dot a}\tilde Q_{1 b \dot b}\,,
\\
&S_{2 a \dot a}- (\sigma^3)^{b}{}_a(\sigma^2)^{\dot b}{}_{\dot a}S_{2 b \dot b},\quad \tilde S_{2 a \dot a}- (\sigma^3)^{b}{}_a(\sigma^2)^{\dot b}{}_{\dot a}\tilde S_{2 b \dot b}\,.
\fe
and
\ie
&Q_{2 a \dot a}+(\sigma^3)^{b}{}_a (\sigma^2)^{\dot b}{}_{\dot a}Q_{2 b \dot b},\quad \tilde Q_{2 a \dot a}+ (\sigma^3)^{b}{}_a (\sigma^2)^{\dot b}{}_{\dot a}\tilde Q_{2 b \dot b}\,,
\\
&S_{1 a \dot a}+ (\sigma^3)^{b}{}_a(\sigma^2)^{\dot b}{}_{\dot a}S_{1  b \dot b},\quad \tilde S_{1 a \dot a}+ (\sigma^3)^{b}{}_a(\sigma^2)^{\dot b}{}_{\dot a}\tilde S_{1 b \dot b}\,.
\fe
respectively.
Equivalently, these supercharges are specified in terms of the constant spinors by 
\ie
\ep_s= \CC_{1279} \ep_s,\quad \ep_c= \CC_{1279} \ep_c\,.
\fe

In particular note that the two types of half-BPS surface operators intersect at two points $x_3=\pm 1$ and $x_1=x_2=x_4=0$. The common supercharges generate four copies of $\mf{psu}(1|1)$ where the first two factors are centrally extended by 
\ie
J_{12}-J_{\dot 1 \dot 2}+P_{1 1}-K_{2 2}-i(R_{80}-R_{56}-R_{79})\,,
\fe
and the last two factors are centrally extended by
\ie
J_{12}-J_{\dot 1 \dot 2}+P_{2 2}-K_{1 1}-i(R_{80}+R_{56}-R_{79})\,.
\fe
 In terms of the constant spinors, they are given be
\ie
\ep_s= \CC_{1279} \ep_s,\quad \ep_c= -\CC_{456} \ep_s\,.
\fe 

\subsubsection{Boundaries and interface defects}
At $d=3$, the worldvolume submanifold $\cD_3$ is generated by flowlines of $M_\perp$ through a 2d slice   $\pi^{\rm in}(\cD_3)$ in $\bar B^3$ which either lies entirely in the interior  $B^3$, or it intersects with the boundary $S^2_{\rm YM}$ along a curve.   We will denote the two types of defects by $\cD_3^{\rm I}$ and $\cD_3^{\rm II}$ respectively. They have the topology $\Sigma\times S^1$ for some Riemann surface $\Sigma$ or $D^3_\infty$. We will focus on the latter case here with the interface (or boundary) along the hyperplane at $x_1=0$. This boundary/interface preserves the ${1\over 2}$-BPS  symmetry algebra 
\ie
\mf{osp}(4|4,\mR) \supset \mf{so}(3)_{567} \oplus  \mf{so}(3)_{890} \oplus  \mf{so}(3,2)_{\rm conf}
\fe
where $\mf{so}(3,2)_{\rm conf}$ is the conformal group acting on the hyperplane at $x_1=0$, and $\mf{so}(3)_{567} \oplus  \mf{so}(3)_{890}$ gives to a maximal subalgebra of the full $\mf{so}(6)_R$ symmetry. There is a family of such  $\mf{osp}(4|4,\mR)$ algebras (with the same bosonic subalgebras) parametrized by $\zeta \in \mC^\star$. The corresponding 
 supercharges   are
\ie
\cQ_{\A a \dot a}=Q_{\A a \dot a}+i \zeta  (\sigma_3)^{\dot \A}_\A\bar Q_{\dot \A a \dot a}\,, 
\quad
\cS_{\A a \dot a}=S_{\A a \dot a}+{i\over \zeta} (\sigma_3)^{\dot \A}_\A\bar S_{\dot \A a \dot a}\,.
\fe
In terms of the constant 16-component spinors, they are specified by
\ie
 (\CC_{1234}+1)(\zeta\CC_{1890}-1)\ep_s=0,\quad  (\CC_{1234}+1)( {1\over \zeta}\CC_{1890}-1)\ep_c=0\,.
\label{bce}
\fe
For the case $\zeta=1$, this becomes
\ie
\CC_{1890} \ep_s=\ep_s,\quad \CC_{1890} \ep_c=\ep_c
\fe
which are clearly compatible with \eqref{Qspinor}.
Hence $\cQ$ is in this subalgebra preserving the BPS boundary condition. Moreover, comparison between \eqref{bce} and supersymmetries of IIB branes (see Section~\ref{sec:halfbpsalg})
implies that the defects can be described by D5-branes along the $234890$  directions or NS5s along the $234567$  directions, intersecting with the D3 branes that lie along the 1234 directions in the 10d IIB spacetime. If we split the six scalar fields of SYM as
\ie
X_i=(\Phi_8,\Phi_9,\Phi_0 ),\quad Y_{j}=(\Phi_5,\Phi_6,\Phi_7 )
\fe
with $i,j=1,2,3$, the D5 brane type boundary condition (sometimes referred to as \textit{generalized} Dirichlet boundary condition since it imposes Dirichlet boundary condition for the gauge fields of the $\cN=4$ vector multiplet) is given by
\ie
&{\rm D5}:~&F_{\m\n}  \left.  \right|_{x_1=0} =  D_1 X_i - {1\over 2} \ep_{ijk}[X_j,X_k]\left.\right|_{x_1=0}= Y_i \left.  \right|_{x_1=0}=0,\quad \CC_{1890}\Psi=-\Psi\,.
\label{D5bc}
\fe
They ensure that the component of the bulk supercurrent normal to the boundary   vanish. We have also imposed conformal (scale) invariance explicitly.

Note that the second equation above is the Nahm equation and this boundary condition is also known as the Nahm (pole) boundary condition for the $\cN=4$ SYM. Near the boundary $x_1=0$, the solutions to the Nahm equation are given by
\ie
X_i = - {t_i\over x_1} + {\rm regular}
\fe
where $t_i$ takes value in the Lie algebra $\mf{g}$ of the gauge group and obeys the $\mf{su}(2)$ commutation rules
\ie
{}[t_i,t_j]=\ep_{ijk}t_k\,.
\fe
Consequently $t_i$ are determined (up to gauge transformations) by homomorphisms $\rho:\mf{su}(2)\to \mf{g}$. For $\mf{g}=\mf{u}(N)$, such a homomorphism is in one-to-one correspondence with a partition $d=[p_1,\dots,p_k]$ of $N$ with $p_1\geq p_2\geq \dots \geq p_k>0$. In particular, $t_i$ can be represented as an $N\times N$ matrix of the block diagonal form
\ie
t_i=t^{p_1\times p_1}_i\oplus t^{p_2\times p_2}_i \oplus \cdots \oplus t^{p_k\times p_k}_i
\fe
where each triplet $t^{p_i\times p_i}_i$ with $i=1,2,3$ gives rise to a $p_i$-dimensional irreducible representation of $\mf{su}(2)$ and $t_3^{p\times p}=-{i\over 2}{\rm Diag}[p-1, p-3,\dots, 1-p]$. The special case with $t_i=0$ (associated to the partition $d=[1,1,\dots,1]$) corresponds to the familiar Dirichlet boundary condition. 

The NS5-brane type boundary condition 
\ie
{\rm NS5}:~&F_{1\m}  \left.  \right|_{x_1=0} =   X_i \left.  \right|_{x_1=0}=D_1 Y_i  \left.\right|_{x_1=0}=0,\quad \CC_{1890}\Psi= \Psi
\label{NS5bc}
\fe
is the familiar Neumann boundary condition for the bulk SYM. Once again the component of the bulk supercurrent normal to the boundary vanishes.

The D5-type and NS5-type boundary conditions are related by S-duality as explained in \cite{Gaiotto:2008ak}. In IIB, the SYM scalars $X_i$ and $Y_i$ parametrize the transverse directions to the D3 branes. Consider a general 5-brane that shares the 234 directions with the D3-branes and extend in three other directions among $X_i$ and $Y_i$.
Since the supercharges transform nontrivially under $SL(2,\mZ)$ \eqref{susysl2}, 
for a fixed $\mf{osp}(4|4)$ subalgebra containing $\cQ$, 
and when the 5-brane world-volume directions  transverse to the D3 branes parametrized by coordinates $X'_i$ as in 
\ie
X_i=\cos\theta   X'_i,\quad Y_i=\sin\theta   X'_i
 \fe 
 for some angle $\theta \in [0,\pi/2]$, there is a unique \textit{minimal} half-BPS boundary condition. The case with $\theta={0}$ corresponds to the D5-type boundary condition, while the $\theta={\pi \over 2}$ case corresponds to the NS5-type boundary condition. When 
\ie
\tan \theta={q\over p}
\fe
for co-prime positive integers $p$ and $q$, the boundary condition is given by D3 branes ending on the $(p,q)$ 5-branes that extend in the 234 directions longitudinal to the D3 branes and  another three directions parametrized by $X'_i$ transverse to the D3 branes.

Another generalization of the D5-type and NS5-type boundary conditions is to introduce partial \textit{gauge symmetry breaking} \cite{Gaiotto:2008sa}. For a subgroup of the gauge group $H\subset G$ (we do not assume $G$ is simple), the corresponding Lie algebra of $G$ decomposes as
\ie
\mf{g} =\mf{h}\oplus \mf{h}^\perp 
\fe 
into the Lie algebra of $H$ and its orthogonal complement (which is not a Lie algebra  in general). Then one can consider a mixture of NS5-type boundary condition  \eqref{NS5bc} for the components of the SYM fields in $\mf{h}$ and D5-type boundary condition \eqref{D5bc} for the components in $\mf{h}^\perp$. This defines the symmetry breaking boundary  condition associate to the subgroup $H\subset G$.
\begin{figure}[!htb]
	\begin{minipage}{.4\linewidth}
	\includegraphics[scale=.4]{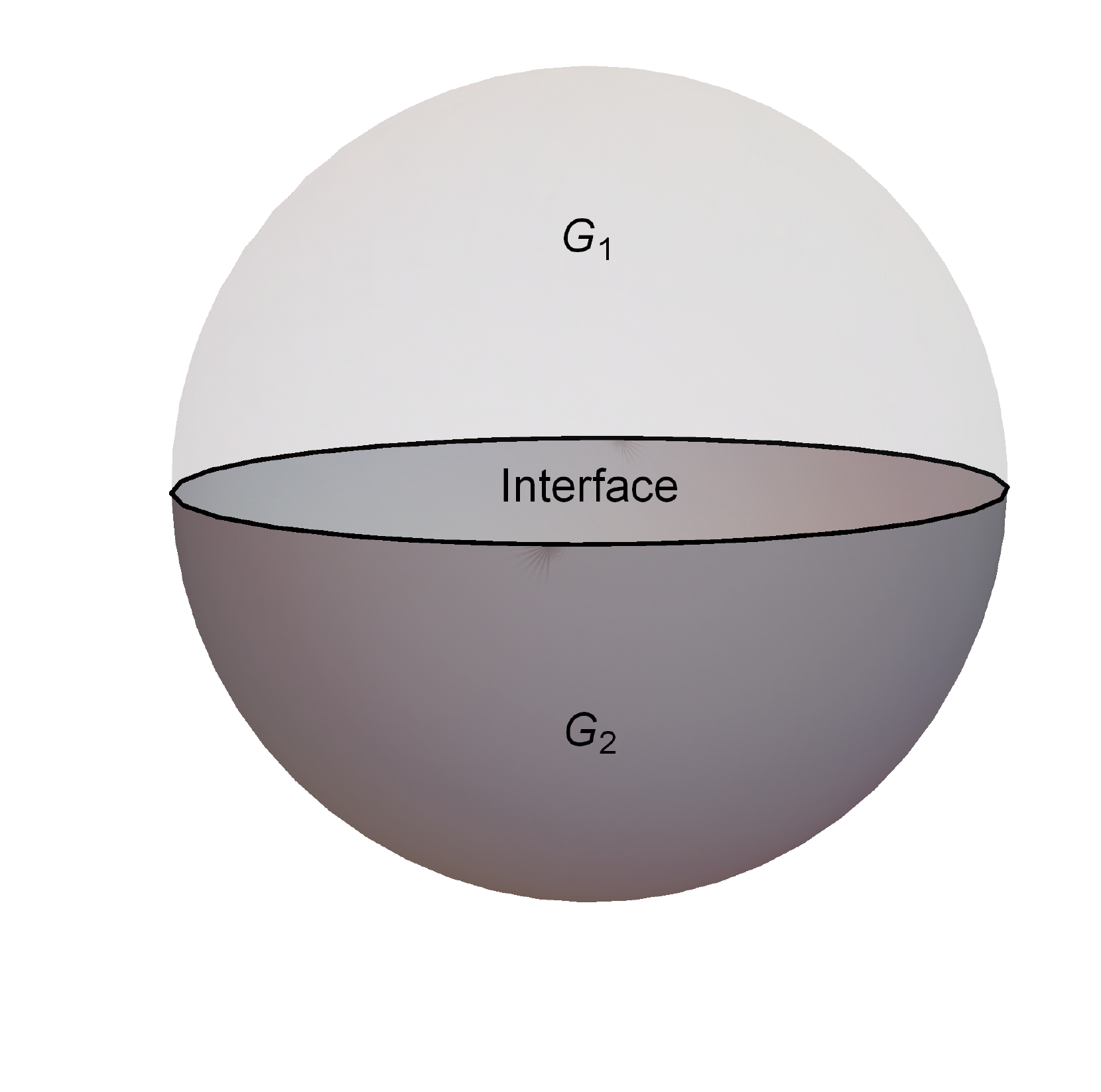}
	\end{minipage}
		\begin{minipage}{.15\linewidth}
		\ie
		&
		\xleftrightarrows[\rm{Folding~by}~\iota_{\rm fold}]{\text {~Unfolding~}}
	 \\
		 \nonumber
	 \fe
		\end{minipage}
		\begin{minipage}{.4\linewidth}
			\includegraphics[scale=.4]{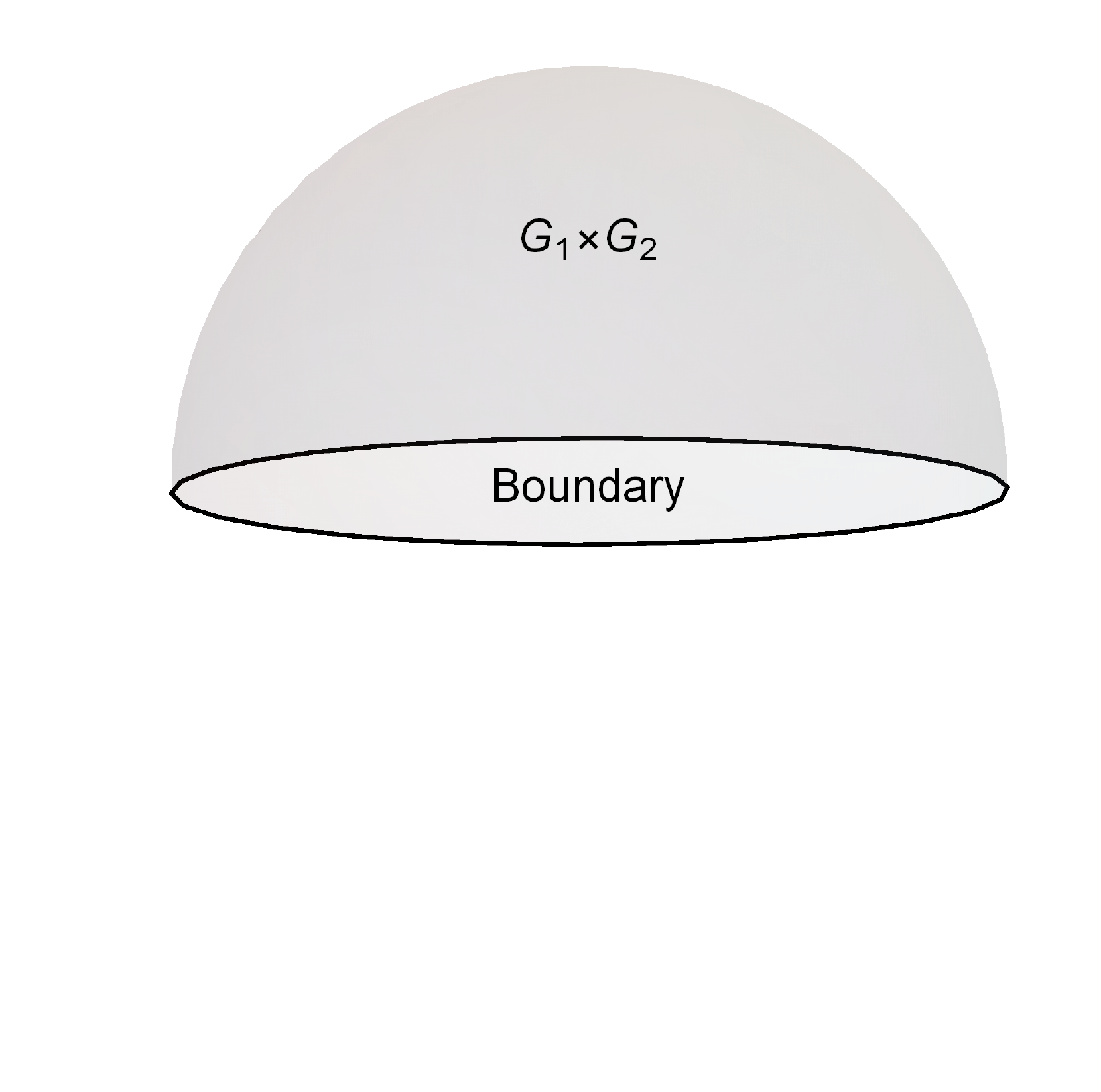}
		\end{minipage}
\caption{The (un)folding trick that relates interface between $G_1$ and $G_2$ SYM and boundary condition for $G_1\times G_2$ SYM. }
\label{fig:fold}
\end{figure} 

So far we have focused on boundaries for the SYM. The generalization to half-BPS interfaces is straightforward thanks to the (un)folding trick \cite{Gaiotto:2008sa} (see Figure~\ref{fig:fold}). The idea is that BPS interfaces in the $\cN=4$ SYM with gauge group $G_1$ on one side $x_1>0$ and another gauge group $G_2$ on the other side $x_1<0$ can be folded to give an auxiliary (tensor-product) SYM with gauge group $G_1\times G_2$ on the half-space $x_1\geq 0$ with a half-BPS boundary condition at $x_1=0$. This is implemented by flipping the $G_2$ factor using a $\mZ_2$-automorphism $\iota_{\rm fold}$ of the superconformal algebra $\mf{psu}(2,2|4)$ that acts by
\ie
\iota_{\rm fold}:~(x_1,x_2,x_3,x_4) \to (-x_1,x_2,x_3,x_4),\quad (X_i,Y_i)\to (-X_i,Y_i).
\label{folding}
\fe
Similarly one can unfold the $G_1\times G_2$ SYM on half-space with BPS boundary conditions to obtain an interface between the $G_1$ and $G_2$ SYM. 
For example, the transparent interface in the $G$ SYM corresponds to unfolding the $G\times G$ SYM with the partial symmetry breaking boundary condition that preserves the diagonal subgroup $G_{\rm diag}\subset G\times G$. 
 
\subsection{Defect CFTs and  networks}
We have  restricted to \textit{minimal} defects of the $\cN=4$ SYM in the previous section. One may also couple these extended objects to localized degrees of freedom on the defects. To make sure such additions contribute and further enrich the $\cQ$-cohomology, one needs to check that the defect theory and defect-bulk coupling respect  the $\mf{su}(1|1)$ symmetry generated by $\cQ$. For example, we can stack a 1d $\cN=4$ quantum mechanics on the ${1\over 8}$-BPS Wilson loops, couple a 2d $\cN=(4,4)$ SCFT to the surface operators, and place a 3d $\cN=4$ SCFT on half-BPS boundaries or interfaces.\footnote{These defect theories  preserve the maximal symmetry of the corresponding defect operator in the $\cN=4$ SYM. One can consider more general defect theories with less supersymmetry as long as  $\cQ$ is included in the supersymmetry algebra.} The operator spectrum of these additional defect theories then gives rise to new nontrivial classes in the $\cQ$-cohomology. Luckily when restricted to these lower dimensional defect theories, the $\cQ$-cohomology has been well-studied. 

For example for the type $\cD_2^{\rm II}$ half-BPS surface defect (see \cite{Benini:2016qnm} for a review), the supercharge $\cQ$ (and the $\mf{su}(1|1)$ algebra) is preserved by any $\cN=(2,2)$ SCFT on the defect worldvolume which we can take to be an $S^2\subset S^4$ related a Weyl transformation to  $\mR^2\subset \mR^4$ as described in the previous section. The $\cQ$-cohomology is then enriched by local chiral and anti-chiral operator insertions at the poles of the $S^2$ (correspondingly the points $x_\m=(0,0,\pm 1, 0)\in \mR^4$), as well as Wilson loops along latitudes of the $S^2$ (correspondingly circles in the $(x_3,x_4)$ plane centered at $x_\m=(0,0,\pm 1, 0)$ in $\mR^4$).

Below we provide more details for the 3d boundary/interface defect SCFT which we will use to extract defect OPE data in the $\cN=4$ SYM in Section~\ref{sec:applications}.

\subsection{Topological quantum mechanics on the half-BPS boundary}
\label{sec:TQMonthebdy}
In \cite{Beem:2016cbd}, it was shown that 3d $\cN=4$ SCFTs contain nontrivial 1d  sectors living on a line $\mR_{\rm TQM}\subset \mR^3$ (up to a conformal transformation) described by a topological quantum mechanics \cite{Dedushenko:2016jxl}. Since the half-BPS boundary condition preserves the $\cN=4$ superconformal algebra $\mf{osp}(4|4)$, one naturally expects the 2d sector on $HS^2_{\rm YM}$ in the bulk 4d SYM to be enriched by coupling to a 1d sector (possibly with local degrees of freedom) living on the boundary $S^1$ which we will refer to as $S^1_{\rm TQM}$
\ie
S^1_{\rm TQM}:~ \{x_4=x_1=0,~x_2^2+x_3^2=1 \}   
\fe
Indeed, we will see that the cohomology with respect to the supercharge $\cQ$ restricted to the boundary coincides with the 1d topological sector of the boundary 3d $\cN=4$ SCFT as in \cite{Chester:2014mea,Beem:2016cbd}. 

In terms of standard 3d $\cN=4$ terminology, we identify the $\mf{so}(3)_{567}\times \mf{so}(3)_{890}$ R-symmetry of $\mf{osp}(4|4)$ with the Coulomb branch and Higgs branch R-symmetries $\mf{su}(2)_C\times \mf{su}(2)_H$ respectively. The  1d sector of a 3d $\cN=4$ SCFT is defined using a subalgebra of the 3d superconformal algebra $\mf{osp}(4|4)$ (up to an automorphism)\footnote{A 3d $\cN=4$ SCFT in general can have two 1d TQM sectors for Higgs and Coulomb branches respectively \cite{Dedushenko:2016jxl,Dedushenko:2017avn,Dedushenko:2018icp}. Here for a given choice of $\cQ$ only one of such 1d sectors is relevant for studying the $\cQ$-cohomology, which is the Higgs branch TQM accordingly to our convention.}
\ie
\mf{psu}(1,1|2)_H \times  \widehat{\mf{so}(2)}_\perp \supset \mf{so}(2,1)_{\rm conf} \times \mf{su}(2)_H\,\times \widehat{\mf{so}(2)}_\perp
\label{TQMalg}
\fe
where $\widehat{\mf{so}(2)}_\perp$ implements a central extension of the $\mf{psu}(1,1|2)_H$ algebra. Below we will identify the generators within the bulk 4d superconformal algebra $\mf{psu}(2,2|4)$ (see also Appendix~\ref{app:3dsca}). Here $\mf{so}(2,1)_{\rm conf}$ is the conformal symmetry of $S^1_{\rm TQM}$ generated by
\ie
L_0= i M_{23},\quad L_{+}= {i\over 2} (P_2+K_2+i P_3+i K_3),\quad  L_{-}= -{i\over 2} (P_2+K_2 -i P_3-iK_3)
\fe
that satisfies
\ie
&[L_0, L_\pm ]=\pm L_\pm,\quad [L_+,L_-]=2L_0\,.
\fe
$\mf{su}(2)_H=\mf{so}(3)_{890}$ is the Higgs branch $R$-symmetry, and $\widehat{\mf{so}(2)}_\perp$ is the combination of transverse rotation to the $S^1_{\rm TQM}$ and the Cartan generator $T_3^C=-i R_{56}$ of the Coulomb branch $R$-symmetry. The fermionic generators $\mathsf{Q}_a,\mathsf{S}_a,\tilde{\mathsf{Q}}_b,\tilde{\mathsf{S}}_b$ of $\mf{psu}(1,1|2)_H $ transform as doublets under the $\mf{su}(2)_H$ R-symmetry. There explicit forms in terms of the 4d supercharges are  
\ie 
&\mathsf{Q}_1 =
{1\over 2\sqrt{2}}
\left(
\sigma_3^{\A \dot a }\cQ_{\A 1\dot a}+i\sigma_2^{\A \dot a }\cQ_{\A 1\dot a}
+
i\D^{\A \dot a }\cS_{\A 1\dot a}+ \sigma_2^{\A \dot a }\cS_{\A 1\dot a}
\right)\,,
\\
&\mathsf{Q}_2 =
{-i\over 2\sqrt{2}}
\left(
\D^{\A \dot a }\cQ_{\A 1\dot a}-\sigma_1^{\A \dot a }\cQ_{\A 1\dot a}
+
i\sigma_3^{\A \dot a }\cS_{\A 1\dot a}+i\sigma_1^{\A \dot a }\cS_{\A 1\dot a}
\right)\,,
\\
&\tilde{\mathsf{Q}}_1 =
-{1\over 2\sqrt{2}}
\left(
\sigma_3^{\A \dot a }\cQ_{\A 2\dot a}+i\sigma_2^{\A \dot a }\cQ_{\A 2\dot a}
+
i\sigma_3^{\A \dot a }\cS_{\A 2\dot a}-\sigma_2^{\A \dot a }\cS_{\A 2\dot a}
\right)\,,
\\
&\tilde{\mathsf{Q}}_2 =
{ i\over 2\sqrt{2}}
\left(
\D^{\A \dot a }\cQ_{\A 2\dot a}-\sigma_1^{\A \dot a }\cQ_{\A 2\dot a}
+
i\D^{\A \dot a }\cS_{\A 2\dot a}-i\sigma_1^{\A \dot a }\cS_{\A 2\dot a}
\right)\,,
\\
&\mathsf{S}_1 =
-{1\over 2\sqrt{2}}
\left(
\D^{\A \dot a }\cQ_{\A 2\dot a}+ \sigma_1^{\A \dot a }\cQ_{\A 2\dot a}
+
i\sigma_3^{\A \dot a }\cS_{\A 2\dot a}-i \sigma_1^{\A \dot a }\cS_{\A 2\dot a}
\right)\,,
\\
&\mathsf{S}_2 =
{i\over 2\sqrt{2}}
\left(
\sigma_3^{\A \dot a }\cQ_{\A 2\dot a}-i \sigma_2^{\A \dot a }\cQ_{\A 2\dot a}
+
i\D_3^{\A \dot a }\cS_{\A 2\dot a}-\sigma_2^{\A \dot a }\cS_{\A 2\dot a}
\right)\,,
\\
&\tilde{\mathsf{S}}_1 =
-{1\over 2\sqrt{2}}
\left(
\D^{\A \dot a }\cQ_{\A 1\dot a}+ \sigma_1^{\A \dot a }\cQ_{\A 1\dot a}
+
i\D^{\A \dot a }\cS_{\A 1\dot a}+i\sigma_1^{\A \dot a }\cS_{\A 1\dot a}
\right)\,,
\\
&\tilde{\mathsf{S}}_2 =
{ i\over 2\sqrt{2}}
\left(
\sigma_3^{\A \dot a }\cQ_{\A 1\dot a}-i\sigma_2^{\A \dot a }\cQ_{\A 1\dot a}
+
i\sigma_3^{\A \dot a }\cS_{\A 1\dot a}+\sigma_2^{\A \dot a }\cS_{\A 1\dot a}
\right)\,,
\fe
and they satisfy the anti-commutation rules
\ie
&\{\sf{Q}_a,\tilde{\sf{Q}}_b\}=\ep_{ab}L_+ ,\quad 
\{\sf{S}_a,\tilde{\sf{S}}_b\}=\ep_{ab}L_-  \,,
\\
&\{\sf{Q}_a,\sf{S}_b\}=\ep_{ab}(L_0-\hat M_\perp)+\tau^i_{ab} T_i^H,\quad 
\{\tilde{\sf{Q}}_a,\tilde{\sf{S}}_b\}=-\ep_{ab}(L_0+\hat M_\perp)-\tau^i_{ab} T_i^H \,,
\fe
with $\tau^i=(\sigma^3,\sigma^1,\sigma^2)$. The 1d sector of \cite{Beem:2016cbd} is defined by  choosing nilpotent supercharges
\ie
\cQ^H_1={1-i\over 2}\left(
\sf{Q}_2 +  \C \tilde{\sf{S}}_1
\right),\quad 
\cQ^H_2={1-i\over 2}\left(
\tilde{\sf{Q}}_2+ {1\over \C} \sf{S}_1 
\right)
\label{TQMnilp}
\fe
with $\C \in \mC^\star$. They satisfy
\ie
\{\cQ^H_1,\cQ^H_1\}=\{\cQ^H_2,\cQ^H_2\}=0,\quad \{\cQ^H_1,\cQ^H_2\}=i \hat M_\perp  
\fe
and their anti-commutators with other fermionic generators in $\mf{osp}(4|4)$ give rise to $\cQ^H_{1,2}$-exact twisted translations
\ie
\hat L_0\equiv L_0 +T_{1}^H,\quad \hat L_+\equiv L_+ +T_{2}^H+i T_3^H,\quad \quad \hat L_-\equiv L_- +T_{2}^H-i T_3^H \,.
\fe
We can use either $\cQ_1^H$ or $\cQ_2^H$ to define the 1d sector by taking its cohomology in the full 3d operator algebra. As explained in \cite{Beem:2016cbd}, the cohomology of the $\cQ^H_1$ and $\cQ^H_2$ (or any combination) actually coincides. 

 Now to make connection to the bulk supercharge $\cQ$, we simply note that, comparing to the definition \eqref{LocQdef} of $\cQ$, we find that on the boundary
\ie
\cQ=\cQ^H_1+\cQ^H_2 
\fe
with  $\C=1$ in \eqref{TQMnilp}.

The basic observable in the 1d sector consists of Higgs branch chiral primary operators $\cO_{a_1a_2\dots a_{2j}}$ which are scalar operators of $SU(2)_H$ spin $j\in \mZ/2$ and scaling dimension $\Delta=j$. They give rise to cohomology classes of $\cQ^H_{1,2}$  defined by the twisted translation of the $SU(2)_H$ highest weight state $\cO_{11\dots 1}$ on the $S^1_{\rm TQM}$ with angular coordinate $\varphi\in [0,2\pi]$,
 \ie
 \hat \cO_j(\varphi)\equiv e^{i \varphi \hat L_0}\left.\cO_{11\dots 1}\right|_{\varphi=0} e^{-i \varphi \hat L_0}
 =\left.\cO_{a_1\dots a_{2j}}\right|_{\varphi=0} u^{a_1}\dots u^{a_{2j}}
 \fe
 Note that the $R$-symmetry polarization $u^a=(\cos {\varphi\over 2},\sin {\varphi\over 2})$ depends on the position on the $S^1_{\rm TQM}$.
 
The $\cQ^H_{1,2}$-cohomology also contain half-BPS loop operators along the $S^1_{\rm TQM}$. For example, if the 3d boundary theory includes an $\cN=4$ vector multiplet which contains the gauge fields $a$ and a triplet of dimension one scalars $\phi_{(\dot a\dot b)}$, one can define the Wilson loop
\ie
W^{3d}_R \equiv {1\over d_R} \tr_R {\rm P }e^{i\oint_{S^1_{\rm TQM}} d\varphi\left(a_\varphi + i\phi_{\dot 1\dot 2} \right)}
\fe
in the $\cQ^H_{1,2}$-cohomology.

Before ending this section, we note that here  for convenience we have chosen to identify the $\mf{so}(3)_{890}$ subalgebra of the bulk R-symmetry $\mf{so}(6)_R$ with the Higgs branch R-symmetry on the boundary. We could of course instead identify $\mf{so}(3)_{890}$ with the Coulomb branch R-symmetry on the boundary (related by mirror symmetry). In that case, the relevant $\cQ$-cohomology in the boundary operator algebra includes monopole operators as well as vortex loops wrapping the $S^1_{\rm TQM}$.

\section{Two-dimensional Defect-Yang-Mills from Localization}
\label{sec:loctodym}
In this section, we carry out the supersymmetric localization  for the defect network observables (see Figure~\ref{fig:network}) introduced in the previous section, put now on $S^4$ using a Weyl transformation from $\mR^4$
\ie
ds^2=e^{2\Omega}{dx^2},\quad e^{ \Omega}= {1\over 1+{x^2\over 4R^2}}\,.
\label{Weyltransform}
\fe
We will slightly abuse the notation and denote the Weyl-transformed partner of $\cQ$ in \eqref{LocQdef} also by $\cQ$ where the corresponding conformal killing spinor becomes
\ie
\ve =e^{\Omega\over 2}(\ep_s+x^\m \CC_{\hat\m} \ep_c)
\fe
according to \eqref{ksWeyl} with $\lambda=\Omega$. It satisfies the conformal killing spinor equation \eqref{CKSeqn} on $S^4$ with
\ie
\tilde \ve=-{i\over 2R}\CC_{790} \ve\,.
\fe
 
 Previously the localization of 4d $\cN=4$ SYM on $S^4$ with respect to the supercharge $\cQ$ was carried out in \cite{Pestun:2009nn} which identifies an emergent 2d Yang-Mills theory on the $S^2_{\rm YM}$ with the same gauge group as the 4d theory. 
Here we extend the previous work by considering $\cN=4$ SYM on $HS^4$ with BPS boundary condition at the equator $S^3$ (from the Weyl transformation of the hyperplane at $x_1=0$). The case with an interface on $S^4$ is related by the (un)folding trick (see Figure~\ref{fig:fold}).

 As explained in section~\ref{sec:TQMonthebdy}, restricted to the boundary $S^3$, $\cQ$ coincides with the supercharge defining the 1d topological sector in the boundary 3d $\cN=4$ theory. The localization of a 3d  $\cN=4$ SCFT (with $\cN=4$ Lagrangian) on $S^3$ with respect to such   supercharge  gives rise to a 1d topological gauged quantum mechanics on the $S^1_{\rm TQM}$ \cite{Dedushenko:2016jxl}. By putting together these results, we will find   a coupled 2d/1d quantum system that captures the dynamics of general defect observables of the 4d SYM in the $\cQ$-cohomology. We refer to this effective 2d/1d theory as the defect-Yang-Mills (dYM).

\subsection{Off-shell supersymmetric boundary conditions}
In general, when the supercharge (here given by $\cQ$) defining the supersymmetric observables has an off-shell realization in the path integral, one can show that the full path integral localizes to the BPS locus with respect to $\cQ$ by a standard procedure. The resulting effective theory governing the dynamics of the BPS locus is typically much simpler than the original path integral. It often takes the form of a zero-dimensional matrix model, or as we will see here, as a coupled system of two- and one-dimensional quantum field theories, namely the defect-Yang-Mills.

We start by explaining in more detail the setup of 4d $\cN=4$ SYM on  $HS^4$ with  boundary conditions preserving off-shell supersymmetry.  

The supersymmetry variation of the SYM action \eqref{SYMos} is nonzero when the spacetime manifold has a boundary. Here we find using \eqref{SUSYos} the following boundary variation at $x_1=0$ (recall that we have set $R={1\over 2}$)
\ie
\D_\ve S_{\rm SYM}
=&{1\over 2  g_4^2}\int_{\partial HS^4} d^3 x\, \sqrt{\C}  \,  \tr
\Bigg(
\Psi \CC^{\hat 1} \n_m K^m+2 \tilde\ve \CC_A{}^{\hat 1}\Psi \Phi^A
+{1\over 2}  F_{MN} \ve\CC^{MN{\hat 1}} \Psi
+ e^\Omega F^{1 N} \Psi \CC_N \ve
\Bigg)\,,
\label{bdyvar}
\fe
where $\C$ is the induced metric on the boundary $S^3$. To preserve an off-shell supercharge parametrized by the conformal Killing spinor $\ve$, we need to specify the boundary conditions on the SYM fields (including the auxiliary scalars $K_m$) such that $\D_\ve S_{\rm SYM}=0$.\footnote{Another possibility is to include appropriate boundary term  to cancel $\D_\ve S_{\rm SYM}$ such as the one considered in \cite{Mezei:2013gqa}.}

Our chosen supercharge $\cQ$ is associated to the conformal killing spinor $\ve$ satisfying 
\ie
\CC_{1890} \ve =\ve,\quad \tilde \CC_{1890} \tilde\ve = \tilde \ve 
\fe
at $x_1=0$. To identify the boundary conditions for $K_m$, we need to determine the auxiliary pure spinors $\n_m$ at the boundary. There are 7 independent solutions to the pure spinor constraints  \eqref{PSpinor} that are rotated into each other by an $SO(7)$ transformation that acts on the $m$ index. One convenient set of  $\n_m$ that satisfies \eqref{PSpinor} is given by
\ie
\n^m=\{
 \CC^{   7a}\ve,\CC^{\hat 17}\ve,\CC^{\hat  1a}\ve
\}\,,\qquad a=8,9,0 
\label{psv1}
\fe
at $x_1=0$. Consequently, the auxiliary spinors satisfy
\ie
\CC_{1890}\n_{m}=\begin{cases}
\n_{m} & {\rm for}~1\leq m \leq 4\,,
\\
-\n_{m}& {\rm for}~5\leq m \leq 7\,.
\end{cases}
\fe
Then from the supersymmetry transformation rules \eqref{SUSYos}, the SYM fields $(A_\m,\Phi_A,\Psi,K_m)$ can be regrouped into 3d off-shell super-multiplets that are closed under the $\cN=4$ subalgebra $\mf{osp}(4|4)$,\footnote{Here we follow the convention of \cite{Gaiotto:2008sa} when referring to the 3d $\cN=4$ decomposition  of the 4d $\cN=4$ vector multiplet. In particular, the bottom components of the hyper-multiplet here transform as ${\bm 3}\oplus {\bm 1}$ under the $SU(2)_H$ R-symmetry subgroup.}
\ie
&{\rm hyper~multiplet}~:\Psi^-, A_1, X_i,K_i, K_4
\\
&{\rm vector~multiplet}~:
\Psi^+, A_{2,3,4}, Y_i,K_{i+4}
\label{4d3dsplit}
\fe
where $i=1,2,3$ above,  $\Psi^\pm$ are components of the gaugino $\Psi$ with eigenvalue $\pm 1$ under $\CC_{1890}$ respectively. 

The basic half-BPS supersymmetric boundary conditions of the 4d $\cN=4$ SYM are specified by assigning Neumann-like  and Dirichlet boundary conditions for the hyper and vector multiplets respectively in \eqref{4d3dsplit} \cite{Gaiotto:2008sa}. Depending on whether the   vector or the hyper multiplet satisfies the Dirichlet boundary condition, we have the D5-type boundary condition
\ie
 F_{\m \n}   |_{x_1=0}=  e^{-\Omega} D_1   X_i-{1\over 2}\ep_{ijk}[X_j,X_k]  |_{x_1=0}= Y_k  |_{x_1=0}= \Psi_+ |_{x_1=0}=  K_{i+2} |_{x_1=0}=0
\label{osD5}
\fe
and the NS5-type boundary condition
\ie
F_{1\n} |_{x_1=0}=  X_i |_{x_1=0} =D_1 Y_k |_{x_1=0}=\Psi_- |_{x_1=0}=K_{i} |_{x_1=0}=K_4 |_{x_1=0}=0
\label{osNS5}
\fe
For either case the boundary variation \eqref{bdyvar} vanish identically.

We can also shift the boundary values of $Y_i$ in \eqref{osD5}, which corresponds to  non-conformal generalizations of the Dirichlet boundary condition. Note that the BPS condition requires $[Y_i, Y_j]=0$, so for example we can take (non-conformal but still preserve the localizing supercharge $\cQ$)
\ie
Y_3  |_{x_1=0}=K_5 |_{x_1=0}=a
\label{osDir}
\fe
and all other fields as in \eqref{osD5}.
Note the familiar Dirichlet boundary condition is a special case of the D5-type boundary condition above, with $X_i$ also given by a commuting triple, or equivalently the corresponding Young diagram is of the type $[1,1,\dots,1]$.\footnote{They correspond to Ishibashi boundary states in the Toda CFT via the AGT correspondence \cite{LeFloch:2017lbt}. They are building blocks for brane-like boundary conditions that satisfies the Cardy condition: the Neumann boundary condition (with boundary Wilson lines) corresponds to the identity (general) ZZ brane, the symmetry breaking conditions with FI parameters and boundary Wilson lines correspond to FZZT branes. } They have simple relations to the Neumann boundary condition by gauging the boundary gauge symmetries and we will see how this is reflected in the localized theory. 

These basic boundary conditions can also be modified while preserving the 3d $\mf{osp}(4|4)$ supersymmetry by coupling to 3d $\cN=4$ SCFTs (as boundary conformal matter). The systematic analysis of general supersymmetry boundary conditions for the 4d $\cN=4$ SYM can be done following the recent work of \cite{Dedushenko:2018icp,Dedushenko:2018tgx} by specifying supersymmetric polarizations of the classical phase space on the boundary.

Now that we understand how to realize the off-shell supercharge $\cQ$ in the SYM on $HS^4$, we are now ready to use localization to compute 
\ie
\la \cO \ra_{\rm unnormalized} \equiv  \int \left.DAD\Psi\right|_{HS^4_{\rm bc}}\,  e^{-S_{\rm SYM}} \cO
\fe
where $\cO$ denotes a general observable in the $\cQ$-cohomology and ``bc'' denotes one of the compatible supersymmetric boundary conditions.

\subsection{The BPS equations on $HS^4$ and solutions}
The power of the supersymmetry localization lies in the fact that we can turn on $\cQ$-exact deformations (known as the localizing term) in the path-integral as
\ie
S_{\rm SYM} \to S_{\rm SYM} + t \D_\ve V\,.
\fe
Since the $\cQ$-closed observables are not affected by such deformations, we are free to take the limit $t\to  \infty$. Supposing the localizing term is positive definite in the bosonic fields on an appropriate integration contour,  the path integral localizes to the  configurations such that $\D_\ve V=0$ identically. Here we follow \cite{Pestun:2009nn} and choose the localizing term as
\ie
V=\int_{HS^4} d^4x \sqrt{g}  \tr (\bar\Psi\D_\ve \Psi)
\fe
in terms of the gaugino $\Psi$, 
then the localization lands on the BPS locus defined by
\ie
 \D_\ve  \Psi=0\,.
\label{BPSeqn}
\fe
To analyze the corresponding BPS equations on $HS^4$, it is useful to the introduce the coordinates (see also Appendix~\ref{app:coords})
\ie
ds^2=R^2( d\zeta^2 +\cos^2\zeta d\tau^2 +\sin^2\zeta (d\theta^2+ \sin^2 \theta d\phi^2))
\fe
which makes manifest the singular fibration of $S^1_\tau  \times HS^2$ over a segment $I_\zeta$ given by $\zeta \in [0,\pi/2]$. Here the $S^1_\tau$ is parametrized $\tau \in [0,2\pi]$ and the $HS^2$ is parameterized by $\theta \in [0,\pi/2]$ and $\phi\in [0,2\pi]$. The $S^1_\tau$ fiber shrinks at one end of $I_\zeta$ whereas the $HS^2$ shrinks at the other end. These coordinates are chosen such that $\cQ^2$ generates translation in the $S^1_\tau$ direction and the boundary $S^3$ of $HS^4$ at $x_1=0$ in the original coordinates gets mapped to $\theta={\pi \over 2}$ here. Then the 2d/1d sector resides on the great $HS^2$ at $\zeta={\pi \over 2}$ which we call $HS^2_{\rm YM}$ with boundary $S^1_{\rm TQM}$.

As explained in \cite{Pestun:2009nn}, among the sixteen \textit{complex} components of the BPS equations \eqref{BPSeqn}, nine of them impose the covariantly-constant condition of the SYM fields along the $\tau$ direction as\footnote{These conditions also come immediately from the equations $\D^2 A_M=\D^2 K_m=0$.}
\ie
{}[\cD_\tau, \cD_\m]=0,~[\cD_\tau, \Phi_{7,8,9,0}]=[\cD_\tau, \hat\Phi_{5,6}]=0,\quad [\cD_\tau, K_m]=-{R\over 8}M_{mn} K^n\,,
\label{covtaucst}
\fe
where the $\hat \Phi_{5,6}$ are twisted combinations of two real scalars $\Phi_{5,6}$
\ie
\hat \Phi_5 = \cos \tau \Phi_5 +\sin \tau \Phi_6,\quad \hat \Phi_6 = \ve \CC_I \ve \Phi^I=  \sin \tau \Phi_5 -\cos \tau \Phi_6\,,
\fe
and $D_\m$ denotes the twisted connection 
\ie
\cD_\tau = D_\tau - iR \cos \zeta  \hat  \Phi_5,\quad \cD_{\zeta,\theta,\phi}=D_{\zeta,\theta,\phi}\,.
\fe
Finally, $M_{mn}$ generates an $SO(7)$ rotation of the auxiliary fields $K_m$ by
\ie
M_{mn}=\n_{[m} \CC^\m D_\m \n_{n]}\,.
\fe

Consequently, to study the BPS locus, it suffices to restrict our attention to the base of the $S^1_\tau$ fibration given by a half three-ball $HB^3$ at $\tau=0$. To study the remaining seven BPS equations, it's convenient to make use of the Weyl invariance of the SYM 
and map the BPS equations to those on the warped geometry $HB^3 \times_w S^1_\tau $ such that the metric on the base $HB^3$ is flat (with coordinates $\tilde x_i$)
\ie
ds^2=d\tilde x ^2+{1\over 4}(1-|\tilde x|^2)^2d\tau^2~~{\rm where}~~|\tilde x|\leq 1,~\tilde x_1\geq 0\,,
\fe
related to the   original metric by
\ie
ds^2_{S^4}={ds^2\over (1+|\tilde x|^2)^2}\,.
\label{HS4toS1HB3}
\fe
Then the boundary $S^3$ is now located at $\tilde x_1=0$, and the $HS^2_{\rm YM}$ at $|\tilde x|=1$.

We refer the readers to Appendix~\ref{app:coords} for explicit relations between the various coordinate systems. In particular note that at $\tau=0$, $\tilde x_i=x_i$ where $x_i$ for $i=1,2,3$ are part of the stereographic coordinates. Below we will focus on the BPS equations on the $\tau=0$ base, thus for convenience, we will simply use $x_i$ for coordinates on $HB^3$. For convenience, we define another set of auxiliary spinors which at $\tau=0$ are given by 
\ie
\tilde \n_i= \CC_{8i} \ve,~\tilde\n_4=\CC_{86}\ve,~~\tilde\n_{i+4} =\{\CC_{87}\ve,\CC_{89}\ve,\CC_{80}\ve \} \,,
\label{psv2}
\fe
and related to the $\n_m$  introduced earlier in \eqref{psv1} by an $SO(7)$ rotation. We also redefine the auxiliary scalar fields $K_m$ as $\tilde K_m$ accordingly such that $K_m \n^m =\tilde K_m \tilde \n^m$.

The \textit{real} components of the remaining nontrivial seven BPS equations from \eqref{BPSeqn} (at $\tau=0$) 
can then be written in the following simplified form \cite{Pestun:2009nn}
\ie
& {1-x^2\over 1+x^2} \left(
D_k \Phi_8-{1\over 2}[\phi_i,\phi_j]
\ep_{ijp} T_{pk}
\right)+
{1\over 2}F_{ij}\ep_{ijk}-
[\Phi_6, \phi_j] T^{-1}_{jk}
 -{2x_k\over 1+x^2} \Phi_8 
=0\,,
\\
& {1-x^2\over 1+x^2} [\Phi_6,\Phi_8]+ D_i\phi_jT_{ij}^{-1}  -{2 x_j \over 1+x^2}\phi_j=0\,,
\\
& {1-x^2\over 1+x^2} \left(
[\Phi_8,\phi_i]  T_{ik} +D_i\phi_j (\ep_{ijp}T_{pk}- \ep_{kjp} T_{pi}+\ep_{ijk} )
\right)
+
D_k \Phi_6 
+  {2\over 1-x^2}x_k \Phi_6
-{2  x_j\over 1+x^2}\ep_{ijk}\phi_i=0\,.
\label{BPSeqnsim}
\fe
Recall that
\ie
(\phi_1,\phi_2,\phi_3)=(\Phi_7,\Phi_9,\Phi_0)\,.
\fe
For convenience we have introduced  the matrix
\ie
T_{ij}\equiv  \D_{ij}+ {2 x_i x_j \over   1-x^2 }\,,
\fe
and its inverse
\ie
T^{-1}_{ij}\equiv  \D_{ij}- {2 x_i x_j \over   1+x^2 }\,.
\fe
The \textit{imaginary} components of the  seven BPS equations determine the auxiliary fields $K_m$ in terms of the physical SYM fields:
\ie
-{i\over 2}  \tilde K_i (1-x^2)=& \ep_{ijk} x_j D_k \Phi_6 -x_j D_j \phi_i  +x_i D_j \phi_j -\phi_i \,,
\\
-{i\over 2}\tilde  K_4 (1-x^2)=&\Phi_8-{1\over 2}\ep_{ijk} x_i( F_{jk}-[\phi_j,\phi_k])\,,
\\
-{i\over 2}\tilde  K_{4+i} (1-x^2)=&  x_j F_{ji}+  \ep_{ijk}x_j [\Phi_6,\phi_k] \,.
\label{Kgen}
\fe
One can easily check that the BPS boundary conditions \eqref{osD5}, \eqref{osNS5} and \eqref{osDir} specified in the previous section are compatible with the  BPS equations \eqref{BPSeqn}. For Nahm pole boundary condition \eqref{osD5}, we have the following solution to \eqref{BPSeqnsim}
\ie
X_a= -{1\over x_1}t_a ,~
\tilde  K_4={2i\over x_1(1-x^2)}t_1,~
\tilde  K_{4+i}=0,~
\tilde  K_i=0\,,
\fe
with all other fields vanishing.  For the general Dirichlet boundary condition \eqref{osDir} we present explicit solutions to the BPS equations \eqref{BPSeqnsim} below. 
 
In the absence of disorder type defects such as 't Hooft lines, surface operators or Nahm pole boundary conditions, we can focus on smooth solutions of \eqref{BPSeqn} which requires setting $\Phi_6=\Phi_8=0$ \cite{Pestun:2009nn}. With this restriction, the BPS equations \eqref{BPSeqn} can be further simplified
\ie
& -{1\over 2}{1-x^2\over 1+x^2} 
[\phi_i,\phi_j]
\ep_{ijp} T_{pk}+{1\over 2}F_{ij}\ep_{ijk} 
=0\,,
\\
&   D_i\phi_jT_{ij}^{-1}  -{2 x_j \over 1+x^2}\phi_j=0\,,
\\
& {1-x^2\over 1+x^2} \left(
D_i\phi_j (\ep_{ijp}T_{pk}- \ep_{kjp} T_{pi}+\ep_{ijk} )
\right)
-{2  x_j\over 1+x^2}\ep_{ijk}\phi_i=0\,,
\label{simBPSeqn}
\fe
and the expressions for the auxiliary fields also simplify to
\ie
{i } \tilde K_i =&    x_j D_j \phi_i +x_j D_i \phi_j  -x_i D_j \phi_j +2\phi_i\,,
\\
{i }\tilde  K_4  =&   0\,,
\\
{i }\tilde  K_{4+i}  =&  x_j F_{ij} +x_j[\phi_i,\phi_j]\,,
\fe
after using \eqref{Kgen} and \eqref{simBPSeqn}. To understand the moduli space of solutions to \eqref{simBPSeqn}, it is useful to define the twisted scalar fields 
\ie
\tilde \phi_i \equiv \phi_i T^{-1}_{ij}
\fe
and then \eqref{simBPSeqn} become
\ie
&F_{ij}-[\tilde \phi_i,\tilde \phi_j]=0
,\quad
\epsilon_{ijk}D_i \tilde{\phi}_j=0
,\quad 
(1+x^2) D_i \tilde \phi_i+2{3+x^2\over 1-x^2} z_i \tilde \phi_i=0\,.
\fe
If we further define the twisted complex connection (not to be confused with the emergent gauge field $\cA$ for the 2d YM)
\ie
\tilde\cA \equiv A + i\tilde \phi_i dx^i\,,
\fe
the above equations simply says $\tilde\cA$ is a flat $G_\mC$-connection
\ie
F(\tilde\cA)\equiv d\tilde\cA + \tilde\cA \wedge \tilde\cA =0\,,
\fe
with a partial gauge fixing condition 
\ie
d_{\tilde\cA} *_h \tilde \phi =0\,,
\label{partialgf}
\fe
where the Hodge dual is take with respect to the following Weyl flat metric  on the $HB^3$  
\ie
h_{ij}={(1+x^2)^2\over (1-x^2)^4}\D_{ij}\,.
\fe
The solutions are flat connections parameterized by functions $g_\mC:B^3\to G_\mC$ as
\ie
\tilde\cA =g_\mC^{-1} d g_\mC\,.
\fe
To eliminate the residual gauge redundancy, we can implement a complete gauging fixing (that includes \eqref{partialgf}) by
\ie
d_{A_\mC} *_h (g_\mC^{-1} d g_\mC)=0\,,
\fe
which amounts to an elliptic second order differential equation. Consequently the solutions are parametrized by the values of $g_\mC$ on the boundary $HS^2_{\rm YM} \cup B^2$ of $HB^3$. In summary, the moduli space of solutions to \eqref{simBPSeqn} is completely determined by the boundary values of $g_\mC$ or equivalently, the values of $\tilde \phi_i$ on $HS^2_{\rm YM} \cup B^2$. 

In particular, the usual Dirichlet boundary conditions \eqref{osDir} corresponds to a Neumann type boundary condition for $g_\mC$ on the $x_1=0$ boundary $B^2$ of the $HB^3$, such that
\ie
\phi_1={a\over 1+x^2},\quad \tilde K_1=-K_5=-{a\over 1+x^2}\,.
\fe
Note that we have taken into the Weyl factor $(1+x^2)$ compared to \eqref{osDir}. Here $a$ denotes a Cartan element of the gauge algebra $\mf{g}$.

\subsection{The 2d action on the BPS locus }
Now that we understand the space of solutions to the BPS equations \eqref{BPSeqn}, we need to determine the effective action on this BPS locus. In the case of $\cN=4$ SYM on $S^4$, this was done in \cite{Pestun:2009nn}. Here we give a streamlined derivation for the theory on $HS^4$ with appropriate boundary conditions. Note that we will not assume the regularity condition $\Phi_6=\Phi_8=0$ in the derivation until the very end keeping in mind future applications to include disorder defect operator insertions. Using the covariant constantness of the SYM fields in the $\tau$ direction on the BPS locus \eqref{covtaucst}, the SYM action reduces to an integral over the base $HB^3$ \cite{Pestun:2009nn}
\ie
S_{HB^3}=&-{2\pi \over  g_4^2} \int_{HB^3} d^3 x \,\bigg(
 {1-x^2\over 2}\tr\left (
  {1\over 2}F_{MN}F^{MN}
+{2\over 1-x^2}\Phi_A\Phi^A
- K_mK^m
\right )
+D_i\tr\left ( x_i {1-x^2\over 1+x^2}\Phi_A \Phi^A\right ) \bigg)
\label{SHB3}
\fe
where we have kept the total derivative terms.

We define $\CC_\tau \equiv  {v^\m \CC_\m \over v^\n v_\n}$ such that $\ve \CC_\tau \ve=1$ and consequently  $\n_m \CC_\tau \n_n =\D_{mn}$.
Then using \eqref{SUSYos} and setting $\D \Psi=0$, we can determine the contributions from the auxiliary fields to $S_{HB^3}$,
\ie
-K_mK^m=&\left ({1\over 2} F_{PQ}\ve \tilde \CC^{PQ}+2\Phi^B \tilde  \ve   \tilde \CC_B \right) \CC_\tau \left(
{1\over 2}F_{MN}\CC^{MN} \ve -2 \Phi^A \tilde \CC_A  \tilde \ve
\right)\,.
\fe
At $x_4=0$, we have 
$\CC_\tau \equiv -\CC_{\hat 4} {1\over 1-x^2}$   and $\tilde \ve \CC_\tau \tilde \ve = {1\over 1-x^2}$, consequently 
\ie
-K_mK^m=& 
-4\Phi^A \Phi^B \tilde  \ve   \tilde \CC_B \CC_\tau \tilde \CC_A  \tilde \ve +{1\over 4}F_{PQ}F_{MN}\ve \tilde \CC^{PQ} \CC_\tau \CC^{MN}\ve 
+
2\Phi^B F_{MN}\tilde \ve   \tilde  \CC_B \CC_\tau \CC^{MN}\ve  
 \\
=& {4\over 1-x^2}\Phi^A\Phi_A-{1\over 2}F_{MN} F^{MN}-{1\over 4 (1-x^2)} F_{PQ}F_{MN}\ve \CC^{MNPQ\hat 4}\ve 
\\
&-{2\over 1-x^2}\Phi_B F_{MN}\tilde \ve   \tilde \CC^{BMN \hat 4}\ve +{4\over 1-x^2}\Phi^B F_{NB}\tilde  \ve   \CC^{ N \hat 4}\ve 
\fe
where we have used the equations \eqref{covtaucst} at $\tau=0$. Plugging the above back into \eqref{SHB3}, and using
\ie
{1\over 4} \tr ( F_{PQ}F_{MN})\ve \CC^{MNPQ\hat 4}\ve 
 =&\,
\ve \CC^{iA jB \hat 4}\ve D_i \tr (\Phi_A F_{jB} )
+{1\over 3} \ve \CC^{i ABC \hat4} \ve D_i\tr (\Phi_A F_{BC})
+
\ve \CC^{i jk A \hat4}\ve D_i\tr (\Phi_A F_{jk})\,,
\fe
with
\ie
& D_i \bigg(\ve \CC^{iA jB \hat 4}\ve   \tr (\Phi_A F_{jB} )
+{1\over 3} \ve \CC^{i ABC \hat4} \ve \tr (\Phi_A F_{BC})
+
\ve \CC^{i jk A \hat4}\ve  \tr (\Phi_A F_{jk}) 
\bigg) 
\\
=&{1\over 4} \tr ( F_{PQ}F_{MN})\ve \CC^{MNPQ\hat 4}\ve 
+
2\tr ( \Phi_A F_{MN})\tilde \ve \CC^{AMN\hat 4}\ve \,,
\fe
and
\ie
D_i \tr (\Phi^B \Phi_B \tilde \ve \CC^{i \hat 4} \ve)
=&
2 \tr (\Phi^B F_{iB}) \tilde \ve \CC^{i \hat 4} \ve
+
3 \tr (\Phi^B \Phi_{B})  \,,
\fe
we can rewrite the action on $HB^3$ as a total derivative
\ie
S_{HB^3}=& { \pi \over  g_4^2} \int_{HB^3} d^3 x
D_i 
\bigg(\ve \CC^{iA jB \hat 4}\ve   \tr (\Phi_A F_{jB} )
+{1\over 3} \ve \CC^{i ABC \hat4} \ve \tr (\Phi_A F_{BC})
+
\ve \CC^{i jk A \hat4}\ve  \tr (\Phi_A F_{jk}) 
\\
&
-2 \Phi^B \Phi_B \tilde \ve \CC^{i \hat 4} \ve
-2 x_i {1-x^2\over 1+x^2}\Phi_A \Phi^A 
\bigg) \,.
\fe
Thus the integral becomes
\ie
S_{HB^3}=S_{HS^2}+S_{B^2}\,,
\fe
with two boundary term at $|x|=1$, namely on the $HS^2_{\rm YM}$, 
\ie
S_{HS^2}=&
 { \pi \over  g_4^2} \int_{HS^2}  
x_i
\bigg(\ve \CC^{iA jB \hat 4}\ve   \tr (\Phi_A F_{jB} )
+{1\over 3} \ve \CC^{i ABC \hat4} \ve \tr (\Phi_A F_{BC})
+
\ve \CC^{i jk A \hat4}\ve  \tr (\Phi_A F_{jk}) 
\\
&
-2 \Phi^B \Phi_B \tilde \ve \CC^{i \hat 4} \ve
\bigg) \,,
\fe
and at $x_1=0$ 
\ie
S_{B^2}=&-{ \pi \over g_4^2} \int_{B^2}  
\bigg(\ve \CC^{1 A jB \hat 4}\ve   \tr (\Phi_A F_{jB} )
+{1\over 3} \ve \CC^{1  ABC \hat4} \ve \tr (\Phi_A F_{BC})
+
\ve \CC^{1 jk A \hat4}\ve  \tr (\Phi_A F_{jk}) 
\\
&
-2 \Phi^B \Phi_B \tilde \ve \CC^{1 \hat 4} \ve
\bigg) \,.
\label{SB2}
\fe
Using at $|x|=1$ the explicit forms of the following spinor bilinears
 \ie
 &x_i \tilde \ve \CC^{i \hat 4} \ve=1\,,
 \\
 &x_i \ve \CC^{i ABC \hat 4}\ve \tr (\Phi_A F_{BC})=4i \ep^{ijk}\tr (\phi_i \phi_j \phi_k)-6 \ep_{ijk} x_k \tr (\Phi_8 \phi_i \phi_j)\,,
 \\
 &x_i \ve \CC^{i A j B  \hat 4}\ve\tr (\Phi_A F_{iB})= 2 x_j \tr (\phi_j D_i \phi_i-\phi_i D_i \phi_j)\,,
 +
 2i(\D_{ij}-x_i x_j) \tr(\Phi_8 D_i\phi_j )
 -
 2i  (\D_{ij}-x_i x_j) \tr (D_i \Phi_8 \phi_j)\,,
 \\
 &\ve \CC^{1 jk A \hat4}\ve  \tr (\Phi_A F_{jk}) 
 =2 \ep_{ijk} x_i \tr (F_{jk} \Phi_8)
 -2i \ep_{ijk}x_k x_p  \tr ( F_{ij} \phi_p )\,,
 \fe
 we can further simply the boundary term on $HS^2_{\rm YM}$ to
 \ie
 S_{HS^2}
 =&- { \pi \over  g_4^2}\int_{HS^2}  
 \bigg(
  - 2 x_j \tr (\phi_j D_i \phi_i-\phi_i D_i \phi_j)
 -
 2i(\D_{ij}-x_i x_j) \tr(\Phi_8 D_i\phi_j )
 +
 2i  (\D_{ij}-x_i x_j) \tr (D_i \Phi_8 \phi_j)
 \\
&  +2 \phi_i\phi^i -2 \Phi_8^2 +4 i \Phi_8\phi_n+2 \Phi_6^2
 \bigg) \,.
 \fe
Let's introduce the projector 
 \ie
 P_{ij} \equiv  \D_{ij} -{n_i n_j} 
 \fe
 with $n_i \equiv {x_i\over |x|}$ denoting the unit normal to the $HS^2_{\rm YM}$, and the normal combination of scalar fields 
 \ie
 \phi_n\equiv n_i \phi_i\,.
 \fe
 Then assuming all the fields are all non-singular  on $HS^2_{\rm YM}$, the relevant BPS equations take the following simplified forms
 \ie
 P_{ij}D_i \phi_j=\phi_n,\quad \ep_{ijk} x_i(F_{jk}-[\phi_j,\phi_k])=2\Phi_8 \,.
 \fe
 Using these relations, we can further simplify $S_{HS^2}$ and obtain
  \ie
S_{HS^2} 
 =&- { 2\pi \over  g_4^2}\int_{HS^2}  dV_{S^2}
\tr (
  \phi_n^2+2i \phi_n\Phi_8- \Phi_8^2  +\Phi_6^2  
 ) 
+{ 2\pi    \over  g_4^2}\int_{S^1}  d\varphi     \tr(( \phi_n +  i\Phi_8 ) \phi_1 )
  \\
=& -{ 2\pi \over  g_4^2}\int_{HS^2}   dV_{S^2}
\tr (
\hat \phi_n^2  +\Phi_6^2  
) 
+{2 \pi    \over  g_4^2}\int_{S^1}  d\varphi     \tr(\hat \phi_n\phi_1 )\,,
 \fe
 where $dV_{S^2}$ is the volume form on $HS^2$,
 \ie
 dV_{S^2}=-{1\over 2}\ep_{ijk}x^i dx^j dx^k=\sin \theta d\theta d\varphi \,,
 \fe
 and the boundary term on $S^1$ at $x_1=0$ arises from integration by parts. In the last line we have defined the following combination of scalar fields 
 \ie
 \hat \phi_n=& x^i \phi_i + i \Phi_8\,,
 \label{hatphi}
 \fe
 which bears close relation to the emergent 2d Yang-Mills connection $\cA$
 \ie
 \cA=&A+i \epsilon_{ijk} x^j \phi^k dx^i \,.
 \fe
 Indeed its field strength is given by
 \ie
 \cF= d\cA+\cA \wedge\cA= \left (-{1\over 2}\ep_{ijk}x^i F^{jk} 
 +    \ep_{ijk} x^i \phi^j \phi^k 
 +2i \phi_n 
 -i P_{ij} D_i \phi_j
  \right) dV_{S^2}
    =i \hat \phi_n  dV_{S^2}\,,
 \fe
where we have used the BPS equation \eqref{BPSeqn}. In other words,  the action on $HS^2_{\rm YM}$ is given by the 2d Yang-Mills up to boundary terms and a non-interacting term in $\Phi_6$
  \ie
  S_{HS^2} 
   =& S_{\rm YM} -{ 2\pi \over  g_4^2}\int_{HS^2}   dV_{S^2}
  \tr  \Phi_6^2  
  +{2 \pi    \over  g_4^2}\int_{S^1}  d\varphi     \tr(\hat \phi_n\phi_1 )\,
  \label{HS2sim}
  \fe
where
 \ie
 S_{\rm YM} \equiv -{ 4\over  g_{\rm YM}^2}\int_{HS^2}   dV_{S^2} \tr (\star \cF)^2\,
 \fe
  with the 2d Yang-Mills coupling $g_{\rm YM}$ related to the 4d SYM coupling $g_4$ by
  \ie
   g_{\rm YM}^2=-{g_4^2\over 2\pi R^2}\,.
     \fe
 In the following we will show that the boundary term in \eqref{HS2sim} above  cancel together with $S_{B^2}$ in \eqref{SB2} for the Neumann and Dirichlet boundary conditions.

Now let us examine the boundary term \eqref{SB2}. Assuming $\Phi_6=0$ and the rest of the fields are regular on $B^2$, we  have 
 \ie
 S_{B^2} 
   &= -{ \pi \over  g_4^2} \int_{B^2}  
 \bigg( 
 -2(1-x^2)\Phi_8 [\phi_2,\phi_3]
 +D_j((1+x^2) \phi_i T^{-1}_{ij}\phi_1)
 +2i D_i(\phi_1\Phi_8 x_i)
  \\
 &-4i \Phi_8 x_i D_i \phi_1
 +2 F_{23}\Phi_8(1+x^2)
 +4i F_{ij} x^i \phi_k \ep_{1jk}
 -4 i\phi_1 \Phi_8+4 \phi_1 x_i \phi_i
  \\
 &
 -2 \phi_1 D_i\phi_j (1+x^2)T^{-1}_{ij}
 -2\phi_i D_i \phi_1+2\phi_1 D_i \phi_i
 \bigg) \,,
 \fe
 where the spacetime indices $i,j$ are restricted to take values $2,3$ along $B^2$.
 
For NS5-type or Neumann boundary condition \eqref{osNS5}, the boundary action $S_{B^2}$ vanishes identically since every term involves one of the fields $\phi_2, \phi_3, \Phi_8$ which are zero by \eqref{osNS5} on $B^2$. The 4d Neumann boundary condition translates into the Neumann boundary condition for the 2d YM connection
\ie
{\rm Neumann:}\quad\quad \left. \cF_{\theta \phi} \right |_{\theta={\pi\over 2}} =0 \,.
\fe

The Dirichlet boundary condition \eqref{osDir}, due to the Weyl transformation in \eqref{HS4toS1HB3}, corresponds to
\ie
D_i((1+x^2)\phi_1)=0,~[\phi_2,\phi_3]=0,~F_{ij}=0
\fe
at $x_1=0$ with $i=2,3$.  Then we can simplify 
\ie
S_{B^2}
 =& -{ \pi \over  g_4^2} \int_{B^2}  
\bigg( 
  D_j((1+x^2) \phi_i T^{-1}_{ij}\phi_1)
+2i D_i( x_i \phi_1\Phi_8 ) 
+4 \phi_1 x_i \phi_i
\\
&
-2 \phi_1 D_i\phi_j (1+x^2)T^{-1}_{ij}-2 \phi_i D_i \phi_1
 +2\phi_1 D_i \phi_i
\bigg) 
\\
   =&- { 2\pi \over  g_4^2} \int_{S^1}  
       \phi_1 \hat \phi_n\,,
 \fe
 after integration by parts. This precisely cancels the boundary piece in \eqref{HS2sim}. The 4d Dirichlet boundary condition \eqref{osDir} corresponds fixing the holonomy of the 2d YM connection on the boundary $S^1$ of $HS^2_{\rm YM}$\footnote{Note that up to a gauge transformation, $A_i=0$ on the boundary.}
 \ie
{\rm Dirichlet:}\quad\quad \oint_{\theta={\pi\over 2}}  \cA =i\oint_{\theta={\pi\over 2}} d \varphi \, \phi_1= i \pi a\,.
 \fe

\subsection{Summary and comments on the localization computation}
To summarize the computation in the previous subsections, we find convincing evidence that the $\cN=4$ SYM with gauge group $G$ on $HS^4$ with  non-singular BPS boundary conditions  localizes with respect to the supercharge $\cQ$ to 2d YM with the same gauge group on  the $HS^2_{\rm YM}$ with corresponding boundary conditions. 

To prove such a statement rigorously, one would need to compute the one-loop determinant for fluctuations normal to the BPS locus associated to $\cQ$. This is complicated by the fact that the relevant operator is not transversally elliptic everywhere, which is expected since this localization procedure lands on a two dimensional quantum field theory rather than a zero dimensional matrix integral as in \cite{Pestun:2007rz}. However it is believed due to the $\cN=4$ supersymmetry of the setup such determinant factor $\Delta_\cA=1$ \cite{Pestun:2009nn}. In the case without boundaries, this conjecture states that, in the absence of disorder-type defects,
\begin{center}
	4d $\cN=4$ $G$ SYM on $S^4$   $\xrightarrow{\cQ-{\rm localization}}$ 2d $G$ cYM on $S^2$  
\end{center}
with YM action
\ie
S_{\rm YM} \equiv -{ 1\over   g_{\rm YM}^2}\int_{HS^2}   dV_{S^2} \tr (\star \cF)^2,
\quad 
g_{\rm YM}^2=-{g_4^2\over 2\pi R^2}.
\label{YM2}
\fe
where we have taken into account the Weyl transformation from $HB^3\times_w S^1$ back to $S^4$. 

Here cYM denotes the 2d YM theory constrained to the zero instanton sector \cite{Witten:1991we,Witten:1992xu}. This conjecture confirms the perturbative findings involving correlation functions of Wilson loops and local operators \cite{Drukker:2007qr,Drukker:2007dw,Drukker:2007yx,Young:2008ed, Bassetto:2008yf} and has since passed various nontrivial checks involving 't Hooft loops \cite{Giombi:2009ek} and defect correlators on the Wilson loops \cite{Giombi:2018qox,Giombi:2018hsx}.

Here our computation suggests that
\begin{equation*}
\begin{tikzcd}[column sep=6pc]
\text{4d $\cN=4$ $G$ SYM on $HS^4$ with BPS Dirichlet (Neumann) b.c.}   \arrow{d}{\cQ-{\rm localization}}
 \\
\text{2d $G$ cYM on $HS^2$ with Dirichlet (Neumann) b.c.}
\end{tikzcd}
\end{equation*}

In the later sections, we will provide additional evidence for $\Delta_\cA=1$ by comparing the disk partition functions of the 2d cYM to known results from the usual localization of \cite{Pestun:2007rz} as well as AGT \cite{Alday:2009aq}.

The Neumann boundary condition can be enriched by coupling to localized degrees of freedom on the boundary while preserving the $\cQ$ supercharge. Recall the  boundary modes of the  SYM fields furnish a fluctuating 3d $\cN=4$ $G$ vector multiplet, which can be coupled to 3d $\cN=4$ SCFTs with  flavor symmetry $G$ on the Higgs branch via the momentum map multiplets of the 3d SCFT. As explained in Section~\ref{sec:TQMonthebdy}, acting on the boundary 3d fields, $\cQ$ descends to the supercharge studied in \cite{Dedushenko:2016jxl}, consequently  $\cQ$-localization reduces the boundary 3d path integral to that of a 1d topological quantum mechanics (TQM). In the end, we obtain from the 4d/3d setup, a coupled 2d/1d effectively theory, described by the cYM on $HS^2$ coupled to the TQM on the boundary $S^1$. We refer to such a 2d/1d system as the 2d defect-Yang-Mills theory (dYM).

\begin{equation*}
\begin{tikzcd}[column sep=6pc]
\text{4d $\cN=4$ $G$ SYM on $HS^4$ with BPS Neumann b.c. + 3d $\cN=4$ SCFT with $G$ symmetry}   \arrow{d}{\cQ-{\rm localization}}
\\
\text{ 2d  $G$ cYM on $HS^2$ with Neumann  b.c. + 1d TQM with $G$ symmetry}
\end{tikzcd}
\end{equation*}
Using the folding trick \eqref{folding}, our analysis generalizes immediately to 4d $\cN=4$ SYM on $S^4$ with BPS interfaces,
 \begin{equation*}
 \begin{tikzcd}[column sep=6pc]
 \text{4d $\cN=4$ $G_1$ SYM on $HS^4$  + 3d $\cN=4$ SCFT on BPS interface + 4d $\cN=4$ $G_2$ SYM on $\overline{HS}^4$ }   \arrow{d}{\cQ-{\rm localization}}
 \\
 \text{ 2d  $G_1$ cYM on $HS^2$  + 1d TQM on $S^1$+ 2d  $G_2$ cYM on $\overline{HS}^2$ }
 \end{tikzcd}
 \end{equation*}

 Before ending this section, let us comment on the Nahm pole or general D5-type boundary conditions \eqref{osD5} (and the corresponding interfaces from unfolding) in the  context of the previous subsection.
Due to the singularity at $x_1=0$ in \eqref{osD5}, the boundary term $S_{B^2}$ no longer vanishes but the steps leading to \eqref{HS2sim} still applies, consequently, one would hope to still retain a 2d YM effective description on $HS^2_{\rm YM}$ with modified boundary conditions. We leave the study of such boundary conditions for a future publication \cite{nahmp}.
   
\section{Defect Observables in the Defect-Yang-Mills}   
\label{sec:defectindym}
   
   \subsection{Correlation functions in the 2d dYM}
 As explained in the previous section, the $\cN=4$ SYM on $HS^4$ with half-BPS boundary conditions of the Dirichlet and Neumann types at the equator $S^3$ localizes to 2d cYM on $HS^2_{\rm YM}$ with corresponding boundary conditions at the equator $S^1_{\rm TQM}$. Furthermore, by coupling the 4d Neumann boundary condition to 3d $\cN=4$ matter, the resulting 2d cYM is enriched by a gauged topological quantum mechanics on the boundary $S^1_{\rm TQM}$.
   
The partition function of the 2d/1d dYM in general takes the following form
   \ie
   Z_{\rm dYM}=\int_{HS^2_{\rm bc}} D \cA D Q D\tilde Q e^{-S_{\rm YM}-S_{\rm TQM}}\,.
   \label{ZdYM}
   \fe
   where $\cA$ are 2d YM gauge fields on $HS^2_{\rm YM}$. When the boundary condition is of Neumann type, we have in addition $(Q,\tilde Q)$, the twisted combination of hypermultiplet scalars $(q_a,\tilde q_a)$
   \ie
   Q(\varphi)= u^aq_a(\varphi),~\tilde Q=u^a \tilde q_a (\varphi)~{\rm with}~u^a=
   \left(
   \cos {\varphi \over 2}~
     \sin {\varphi \over 2}
   \right)
   \fe
    restricted to the $S^1_{\rm TQM}$ that parametrize the $\cQ$ cohomology among the boundary modes. The bulk 2d YM is defined by the action as in \eqref{YM2}. The TQM, in the case when the boundary 3d $\cN=4$ matter is given by free hypermultiplets, is described by
   \ie
   S_{\rm TQM}=\ell \int d\varphi \tilde Q_{i I} (D_\cA)^i{}_j Q^{j I}\,,
   \label{freehyper}
   \fe
   which couples to the 2d YM through the covariant derivative $D_\cA\equiv d-\cA$ with gauge indices $i,j$ and flavor indices $I,J$. More generally, we can include dynamical gauge fields ${\bm a}_\varphi$ on the boundary in which case the covariant derivative is modified accordingly.
   
   Let us now give the descriptions of the $\cQ$-cohomology defect observables in the dYM. Recall that the disorder-type defects are expected to modify the dYM. For example, the 't Hooft loops correspond to including higher instanton sectors in the 2d cYM \cite{Giombi:2009ek}. Below we will focus on the order-type defects.
For these cases it suffices to specify what the elementary (gauge non-invariant) fields in the 4d SYM and boundary 3d SCFT map to in the 2d/1d dYM:
   \ie\label{obid}
   {\rm On~} S^2_{\rm cYM}~(HS^2_{\rm cYM}):~~&  i (x^i \phi_i+ i \Phi_8)   \to  \star_{2d} \cF
,\quad    A+ {i }\phi^k  \epsilon_{ijk} {x^j} dx^i  \to  \cA\,,
   \\
 {\rm On~} S^1_{\rm TQM}:~~&
   (u^a  q_{a}, u^b \tilde q_b)  \to   (Q,\tilde Q),\quad 
 a_\varphi + i\phi_{\dot 1\dot 2} 
   \to  {\bm a_\varphi} \,.
   \fe
 The gauge-invariant operators built from these elementary fields include the ones studied in the literature as well as additional bilocal operators of the form
   \ie
    \tilde Q_{iI}(\varphi_1) ({\rm P}e^{\int_{\varphi_1}^{\varphi_2}\cA} )^i{}_j Q^{jJ}(\varphi_2),\quad
   \tilde Q_{iI}(\varphi_1) (\star_{2d} \cF)^i{}_j Q^{jJ}(\varphi_2)\,,
   \fe 
   which involves a mixture of bulk and boundary excitations. They are natural extensions of the quasi-topological observables in the 2d YM and topological observables in the TQM: the correlation functions of such observables  are independent of positions $\varphi_{1,2}$ on $S^1$ (as long as they don't cross each other or other insertions).

   The observables involving the 2d gauge field $\cA$ can be computed by standard techniques for 2d YM, with the zero-instanton constraint implemented in the absence of 't Hooft loops.\footnote{For reviews on 2d gauge theories see for example \cite{Blau:1993hj,Cordes:1994fc}. A modern perspective with codimension two defects is discussed in \cite{Nekrasov:2017rqy}.   
} On the other hand, the defect observables that involve $Q,\tilde Q$ on the boundary (interface) can be computed following \cite{Dedushenko:2016jxl}. For illustration, we take the 4d $\cN=4$ SYM with $G=U(N)$ and the boundary hypermultiplets transforming in the fundamental representation. 
   The action for the topological quantum mechanics \eqref{freehyper} is quadratic, so after gauge fixing the gauge field $\cA$ on the $S^1$ to
   \ie
   \left.\cA \right|_{S^1}= {\rm Diag}[\lambda_1,\lambda_2,\dots,\lambda_N]\,,
   \label{gfcA}
   \fe
   we obtain the propagator 
   \ie
   \la  Q_i (\varphi_1)  \tilde Q^j  (\varphi_2)\ra 
   =
   -\D_i^j{s(\varphi_1-\varphi_2)+\tanh (\pi \lambda_i)\over 4\pi} e^{-\lambda_i (\varphi_1-\varphi_2)}\,.
   \label{QQprop}
   \fe 
   The way one computes observables that involve $(Q,\tilde Q)$ is to integrate out $(Q,\tilde Q)$ in \eqref{ZdYM} by performing Wick contractions with \eqref{QQprop}. This gives a boundary (interface) contribution to the measure for $\cA$ in the dYM on the $S^1$. Then one performs the bulk path integral over $\cA$ with these boundary contributions and the partial gauge-fixing \eqref{gfcA} taken into account. The last step typically boils down to a matrix integral with interesting modified potentials compared to the previously encountered ones in SYM computations and they are due to the boundaries (interfaces) here.\footnote{See \cite{Iliesiu:2019xuh,Iliesiu:2019lfc} for an example of this, where a loop defect in 2d YM theory with a noncompact gauge group gives rise to the Schwarzian theory.}

\subsection{Counter-term ambiguities in the 2d YM and higher derivative deformations}
\label{sec:counter}
To fully specify the map between the 4d SYM and the  2d YM that arises from the $\cQ$-localization, there is one potential ambiguity we need to address. It is known that the 2d Yang-Mills theory is defined up to counter-terms associated to the area and curvature of the spacetime manifold $\Sigma$ (with boundaries) \cite{Witten:1992xu}
\ie
S_{\rm counter} =  k_1 g_{\rm YM}^2 \int_\Sigma dV_\Sigma  + k_2  \chi(\Sigma)\,,
\fe
where $dV_\Sigma$ is the volume form used in the definition of the 2d YM on $\Sigma$ and $\chi(\Sigma)$ is the Euler characteristic of $\Sigma$ normalized to be 2 for $\Sigma=S^2$. 

This counter-term ambiguity can be fixed by comparing with the $S^4$ partition function of $\cN=4$ $G$ SYM \cite{Pestun:2007rz}
\ie
Z_{S^4}^{\rm SYM}=
{1\over |W(G)|}  \int_\mf{t} [d a]   \prod_{\A \in {\rm roots} (G)} (\A,a)\,  e^{-   {2\pi   {{\rm Im}\tau}}  (a,a) } \,
\label{S4SYMgen}
\fe
where
\ie
\tau \equiv {4\pi i\over g_4^2}+{\theta\over 2\pi}\,,
\fe
and imposing the consistency condition
\ie
Z_{S^4}^{\rm SYM}=Z_{S^2}^{\rm cYM}.
\fe
Here $a$ parametrize the Cartan subalgebra $\mf{t}$ for $G$, $(\cdot,\cdot)$ denotes the standard Killing form on $\mf{h}$, and $[da]$ is the standard measure on $\mf{t}$ invariant under the Weyl group $W(G)$. For $a=\sum_{i=1}^{r} a_i \A_i$ where $\A_i$ are the simple roots,  $[da ]= \Lambda_{i=1}^{r} da_i \sqrt{\det C_{ij}}$ where $C_{ij}$ is the Cartan matrix for $G$. Below using explicit results in the $G=U(N)$ case, we show
\ie
Z^{\rm cYM}_{\Sigma} =  \left. Z^{\rm YM}_{\Sigma}\right|_{\rm 0-inst}e^{- {1\over 4}|\rho|^2 g_{\rm YM}^2 A(\Sigma)} 
\left( { c_G ( -i{{\rm Im}\tau})^{-\dim (G)/2}}\right)^{\chi(\Sigma)}\,,
\label{fixct}
\fe
where $A(\Sigma)$ is the area of $\Sigma$ with respect to the volume form $dV_\Sigma$, $\rho$ is the Weyl vector associate to $G$ and $c_G$ is Weyl denominator for $G$ given by a product over the positive roots $\Delta_+$
\ie
c_G= \prod_{\A \in \Delta_+} (\rho,\A)\,.
\fe
As shown in \cite{Coquereaux:2010dw} this can be rewritten as a product over factorials of the exponents $e_i$ of $G$ 
\ie
c_G=\omega_G\prod_{i=1}^{\rank G} e_i!
\fe
and $\omega_G=1$ for $G$ simply-laced, and for the rest we have $\omega_{B_r}=1/2^r$, $\omega_{C_r}=1/2^{r(r-1)}$,  $\omega_{G_2}=1/3^3$, and $\omega_{F_4}=1/2^{12}$. In particular for $G=SU(N)$, in terms of the standard basis $e_i$ for $\mR^N$,
\ie
\rho=\sum_{i=1}^N {N+1-2i\over 2}e_i\,,
\fe
and
\ie
C_{A_{N-1}}=G(N+1)\,.
\fe

The partition function of the standard 2d YM on $S^2$ is given by a weighted sum over representations $\lambda$ of $G$\footnote{See \cite{Yamatsu:2015npn} for a nice review on these representation data.}
\ie
Z^{\rm YM}_{S^2}=\sum_\lambda d_\lambda^2 e^{-  \pi R^2g_{\rm YM}^2 c_2(\lambda)}\,.
\fe

For simplicity we focus on the case with $G=U(N)$. The representations  of $U(N)$ are labeled by a sequence of non-increasing integers $\lambda_1\geq\lambda_2\dots \geq \lambda_N $. The sum $\sum_{i=1}^N \lambda_N$ determines the $U(1)$ charge whereas the $SU(N)$ representation corresponds to the Young diagram $\lambda =[\lambda_1-\lambda_N,\dots,\lambda_{N-1}-\lambda_N]$ with non-increasing column lengths. For convenience, we introduce an alternative label of the same representation $\lambda$ by an $N$-tuple of strictly decreasing integers $(\ell_1,\dots,\ell_N)$ with $\ell_j=\lambda_j-j+N$. The dimension and quadratic Casimir of the representation $\lambda$ are  given by
\ie
d_\lambda={\Delta(\ell_i) \over \Delta(N-i)},\quad c_2(\lambda)=-{N(N^2-1)\over 12}+\sum_{i=1}^N\left(\ell_i-{N-1\over 2}\right)^2
\fe
such that for the fundamental representation of $U(N)$ we have $c_2=N$. Here $\Delta$ is the Vandermonde determinant defined for $N$-tuples $x_i$ as,
\ie
\Delta(x_i)\equiv \prod_{1\leq i< j\leq N} (x_i-x_j)\,.
\fe 
Then we have
\ie
Z^{\rm YM}_{S^2}={e^{  {N(N^2-1) \over 12} \pi R^2g_{\rm YM}^2} \over N!} \sum_{\ell_i \in \mZ} \left( {\Delta(\ell_i) \over \Delta(N-i)} \right)^2 e^{-  \pi R^2g_{\rm YM}^2  \sum_{i=1}^N\left(\ell_i-{N-1\over 2}\right)^2}\,.
\fe

The cYM corresponds to the zero instanton sector in the 2d YM \cite{Pestun:2009nn}. Its $S^2$ partition function can be obtained by performing the following integral which implements the projection \cite{Witten:1991we,Bassetto:1998sr} 
\ie
\left.Z^{\rm YM}_{S^2}\right|_{\rm 0-instanton} =&{ 1\over N! \B_{N,g_{\rm YM}}^2}\int \prod_{i=1}^N dz_i\,  \Delta^2(z_i)e^{-  {\pi R^2g_{\rm YM}^2} \sum_{i=1}^Nz_i^2 } \,,
\fe
with 
\ie
\B_{N,g_{\rm YM}}={e^{{N(N^2-1)\over 24} \pi R^2 g_{\rm YM}^2 }\over \prod_{i=1}^{N-1} i!} ={e^{{N(N^2-1)\over 24} \pi R^2 g_{\rm YM}^2 }\over G(N+1)}\,.
\label{Bym}
\fe
Using the strange formula from Freudenthal and de-Vries, which relates the length of the Weyl vector $|\rho|$ to dimension and dual Coxeter number of the Lie group $G$ as $|\rho|^2={h \dim (G)\over 12}$, we can write for $U(N)$
\ie
\B_{N,g_{\rm YM}}={e^{{1\over 2}|\rho|^2 \pi R^2 g_{\rm YM}^2 }\over G(N+1)} \,.
\fe

Note that $g_{\rm YM}^2<0$ and the integral contour above for $z_i$ should be along the imaginary axis
\ie
a_j = {  iz_j  \over  {{\rm Im}\tau}} \in \mR \,.
\fe
Applying the above Wick rotation and taking into account the counterterms in \eqref{fixct} for $\Sigma=S^2$, we have 
\ie
Z^{\rm cYM}_{S^2} =&{ 1\over N!  }\int \prod_{i=1}^N da_i\,  \Delta^2(a_i)e^{-  2\pi {{\rm Im}\tau} \sum_{i=1}^Na_i^2 } 
\label{ZS2}
\fe
in perfect agreement with the familiar result \eqref{S4SYMgen} for $U(N)$ SYM.

We will also be interested in the 2d YM theories deformed by higher derivative interactions from higher degree Casimirs of the gauge group $G$. We find it convenient to introduce auxiliary field $\phi$, and write the deformed YM path integral as \cite{Witten:1992xu}
\ie
Z_\Sigma^{\rm YM'}=\int D A D\phi  \,\exp \left(
 i\int_\Sigma \tr(\phi F) + \sum_{p\geq2} {i^pg_{\rm YM}^{2p}\over 2^p} {t_p \over A(\Sigma)}\int_\Sigma dV_\Sigma  \tr (  \phi^p)
\right) \,,
\label{hdBF}
\fe
where $A(\Sigma)=\int_\Sigma dV_\Sigma$ is the area of $\Sigma$ and $t_2\equiv -{   A(\Sigma)\over g_{\rm YM}^2}$. Note that for the undeformed case, namely when $t_p=0$ for $p>2$, integrating out $\phi$ yields the original YM action \eqref{YM2}. With the deformations turned on, the same procedure gives rise to a combination of higher derivative couplings $\tr ((\star \cF) ^n)$ that depends on $t_p$.

By taking derivatives of the deformed partition function $Z_\Sigma^{\rm YM'}$ with respect to $t_p$, we can access the topological correlators of the Casimir operators $\tr ((\star \cF) ^n)$ in the 2d YM, which maps under \eqref{obid} to correlation functions of ${1\over 8}$-BPS operators in the $\cN=4$ SYM.\footnote{In the undeformed theory, $\left ({g_{\rm YM}^2 i\over 2  }\right)^p \tr(\phi^p)$ is equivalent to $\tr (\star \cF)^p$ up to lower dimensional terms, as a consequence of Dyson-Schwinger type equations.}

As explained in \cite{Witten:1992xu}, upon canonical quantization,  the wave-functions $\Psi(\cA)$ are class functions of the boundary $G$-holonomy $U\equiv {\rm P}e^{\oint \cA}$ and thus can be expanded in $G$-characters $\chi_\lambda(U)$. The gauge invariant operators $\tr ((\star \cF) ^n)$ (or $\tr ( \phi^n)$) act on $\Psi(\cA)$ (or rather $\chi_\lambda(U)$) as Casimirs of $G$, up to normal ordering ambiguities that involve  mixing between $\tr ((\star \cF) ^n)$ and its lower degree cousins. However there's a \textit{preferred} scheme \cite{Witten:1992xu} such that the deformations in \eqref{hdBF} simply translates to a shift in the Hamiltonian $H={g_{\rm YM}^2\over 4}c_2(\lambda)$ by higher Casimirs of the form 
\ie
  \tr (\phi^p) \chi_\lambda(U)  = \tr ((\rho+\lambda)^p)\chi_\lambda(U)\,,
\fe
where we slightly abuse the notation to denote the highest weight associated to the $G$-representation $\lambda$ also by $\lambda$, and in the above equation we have used the Killing form to identify the weight vectors with elements of the Cartan. The partition function of the deformed YM theory takes a rather simple form, which on $S^2$, is given by\footnote{It would be interesting to see explicitly if this scheme arises naturally for the $\cQ$-localization of the 4d $\cN=4$ SYM theory with higher derivative deformations. For the moment, this partition function (and the corresponding matrix model) serves as a trick to compute topological correlators of $\cO_p$ by taking derivatives with respect to $t_p$.} 
\ie
Z_{S^2}^{\rm YM'}=\sum_{\lambda} d_\lambda^2 e^{-\pi R^2 g_{\rm YM}^2 c_2(\lambda)+ \sum_{p\geq 3} i^p 2^{-p} t_p  g_{\rm YM}^{2p} \tr ((\rho+\lambda)^p)}\,.
\fe
The constraint to zero-instanton sector can be implemented by an integral projection as before. Consider the $G=U(N)$ case, for $\lambda$ labelled by $\ell_i$ as before, we have
\ie
(\rho+\lambda)_i=\ell_i -{N-1\over 2}\,,
\fe
thus one easily obtains the matrix model for the deformed 2d cYM\footnote{
We emphasize that this single-matrix model arises for the (deformed) cYM partition function with no insertions. Once we include insertions such as local operators and Wilson loops, we will obtain  multi-matrix models.}
\ie
Z_{S^2}^{\rm cYM'}=  &{ 1\over N!  }\int \prod_{i=1}^N da_i\,  \Delta^2(a_i)e^{-  2\pi {{\rm Im}\tau} \sum_{i=1}^Na_i^2 + \sum_{p\geq 3}    {t_p\over R^{2p}} \sum_{i=1}^N a_i^p} \,.
\label{dcYMS2}
\fe
This agrees with the expressions for the partition function of $\cN=4$ SYM on $S^4$ deformed by chiral couplings $\tau_p$ with $2\Im \tau_p=   -{t_p\over R^{2p}}$ 
\ie
S_{\rm SYM}\to S_{\rm SYM}+ \sum_{p\geq 3}  \left ( i\tau_p \int_{S^4} d^4 x \sqrt{g}  \,\cC_p   +{\rm c.c} \right)
\label{chiralcouplings}
\fe
where $\cC_p$ is a particular combination of the chiral primary $\cO_p$ and its bosonic superconformal descendants, chosen to preserve an $\mf{osp}(4|2)$ subalgebra of $\mf{psu}(4|4)$ on $S^4$ \cite{Gerchkovitz:2016gxx}.

We emphasize that the deformation in \eqref{chiralcouplings} is not $\cQ$-closed. As explained in the end of Section~\ref{sec:review2d}, $\cQ=\cQ_{\cN=2}^+ +\cQ_{\cN=2}^-$ is a combination of supercharges in two different $\mf{osp}(4|2)_\pm$ subalgebras. The above deformation in \eqref{chiralcouplings} is $\cQ_{\cN=2}^+$-closed but not $\cQ_{\cN=2}^-$-closed. The agreement between the resulting matrix model in \cite{Gerchkovitz:2016gxx} from $\cQ_{\cN=2}^+$-localization with that of the deformed YM in \eqref{dcYMS2} suggests that there exists a modification of \eqref{chiralcouplings} by $\cQ_{\cN=2}^+$-exact (and closed) terms such that the modified chiral coupling is $\cQ$-closed. We leave the study of such couplings for future.

\subsection{Comparison to known $HS^4$ partition functions of SYM}
The $HS^4$ partition functions of general $\cN=2$ gauge theories were studied in \cite{Gava:2016oep} and conjectural expressions for the cases with Dirichlet or Neumann BPS boundary conditions were given (see also \cite{Bullimore:2014nla}). These results were later justified in \cite{Dedushenko:2018tgx}. Here we show that these results are consistent with our 2d cYM description on the $HS^2_{\rm YM}$.

We start with the Dirichlet boundary condition in which case  the 2d YM on $HS^2_{\rm YM}$ has a fixed holonomy $U$ on the boundary $S^1$. The disk partition function of ordinary 2d YM is given by\footnote{We refer the reader to \cite{Tachikawa:2016kfc} for a quick summary and to \cite{Cordes:1994fc} for a comprehensive review on 2d Yang-Mills theories.}
\ie
Z^{\rm YM}_{HS^2}(U)=\sum_\lambda e^{- {1\over 2}{\pi R^2g_{\rm YM}^2}  c_2(\lambda)} \chi_\lambda(U)d_\lambda\,.
\fe
For simplicity we focus on the case with $U(N)$ gauge group. 
Parametrizing the $U(N)$ holonomy (up to conjugation) as $U=( e^{i \theta_1},\dots, e^{i \theta_N})$, the character for the representation $\lambda$ is
\ie
\chi_{\lambda}(\theta_i)={\det_{ij} e^{i \theta_i l_j}\over \det_{ij} e^{i \theta_i (N- j)}},~~i,j=1,2,\dots,N\,.
\fe
Putting the above ingredients together, we get the following explicit form of the 2d $U(N)$ YM disk partition function
\ie
Z_{HS^2}(\theta_i)={1 \over N!}\B_{N,g_{\rm YM}}\sum_{\ell_i \in \mZ} \Delta (\ell_i) e^{- {1\over 2}{\pi R^2g_{\rm YM}^2} \sum_{i=1}^N\left(l_i-{N-1\over 2}\right)^2 } {\det_{ij} e^{i \theta_i l_j}\over \det_{ij} e^{i \theta_i (N-j)}}
\label{PSHS2}
\fe
with the constant \eqref{Bym}. 
The constrained 2d YM     corresponds to the zero instanton sector in the 2d YM \cite{Pestun:2009nn}. Its disk partition function can be obtained by performing the following integral which implements the projection \cite{Giombi:2009ek}
\ie
\left. Z^{\rm YM}_{HS^2}\right|_{\rm 0-inst}= {1 \over N!}\B_{N,g_{\rm YM}}
\int \prod_{i=1}^N dz_i\,  \Delta(z_i)e^{- {1\over 2}{\pi R^2g_{\rm YM}^2} \sum_{i=1}^Nz_i^2 }{\det_{ij} e^{i \theta_i (z_j+{N-1\over 2})}\over \det_{ij} e^{i \theta_i (N-j)}}\,.
\fe
Taking into the counter-terms in \eqref{fixct},
\ie
Z^{\rm cYM}_{HS^2}(\theta_i) \equiv 
 {( -i{{\rm Im}\tau})^{-N^2/ 2} \over N!} 
\int \prod_{i=1}^N dz_i\,  \Delta(z_i)e^{- {1\over 2}{\pi R^2g_{\rm YM}^2} \sum_{i=1}^Nz_i^2 }{\det_{ij} e^{i \theta_i (z_j+{N-1\over 2})}\over \det_{ij} e^{i \theta_i (N-j)}}\,.
\fe
Note that $g_{\rm YM}^2<0$ and the integral contour above for $z_i$ should be along the imaginary axis. The result is 
\ie
Z^{\rm cYM}_{HS^2}(\theta_i)
=
i^{ N^2/2}(4\pi)^{-{N(N-1)\over 2}} 
\exp \left(
{-{1 \over 2 \pi R^2 g_{\rm YM}^2} \sum_{i=1}^N \theta_i^2 }
\right){ \Delta(\theta_i)\over \prod_{ i< j} \sin   {   \theta_i-\theta_j \over 2}  }\,.
\label{Zdir}
\fe
After analytically continuing $\theta_i$ to imaginary values and redefine
 \ie
a_j = {  i\theta_j  \over 2\pi }\,,
\fe
 we obtain 
\ie 
Z^{\rm cYM}_{HS^2}(a_i)
=
i^{ N^2/2} 2^{-{N(N-1)\over 2}} 
\exp \left(
{-\pi \Im\tau  \sum_{i=1}^N a_i^2 }
\right){ \Delta(a_i)\over \prod_{ i< j} \sinh     \pi( a_i-a_j)   }\,,
\label{UNd}
\fe
which is in agreement with the  $HS^4$ partition function of $\cN=4$ SYM with Dirichlet boundary condition as derived in \cite{Dedushenko:2018tgx}
\ie
Z^{\rm SYM}_{HS^4}(a_i)=Z^{\rm cYM}_{HS^2}(a_i) 
\fe
up to an $a_i$ independent prefactor.

Recall that the 2d YM theory is defined on arbitrary Riemann surfaces with boundaries. The general partition functions are determined by cutting and gluing. In particular, the $S^2$ partition must be given by gluing two $HS^2$ partition functions and integrating over the boundary holonomy $U$. We expect to same to be true with in the cYM:
\ie
Z_{S_2}^{\rm cYM}= \int [dU]Z_{HS_2}^{\rm cYM}(U)   {Z_{HS_2}^{\rm cYM} (U^{-1}) }\,,
\fe
where $[dU]$ denotes the Haar measure on the gauge group $G$. For $G=U(N)$, we have
\ie
&[dU]= {1\over N!}\prod_{i=1}^N {d\theta_i \over 2\pi} \prod_{i\neq j} (1-e^{i(\theta_i-\theta_j)}) 
=
{2^{N(N-1)}\over N!}\prod_{i=1}^N {d\theta_i \over 2\pi} \prod_{i< j} \left(
\sin {\theta_i-\theta_j\over 2}
\right)^2 \,.
\label{haarSUN}
\fe
Indeed from gluing the $HS^2$ partition functions \eqref{Zdir}, we get
\ie
Z_{S_2}^{\rm cYM}=
{1\over N!}  \int \prod_{i=1}^N dz_i\,  \Delta^2(a_i)e^{-   {2\pi   {{\rm Im}\tau}} \sum_{i=1}^Na_i^2 } \,,
\fe
in agreement with \eqref{ZS2}.

Next we consider the Neumann boundary condition in which case we integrate over the boundary holonomy $U$ with the Haar measure $[dU]$ 
\ie
Z_{HS^2_{\rm N}}^{\rm cYM}
=
\int [dU]  Z_{HS^2_{\rm D}}^{\rm cYM}(U)\,.
\fe
We have introduced the subscripts D and N here to distinguish between the hemisphere partition functions with Dirichlet and Neumann boundary conditions.

For $U(N)$ case, we find the explicit result,
\ie
Z_{HS^2_{\rm N}}^{\rm cYM}
=&
i^{N^2/2}   {\pi ^{-N(N+1)/2}\over2^N N!}
\int  {d\theta_i  }
 { \Delta(\theta_i) \prod_{ i< j} \sin   {   \theta_i-\theta_j \over 2}  } e^{-{1 \over 2 \pi R^2 g_{\rm YM}^2} \sum_{i=1}^N \theta_i^2  }
 \\
 =&
     {  2^{N(N-1)\over 2} i^{-N^2/2} \over N!}  
 \int da_i 
 { \Delta(a_i) \prod_{ i< j} \sinh \pi a_{ij}   } e^{-\pi {\rm Im}\tau  \sum_{i=1}^N a_i^2  }\,.
\label{UNn}
\fe

\subsection{Remarks on $SU(N)$ SYM in relation to AGT}
The AGT correspondence relates observables of $SU(N)$ $\cN=2^*$ SYM on $S^4$ to those in the $A_{N-1}$ Toda theory on a punctured torus $T^2$ where the gauge coupling $\tau$ is identified with the complex structure of the $T^2$ \cite{Alday:2009aq,Alday:2009fs,Alday:2010vg}. Here we review some known results (conjectures) in the literature in relation to what we find from the 2d dYM in the previous sections.

We will focus on the case without squashing, in which case the $S^4$ partition function of the $U(N)$ SYM is given by
\ie
Z^{\cN=2^*}_{U(N)}(\tau,m)={1\over N!} \int d^N a \prod_{i\neq j} a_{ij}  e^{-2\pi {\rm Im}\tau \sum_i a_i^2} {H(ia_{ij})^2\over H(ia_{ij}+m) H(ia_{ij}-m)} |Z_{\rm inst}(q,m,a_i)|^2
\,.\fe
  Here
 $H(z)=G(1+ z)G(1- z)$ captures the one-loop determinant and $Z_{\rm inst}(q,m,a_i)$ is the Nekrasov instanton partition function \cite{Nekrasov:2002qd,Nekrasov:2003rj} at $\ep_1=\ep_2=1$ and $q=e^{2\pi i \tau}$.  The latter is a complicated function at general $m$ but simplifies at special values
\ie
&\left.Z_{\rm inst}(q,m,a_i)\right|_{m=0}=1\,,
\\
&\left.Z_{\rm inst}(q,m,a_i)\right|_{m=\pm1}={1\over \eta(\tau)^N}\,,
\label{simplemass}
\fe
corresponding to the enhanced $\cN=4$ point and the $m_{\rm AGT}=0$ point respectively \cite{Pestun:2007rz, Okuda:2010ke}. 

The $SU(N)$ partition function is conjectured to be given by the $U(N)$ partition function after factorizing the $U(1)$ part  \cite{Alday:2009aq,Alday:2010vg}
\ie
Z^{\cN=2^*}_{U(N)}(\tau,m)=Z^{\cN=2^*}_{SU(N)}(\tau,m)Z^{\cN=2^*}_{U(1)}(\tau,m)\,,
\fe
which receives abelian instanton contributions\footnote{Here we have included a prefactor that comes from counter-terms of the type $\int  d^4 x \sqrt{g}\, \tau R_{\m\n\rho\sigma}R^{\m\n\rho\sigma} +\dots $ and  $\int  d^4 x \sqrt{g}\, \tau m^2 R +\dots$ as well as their complex conjugates. The supersymmetric completion of these counter-terms can be found in \cite{Butter:2013lta} where $\tau$ is treated as the bottom component of a background chiral multiplet and $m$ the bottom component of a background vector multiplet. These counter-terms are chosen here for $Z^{\cN=2^*}_{U(1)}(\tau,m)$ to have definite $SL(2,\mZ)$ properties (similar to the choice in \cite{Pestun:2007rz}).  }
\ie
Z^{\cN=2^*}_{U(1)}(\tau,m)=
q^{(N-1)-N m^2  \over 24}\left[\prod_{i=1}^\infty(1-q^i ) \right]^{ (N-1) -N m^2}
={1\over  \eta^{  N m^2-(N-1) }}\,.
\fe
Therefore
\ie
Z^{\cN=2^*}_{SU(N)}(\tau,m)={1\over N!} \int [da]  {H(ia_{ij})^2\over H(ia_{ij}+m) H(ia_{ij}-m)} \left|  q^{{1\over 2}\sum_i a_i^2} Z_{\rm inst}(q,m,a_i) \over \eta ^{  (N-1)- N m^2 }\right|^2\,.
\label{SUNPF}
\fe
At the special masses in \eqref{simplemass}, the partition functions are  
\ie
\left.Z^{\cN=2^*}_{SU(N)}(\tau,m)\right |_{m=0}=&  { (2\pi)^{{N-1\over 2}}  G(N+2) \over  N!\sqrt{N} (4\pi {{\rm Im}\tau})^{{N^2-1\over 2}}  |\eta|^{2(N-1)}}\,,
\\
\left.Z^{\cN=2^*}_{SU(N)}(\tau,m)\right |_{m=\pm 1}=&  { 1 \over  N!\sqrt{N} (2{{\rm Im}\tau})^{{N -1\over 2}}  |\eta|^{2(N-1)}}\,.
\label{smPF}
\fe

On the $A_{N-1}$ Toda side, we have a CFT with central charge 
  \ie
  c_{A_{N-1}}=N-1+N(N^2-1)Q^2 \stackrel{b=1}{=} (N-1)(4N^2+4N+1)\,,
  \fe
and the partition function $Z^{\cN=2^*}_{SU(N)}(\tau,m)$ corresponds to the one-point function of a semi-degenerate primary $V_\A$ on $T^2_\tau$. Here $\A$ labels the Toda momentum of semi-degenerate type. For general Toda theories and general $\A$, the conformal weights of $V_\A$ is
\ie
h(\A)=\bar h(\A)={(2Q-\A , \A)\over 2} \,,
\fe
where $Q=q \rho$   with $\rho={1\over 2}\sum_{\A \in \Delta^+} \A $ denoting the Weyl vector and $q= b+{1\over b}$ capturing the background charge. Note that for round $S^4$, we take $b=1$.
 For $SU(N)$ case the weights in the fundamental representation are $h_1,\dots,h_N$ defined by
\ie
h_i=e_i-{1\over N}\sum_{j=1}^N e_j\,,
\fe
and the Weyl vector in terms of $h_i$ is
\ie
\rho=\sum_{i=1}^N {N+1-2i\over 2} h_i\,.
\fe
Then the simple puncture on $T^2_\tau$ that engineers the $\cN=2^*$ $SU(N)$ SYM with mass $m$ translates to a Toda operator $V_\A$ with  
  \ie
  \A=N(1+m)h_{1}\,,
  \fe
and conformal weights
  \ie
  h(m)=\bar h(m)= {N(N-1)\over 2} \left(1-m^2\right) \,.
  \label{AGTweights}
  \fe

Recall the torus one-point function of primary operator $\phi_{h,\bar h}$ on $T^2$ satisfies
 \ie
 \la \phi_{h,\bar h} \ra_{-1/\tau}=\tau^h  \bar\tau^{\bar h} \la \phi_{h,\bar h} \ra_\tau
 \fe
under the modular $S$-transformation. By AGT correspondence, the conformal weights \eqref{AGTweights} dictate the modular property of \eqref{SUNPF}. Indeed it is easy to see in \eqref{smPF} that at the special mass $m=\pm 1$, $V_\A$ is identity and the $S^4$ partition function agrees with the partition function of $A_{N-1}$ Toda CFT, while at $m=0$,  $V_\A$ has weights $h=\bar h={N(N-1)\over 2}$ and the $S^4$ partition function transforms accordingly.

Moreover AGT interprets the integral form of  \eqref{SUNPF}  in terms of the conformal block decomposition of the torus one-point function in the Toda theory. The Coulomb branch parameter $a_i$ corresponds to the Toda momentum of the intermediate primaries 
\ie
\A(a)=Q+i \sum_{i=1}^N a_i h_i\,.
\fe
The first factor (one-loop determinant) in the integrand of \eqref{SUNPF} captures the OPE coefficient  involving the external semi-degenerate insertion and propagating non-degenerate primaries $\la V_{N(1+m)h_{1}}  V_{\A(a)} V_{2Q-\A(a)}\ra $ while the last two factors (classical and instanton contributions) give the torus one-point conformal block.
 
The conformal data (OPE coefficients, conformal blocks, boundary conditions etc.) of $A_{N-1}$ Toda CFT are not known in generality but we can be quite explicit in the case of $N=2$. This corresponds to the usual Liouville CFT  which is  solved by bootstrap methods \cite{Zamolodchikov:1995xb,Zamolodchikov:2001ah,Hadasz:2009sw}. The central charge of the relevant Liouville CFT  is 
$c=25$.
 
We focus on the special $\cN=4$ mass $m=0$. The Liouville torus one-point conformal block for an external primary of weight $h_{e}=1$ coincides with that of $h_e=0$,\footnote{This holds for general $c>1$ Liouville CFTs (see for example \cite{Kraus:2016nwo}) and follows from the fact that $\oint dz \,V_1(z)$ commutes with the left-moving Virasoro algebra. We thank Per Kraus for discussions on this point.} namely the Virasoro characters for intermediate operator of weight $h$
\ie
F_h(q)={q^{h-{c-1\over 24}} \over \eta(q)}\,,
\fe
with $h=1+a^2$ in terms of the Liouville momentum $a$.

The one-point function of a general operator $V_{1+ip}$ normalized by the two-point function (we follow the conventions of \cite{Balthazar:2017mxh,Balthazar:2019rnh})
\ie
\la V_{1+ip}(z,\bar z) V_{1+ip'}(0)\ra ={\pi \D(p-p')\over |z|^{4(1+p^2)}}\,,
\fe
in the Liouville CFT is
\ie
\la V_{1+ip} \ra_{\tau}
=& \int_{ \mR^+} d  a {\la V_{1-i a} |{V_{1+ip }}|V_{1+i a}\ra  } \left |{ q^{  a^2 }  \over \eta(q) }\right |^2\,.
\fe
The AGT dictionary suggests this related to the $S^4$ partition function up to an normalization\footnote{This corrects a typo in \cite{Tachikawa:2016kfc}.}
\ie
Z^{\cN=4}_{SU(2)}(\tau)= \lim_{p \to 0}{i\over \Upsilon_1(2+2ip)}  \la V_{1+ip} \ra_{\tau}\,.
\label{liouville1pf}
\fe
To compute the RHS, we need the three-point-function $\la V_{1+i p_1}(0) V_{1+i p_2}(1) V_{1+i p_3}(\infty)\ra$ which is given by the DOZZ formula for $b=1$ 
\ie
\cC(p_1,p_2,p_3) 
=&{1\over \Upsilon_1(1+i(p_1+p_2+p_3))}
\\
\times&
\left[
{2p_1 \Upsilon_1(1+2i p_1) \over \Upsilon_1(1+i (p_2+p_3-p_1)}
{2p_2 \Upsilon_1(1+2i p_2) \over \Upsilon_1(1+i (p_3+p_1-p_2)}
{2p_3 \Upsilon_1(1+2i p_3) \over \Upsilon_1(1+i (p_1+p_2-p_3)}
\right]\,.
\fe
Applied to the OPE coefficient appearing in \eqref{liouville1pf}, it gives
\ie
{i\cC(a,p,-a) \over \Upsilon_1(2+2ip)  }
=&{-8i a^2 p\over \Upsilon_1(-2ip)}
{  \Upsilon_1(1+2i a)  
  \Upsilon_1(1-2i a)  
  \Upsilon_1(1+2i p )  \over  \Upsilon_1(1+i (p+2a)  \Upsilon_1(1+i (p-2a)  \Upsilon_1(1+ip) \Upsilon_1(1-ip)}\,.
 \fe
We take the limit $p \to 0$ and use the identities $\Upsilon_1(1)=1$ and $\Upsilon_1(x)\sim x$ as $x\to 0$,  
\ie
\lim_{p\to 0}{i\cC(a,p,-a) \over \Upsilon_1(2+2ip)  }= 4a^2\,.
\fe
Thus
\ie
\lim_{p \to 0}{i\over \Upsilon_1(2+2ip)}  \la V_{1+ip} \ra_{\tau}
=& 2\int_{-\infty}^\infty  da a^2 \left |{ q^{  a^2 }  \over \eta(q) }\right |^2\,,
\fe
in agreement with \eqref{smPF} for $N=2$ at $m=0$.
 
The AGT correspondence is further enriched by incorporating defects in the gauge theory and in the 2d CFT. In the $SU(2)$ $\cN=4$ SYM case here,  BPS defects in the gauge theory translate to defects in the $c=25$ Liouville CFT. For example, Wilson-'t Hooft line operators are mapped to Verlinde loops, Gukov-Witten surface operators are mapped to insertions of degenerate primary operators on $T^2_\tau$, BPS boundary conditions are mapped to boundary states or branes in the Liouville CFT. In particular, from the general discussion in \cite{LeFloch:2017lbt},
the Dirichlet boundary condition \eqref{osDir} parameterized by the constant boundary value of the scalar  $a$  corresponds to the Ishibashi state $|V_{1+ia}\ra\ra$, while the Neumann boundary condition corresponds to the identity ZZ-brane defined by 
\ie
|{\rm ZZ}\ra=\int_{\mR^+} {da\over \pi} \Psi_{\rm ZZ}(a)|V_{1+ia}\ra\ra\,,
\label{ZZb}
\fe
where $\Psi_{\rm ZZ}(a)$ denotes the corresponding wave function  
  \ie
 \Psi_{\rm ZZ}(a)= 2^{5\over 4} \sqrt{\pi} \sinh(2 \pi a)\,.
\label{ZZwf}
 \fe
To figure out what observable (with boundaries) in the Liouville CFT that the gauge theory hemisphere partition function maps to, as explained in \cite{LeFloch:2017lbt}  (see also \cite{Bawane:2017gjf}), one consider a lift of the $\mZ_2$ involution of the $S^4$ given by $x_1\to -x_1$ to a $\mZ_2$ symmetry of the 6d spacetime manifold $S^4\times \Sigma$ of the $\cN=(2,0)$ $A_1$ SCFT. Here $\Sigma=T^2_\tau$ is the torus with holomorphic coordinate $z$ and a simple puncture at $z=0$, and we take $\tau$ to be purely imaginary (i.e. $\theta=0$). Then the relevant $\mZ_2$ symmetry acts by $z\to \bar z$ and $x_1\to -x_1$. The 6d orbifold $(S^4\times T^2_\tau)/\mZ_2$ has fundamental domains $HS^4\times T^2_\tau$ and $S^4\times A^2_\tau$ where $A^2_\tau$ denotes the annulus with modulus ${\rm Im}\tau$. Using the former fundamental domain, reduction of the 6d theory on $T^2_\tau$ naturally gives the $HS^4$ partition function of the $\cN=4$ $SU(2)$ SYM. In the latter case, reduction of the 6d theory on $S^4$ gives the $c=25$ Liouville CFT on $A^2_\tau$ with boundary states determined by the SYM boundary conditions via the dictionary in \cite{LeFloch:2017lbt}. Equivalence between the two reductions implies, for $\cN=4$ $SU(2)$ SYM with Dirichlet boundary condition parametrized by scalar vev $a$,  the hemisphere partition function equals the annulus one-point-function with the  Ishibashi boundary states $|V_{1+ia}\ra\ra$ on both boundaries and the insertion of  a boundary primary $B_{1+ip}$ on one of them. Therefore, up to a normalization constant $\cN_{\rm D}$, we have
 \ie
Z_{HS^4_{\rm D}} =\cN_{\rm D}\lim_{p \to 0}{i \over \Upsilon_1(-2ip)}  \la \la V_{1+ia }| B_{1+ip} | V_{1+ia } \ra \ra_{A^2_\tau}\,.
\fe
Similarly, for $\cN=4$ $SU(2)$ SYM on $HS^4$ with Neumann boundary condition, 
 \ie
Z_{HS^4_{\rm N}} =\cN_{\rm N}\lim_{p \to 0}{i \over \Upsilon_1(-2ip)}  \la {\rm ZZ} | B_{1+ip} | {\rm ZZ} \ra_{A^2_\tau}\,,
\fe
where $\cN_{\rm N}$ is a normalization constant.
The explicit form of the annulus Liouville correlators can be determined using boundary structure constants and the annulus one-point-blocks but we will not do it here. Here we simply note that from the ZZ wave function \eqref{ZZwf}, we have from the above equations
 \ie
Z_{HS^4_{\rm N}} \propto  \int da  \sinh ^2(2\pi  a)Z_{HS^4_{\rm D}} 
\fe
in agreement with what we have found in the previous section (see \eqref{UNd} and \eqref{UNn}). In other words, the ZZ wavefunction $\Psi_{\rm ZZ}(a)$ squared plays the role of the Haar measure \eqref{haarSUN} for $SU(2)$ that appears in the integral transform between the $HS^4$ partition function with Dirichlet and Neumann boundary conditions.
  
\section{Applications}
\label{sec:applications}
We now apply the 2d/1d defect-Yang-Mills (dYM) to compute defect observables in the $\cN=4$ SYM.

 \subsection{Kinematics of basic bulk-defect correlators}
 The correlation functions involving bulk and defect local operators in a CFT with conformal defects satisfy a number of nontrivial constraints as in the case of correlators with just bulk operators.  The residual (super)conformal symmetry of the setup dictates the allowed structures (conformal blocks) for the bulk-defect correlator. Unitarity constrains the OPE coefficients in the conformal block expansion.  Crossing symmetry, which swaps bulk and defect exchange channels, further constrains the OPE data.

These correlators and corresponding bootstrap constraints are explored in detail for general CFTs in \cite{Liendo:2012hy,Billo:2016cpy} and for the case of $\cN=4$ SYM in \cite{Liendo:2016ymz,Liendo:2018ukf}. Here we will focus on the simplest nontrivial bulk-defect correlators for illustration though our localization method applies more generally.

\subsubsection{One-point functions}
Let's consider a conformal defect along $x_1=0$ in $\mR^4$. The residual $SO(4,1)$ conformal symmetry  demands that the defect one-point function of a bulk operator $\cO$ vanishes unless it's a scalar \cite{Liendo:2012hy}, in which case its one-point function takes the form
\ie
\la\cO(x)\ra_{\mR_+^4} ={h_\cO\over |x^\perp|^{\Delta_\cO}}\,,
\fe
where the position dependence is fixed by the scaling dimension $\Delta$ of $\cO$ and here the perpendicular distance $x^\perp=x_1$. The coefficient $h_\cO$ contains the physical information of the defect CFT (with $\cO$ normalized by its two-point function).

The $\cQ$-cohomology of $\cN=4$ SYM contains ${1\over 8}$-BPS scalar operators at arbitrary locations on the $S^2_{\rm YM}$, but for simplicity let's focus on the ${1\over 2}$-BPS operators $\cO_p$ inserted at   $x^\m=(1,0,0,0)$. Recall that such an operator transform as $[0,p,0]$ under the $SO(6)_R$ R-symmetry and has scaling dimension $\Delta=p$. Then we have
\ie
\la\cO_p\ra_{\mR_+^4}=h_p\,.
\fe
Note that since the half-BPS boundary condition preserves the $SU(2)_H\times SU(2)_C$ R-symmetry subgroup, and the $SO(6)$ representation $[0,p,0]$ contains an $SU(2)_H\times SU(2)_C$ singlet only for even $p$, we conclude $h_p=0$ for $p$ odd.

Since the localization  naturally computes the CFT observables on $HS^4$, let's translate the above correlators on flat space to those on the (hemi)sphere. Thanks to the Weyl symmetry, the one-point function on $HS^4$ is related to that on $\mR_+^4$ by 
\ie
\la\cO(x)\ra_{HS^4} ={h_\cO\over  s_\perp(x)^{\Delta_\cO}}\,,
\fe
where $s(x)$ denotes the chordal distance between $x$ and the equator $S^3$. In the stereographic coordinates\footnote{Compared to the coordinates in \eqref{embtostereo}, we have redefined $x_\m$ by $2R x_\m$ here.}, we have
\ie
s_\perp(x)={2R  |x_1| \over \sqrt{1+x^2}\sqrt{1+x^2-x_1^2}}\,.
\fe
For the ${1\over 2}$-BPS operator $\cO_p$ inserted at the north pole $x^\m=(1,0,0,0)$, we have 
\ie
\la\cO_p\ra_{S^4}={h_p \over (\sqrt{2}R)^{ p} }\,.
\label{OS4}
\fe

\subsubsection{Bulk-defect two-point functions}
There are also nontrivial correlation functions between local operators $\cO$ in the bulk and $\cS$ on the defect. The simplest nontrivial example is given by the two-point function of scalar operators
\ie
\la \cO(x) \cS(0,\vec y) \ra_{\mR^4_+}={c_{\cO \cS}\over |x-y|^{2\Delta_\cS} |x_\perp|^{\Delta_\cO-\Delta_\cS}}\,,
\fe
whose form is fixed by the conformal symmetry $SO(4,1)$ \cite{Billo:2016cpy} and the dynamical information is contained in the coefficient $c_{\cO \cS}$. 

Such correlation functions in the $\cQ$-cohomology consists of the bulk ${1\over 8}$-BPS operators on the $HS^2_{\rm YM}$ and boundary ${1\over 4}$-BPS operators on the $S^1_{\rm TQM}$. For simplicity, we take $\cO$ to be the ${1\over 2}$-BPS operator $\cO_p$ inserted at $x^\m=(1,0,0,0)$ as in the previous subsection, and $\cS$ to be a ${1\over 2}$-BPS operator $\cS_r$ at $y^\m=(0,1,0,0)$ in the $SU(2)_H\times SU(2)_C$ representation $[2r,0]$ and has dimension $\Delta=r$. Then we have
\ie
\la \cO_p \cS_r  \ra_{\mR^4_+}={c_{p,r}\over 2^{r} }\,.
\fe
Once again, the $SU(2)_H\times SU(2)_C$ symmetry of the defect implies that $c_{p,r}=0$ if $p+r$ is odd. 

The corresponding correlator on $HS^4$ is given by 
\ie
\la \cO(x) \cS(0,\vec y) \ra_{HS^4}={c_{\cO \cS}\over s(x,y)^{2\Delta_\cS}  s_\perp(x) ^{\Delta_\cO-\Delta_\cS}}\,,
\fe
for general scalar operators. Here the chordal distance between two points is defined by
\ie
s(x,y)={2R |x-y| \over \sqrt{1+x^2}\sqrt{1+y^2}}\,.
\fe
Specializing to the BPS operators $\cO_p$ at $x^m=(1,0,0,0)$ and $\cS_r$ at $y^m=(0,1,0,0)$, we have
\ie
\la \cO_p \cS_r  \ra_{HS^4}={c_{p,r}  \over (\sqrt{2}R)^{p+r} }\,.
\fe

\subsection{D5 interface and large $N$ limit}
In this section, we consider defect observables in the $\cQ$ cohomology of the $\cN=4$ $SU(N)$ SYM in the large $N$ limit, which are dual to probe branes in the dual IIB string theory on $AdS_5\times S^5$. In particular, we focus on a class of interfaces introduced in \cite{DeWolfe:2001pq}, described by $\cN=4$ SYM coupled to a fundamental hypermultiplet on a codimension one hyperplane.\footnote{A localization computation of the free energy for the large $N$ theory  was done in \cite{Robinson:2017sup}.} In IIB string theory, such an interface is engineered by a single D5-brane intersecting with the $N$ D3-branes along three longitudinal directions. This interface can be further generalized, while preserving the same SUSY, by including additional parallel D5-branes as well as turning on worldvolume fluxes on the D5-branes. We leave the study of such defects to \cite{nahmp}.
  
The dYM associated to the given defect setup is described by \eqref{ZdYM} 
where the TQM sector is described by the following action for twisted anti-periodic scalars in the hypermultiplet $(\tilde Q_{i},Q^{j})$
 \ie
S_{\rm TQM}=&\ell \int d\varphi \, \tilde Q_{  i} (D_\cA)^i{}_j Q^{j  }\,.
\fe
Here $i,j=1,\dots,N$ are $SU(N)$ fundamental indices and the gauge covariant derivative is given by
\ie
(D_\cA)^i{}_j=(d-m)\D^i{}_j-\cA^i{}_j \,,
\fe
where we have included the mass parameter $m$ for the $U(1)$ flavor symmetry of the gauged hypermultiplet.

If we integrate out the 2d/1d fields in \eqref{ZdYM}, we obtain the matrix model
  \ie
  Z(\tau, m)= \int [d a ]   {
  	e^{-2\pi  {\rm Im}\tau  \sum_{i=1}^N a_i^2}  
  	\over  \prod_{i=1}^N   2\cosh(\pi (a_i+m ))}\,,
  \fe
  with  measure
  \ie
  &[da]\equiv \prod_{i=1}^N da_i \prod_{i<j} (a_i -a_j)^2\, \D\left(\sum_{i=1}^N a_i\right )\,.
  \fe
The dYM description \eqref{ZdYM} allows us to compute more general observables in the $\cQ$ cohomology. As we will see, for some simple examples, the (un-integrated) correlation functions can be related explicitly to derivatives of the deformed partition function $Z(\tau,m)$. This is useful since the quantity  $Z(\tau,m)$ admits simple solutions via standard large $N$ methods \cite{Brezin:1977sv} which we review in the next subsection.

\subsubsection{A quick review of the large $N$ matrix model}   
The relevant matrix model for $SU(N)$ $\cN=4$ dSYM (SYM with interface defect) is
\ie
Z_{\rm dSYM}(\tau, t_p,m)=\int \prod_{i=1}^N da_i \D  (\sum_{i=1}^N a_i) e^{-N^2 F(a_i,m,\tau,\tilde  t_p)}\,,
\label{mm}
\fe
where $F$ denotes the effective free energy\footnote{In this section, we set the radius of the $S^4$ and $S^2_{\rm YM}$ to be $R=1$.}  
\ie
N^2 F=-\sum_{i<j} 2\log |a_i-a_j|+{8\pi ^2 N\over \lambda} \sum_i a_i^2- t_p \sum_i a_i^p+\sum_i \log( 2 \cosh(\pi(a_i+m )))\,.
\fe
Recall $t_p$ is a constant source for bulk BPS operator $\cO_p$ and $m$ is a constant source for defect BPS operator $\cS$. The same matrix model arises from the localization of the $\cN=2^*$ SYM on $S^4$ with mass and chiral deformations that preserve the $\cN=2$ supersymmetry (see Section~\ref{sec:counter}).

In the large $N$ limit, we expect the eigenvalues to be dense thus introduce the normalized eigenvalue density $\rho(x)$  by 
\ie
\rho(x)\equiv {1\over N}\sum_{i=1}^N \D(x- a_i),\quad 
\int_{-\infty}^\infty dx \,  \rho(x)=1\,,
\label{rhonm}
\fe
such that the free energy can be written in terms of the following integrals
\ie
{F }=&- \int dx dy\rho(x)\rho(y)\log |x-y|+  \int dx \rho(x) V(x)+{1\over N} \int dx \rho(x) \log 2 \cosh(\pi(x+m))\,,
\label{largeNF}
\fe
with  potential
\ie
V(x)
=\sum_{p\geq 2} g_p x^p \,.
\fe
 where 
\ie
g_{p>2}=-{   t_p\over N}={  2  {\rm Im\,} \tau_p\over N},\quad g_2={8\pi^2\over \lambda}\,.
\fe

The matrix model \eqref{mm} then becomes an integral over the normalized distributions $\rho$.  
Note that the last term (due to the defect modes) in \eqref{largeNF} is suppressed by ${1\over N}$. Therefore to capture the leading effects of the defect, it suffices to use the saddle point distribution from varying the first two terms in \eqref{largeNF} with respect to $\rho$. The saddle point equation is
\ie
\dashint   {\rho(y) dy \over x-y}={1\over 2}V'(x)\,,
\label{saddleqn}
\fe 
where $\dashint $ denotes the principal value integral. This equation can be solved by the method of resolvents and the solution has compact support on an interval $[\m_-,\m_+]$,
\ie
\rho (x)={M(x)\over 2\pi}\sqrt{(\m_+-x)(x-\m_-)}\quad {\rm for~} \m_-\leq x \leq \m_+
\label{saddlerho}
\fe
where $M(x)=\sum_{k\geq 0} c_k x^k$ is a polynomial. Both $c_k$ and $\m_\pm$ are completely determined by the potential $V(x)$ which we explain below. 

The resolvent $\omega(x)$ is an analytic function on the complement of $[\m_-,\m_+]$ in $
\mC$ defined by
\ie
\omega(x)= \int_\mR  dy  {\rho(y) \over x-y}\,,
\label{resolprop}
\fe
and satisfies the asymptotic condition 
\ie
\omega(x) \to {1\over x}~{\rm for~} |x|\to \infty\,.
\label{omegaasymp}
\fe
Close to the branch-cut singularity in $\mR$, from \eqref{saddleqn} we have
\ie
\omega(x\pm i\epsilon)={1\over 2} V'(x)  \mp \pi i \rho(x)\,.
\fe
To determine $\rho(x)$ from the potential $V(x)$, we start by writing the following principal valued integral on the real axis
\ie
&{1\over 2\pi^2}\dashint_{\m_-}^{\m_+ }dx {\sqrt{(\m_+-y)(y-\m_-)}\over \sqrt{(\m_+-x)(x-\m_-)}}{V'(x)\over x-y}
=
{1\over 2\pi^2} 
\lim_{\ep\to 0}\dashint_{\m_-}^{\m_+ } dx {\sqrt{(\m_+-y)(y-\m_-)}\over\sqrt{(\m_+-x)(x-\m_-)}}{\omega(x + i\epsilon)+\omega(x - i\epsilon)\over x-y}
\fe
which can be rewritten as a contour integral over a counter-clock wise contour $\cC_\ep$ surrounding the segment $[\m_-,\m_+]$,
\ie
&{1\over 2\pi^2}\lim_{\ep\to 0}\oint_{\cC_\ep} dx {\sqrt{(\m_+-y)(y-\m_-)}\over \sqrt{(\m_+-x)(x-\m_-)}}{\omega(x) \over y-x}
+{i\over 2\pi}\lim_{\ep\to 0} (\omega(y + i\epsilon)-\omega(y - i\epsilon))\,.
\fe 
Deforming the contour $\cC_\ep$ to $\infty$ and using analyticity and boundedness \eqref{resolprop} of the resolvent, the first term above vanishes and second terms gives $\rho(y)$, so we have 
\ie
\rho(y)={1\over 2\pi^2}\dashint_{\m_-}^{\m_+ }dx {\sqrt{(\m_+-y)(y-\m_-)}\over \sqrt{(\m_+-x)(x-\m_-)}}{V'(x)\over x-y} \,,
\label{rhosol}
\fe
and a similar argument gives the contour formula for the resolvent
\ie
\omega(y)={1\over 4\pi i}\oint_{\cC_\ep}dx {\sqrt{(\m_+-y)(y-\m_-)}\over \sqrt{(\m_+-x)(x-\m_-)}}{V'(x)\over x-y}\,.
\label{resolventsol}
\fe
The integral in \eqref{rhosol} determines the coefficients $c_k$ for the  polynomial $M(x)$ in \eqref{saddlerho} in terms of the ends of the segment $\m_\pm$,
\ie
c_k=&\sum_{n=k+2}^\infty  \sum_{r=0}^{n-k-2} n g_n   b_r b_{n-k-2-r} \m_+^r \m_-^{n-k-r-2}   \,,
\label{ck}
\fe
where \ie
b_k \equiv {\Gamma\left (k+{1\over2 }\right)\over 2\sqrt{\pi } k!}\,,
\fe
and we have used the following identity in \cite{Rodriguez-Gomez:2016ijh}
\ie
{1\over  \pi } \dashint_{-a}^b dy {1\over \sqrt{(y-\m_-)(\m_+-y)}}{y^{n-1}\over y-x}
= \sum_{k=0}^{n-2} \sum_{r=0}^{n-k-2}   b_r b_{n-k-2-r} \m_+^r \m_-^{n-k-r-2}   x^k\,.
\fe
The asymptotic condition \eqref{omegaasymp} for $\omega(x)$ applied to \eqref{resolventsol} demands
\ie
&\oint_{\cC_\ep} dx {V'(x)\over  \sqrt{(\m_+-x)(x-\m_-)}}=0\,,\\
&\oint_{\cC_\ep} dx {x V'(x)\over  \sqrt{(\m_+-x)(x-\m_-)}}=-4\pi  \,,
\fe
which determines $\m_\pm$ in terms of $V(x)$.

For illustration, the Gaussian matrix model which describes the undeformed SYM on $S^4$ has 
\ie
V={8\pi^2\over \lambda} x^2 \,,
\fe
and the solution for $\rho$ is the familiar Dyson-Wigner semi-circle distribution
\ie
\rho_{0}(x)={2\over \pi \m^2}\sqrt{\m^2-x^2},\quad \mu={\sqrt{\lambda}\over 2\pi}\,.
\fe

Finally using the saddle point equation, we have to leading orders in the large $N$ limit 
\ie
\log 
Z_{\rm dSYM}(\tau, t_p,m) 
=- N^2 \int dx \rho(x) \left[
{1\over 2} V(x) -\log x 
+{1\over N}  \log( 2 \cosh(\pi(x+m)))
\right]
+\cO(N^0)\,.
\label{fulllogZ}
\fe
In the later sections, by taking derivatives with respect to the background sources, we can access correlation functions of bulk and defect operators in the large $N$ limit.

\subsubsection{Interface one-point function}

We would like to compute one-point functions of half-BPS operators $\cO_{p}$ in $\cN=4$ SYM with a D5-brane interface. In the 2d/1d dYM theory, using the dictionary \eqref{obid}, this is given by
\ie
\la \cO_p \ra_{\rm dSYM}
=&
{(-i)^p\over Z_{\rm dYM}}\int D \cA D Q D\tilde Q \, \tr(\star \cF)^p e^{-S_{\rm YM}(\cA)-S_{\rm TQM}(Q,\tilde Q,\cA)}
\fe
which is zero unless $p\in 2\mZ$ due to the $\mZ_2$ symmetry $\cA \to -\cA$ of the dYM theory (at $m=0$). Thus we focus on the operators $\cO_{p}$ with $p\in 2\mZ_+$. By introducing the background couplings $ t_p$, we can rewrite $\la \cO_p\ra_{\rm dSYM}$ as a derivative of the dYM partition function
\ie
\la \cO_{p}\ra_{\rm dSYM}
=&
{1\over   Z_{\rm dYM}}\left.\pa_{  \tilde t_{p}}\right|_{  \tilde t_{p}=0}\int D \cA D Q D\tilde Q \,   e^{-S_{\rm YM}(\cA)+{1\over 4\pi}\int  (-i)^p \tilde t_{p}\tr(\star \cF)^{p} - S_{\rm TQM}(Q,\tilde Q,\cA)}\,.
\fe
As explained in Section~\ref{sec:counter}, up to mixing with lower dimensional operators which we address later, the two-dimensional part of the above path integral gives rise to a single matrix model with a polynomial potential of degree $p$, and the potential is further modified by the path integral over the TQM fields $(Q,\tilde Q)$. The full matrix model is of the form \eqref{mm} and
\ie
\la \hat\cO_{p}\ra_{\rm dSYM}=&   \left.\pa_{  t_{p}} \log Z_{\rm dSYM} \right|_{  t_{p>2 }=0}\,.
\fe
Here $\hat\cO_{p}$ differs from $\cO_{p}$ by mixing with $\cO_{p'}$ for even $p'<p$. We do not need the explicit mixing relation here, since later we will normalize these operators once and for all by their two point functions. To ease the notation, we will not distinguish between $\hat\cO_{p}$ and $\cO_{p}$ below.

 In the large $N$ limit we have
\ie
\log Z_{\rm dSYM}(\tau, t_p) 
=- N^2 \int dx \rho(x) \left[
{1\over 2} V(x) -\log x 
+{1\over N}  \pi |x|
\right]
+\cO(N^0)
\fe
where we have  made the replacement $\log( 2 \cosh \pi(x))\to |x|$ since $x$ is of order $\lambda^{1\over 2}$ and we are interested in the leading terms in the ${1\over \lambda}$ expansion.

Using the formulae in the previous section, we find for even $p$,
\ie
\la    \cO_p \ra_{\rm dSYM}=  \pa_{t_p}\log 
\left. Z_{\rm dSYM}(\tau,t_p) \right|_{   t_{p>2}=0} 
=
 \frac{N  \left(\frac{\lambda }{4 \pi^2  }\right)^{p\over 2} \Gamma \left(\frac{p+1}{2}\right)}{\sqrt{\pi } \Gamma \left(\frac{p}{2}+2\right)}
-
\frac{2 \left(\frac{\lambda }{4 \pi^2}\right)^{\frac{p+1}{2}} \Gamma \left(\frac{p+1}{2}\right)}{\sqrt{\pi}  (p+1) \Gamma \left(\frac{p}{2}\right)}
 + \cO(N^{-1})\,.
 \label{1pforig}
\fe
It is well-known that the operators $\cO_p$  as defined are not orthogonal for different $p$ (in particular it mixes with the identity operator) due to curvature of the $S^2$ (or equivalently the supersymmetric background on $S^4$) \cite{Gerchkovitz:2016gxx,Rodriguez-Gomez:2016ijh,Rodriguez-Gomez:2016cem,Binder:2019jwn}. To unmix them, one computes  their connected two-point functions using the matrix model  without the defect insertion given by $Z^{\rm cYM'}_{S^2}$ in \eqref{dcYMS2},
\ie
\la  \cO_p    \bar \cO_q\ra_{\rm SYM} = \pa_{\tau_p} \pa_{\bar \tau_q} \left.\log Z_{\rm SYM}(\tau,g_p)\right|_{\tau_p=0}= \pa_{t_p} \pa_{t_q} \left.\log Z_{\rm SYM}(\tau,g_p)\right|_{t_{p>2}=0}\,.
\fe
One finds for $p,q$ even and nonzero,
\ie
\la   \cO_p   \bar \cO_q\ra_{\rm SYM} =\frac{\left(\frac{\lambda }{4 \pi^2 }\right)^{\frac{p+q}{2}} 2 \Gamma \left(\frac{p+1}{2}\right)\Gamma \left(\frac{q+1}{2}\right)}{\pi  (p+q) \Gamma \left(\frac{p}{2}\right) \Gamma \left(\frac{q}{2}\right)},\quad 
\la \cO_p \ra_{\rm SYM}= \frac{N  \left(\frac{\lambda }{4 \pi ^2 }\right)^{p\over 2} \Gamma \left(\frac{p+1}{2}\right)}{\sqrt{\pi } \Gamma \left(\frac{p}{2}+2\right)}\,.
\fe
Following \cite{Binder:2018yvd}, we define the unmixing vector $v^p_n$ with  $p,n$ even positive integers,
\ie
v^p_n= 
 \frac{n   i^{n}}{2^n \sqrt{\pi}}  {\Gamma \left(\frac{1-p}{2}\right) \Gamma \left(\frac{n+p}{2}\right)  \over {\Gamma \left(\frac{p+2}{2}\right) \Gamma \left(\frac{n-p+2}{2} \right)}  }\left(\frac{\lambda }{4 \pi^2 }\right)^{\frac{n-p}{2}}\,,
\fe  
then the renormalized operators for $n$ even
\ie
   \cO_n^{\rm ren}\equiv  {1\over \sqrt{n}}\left ( 16\pi^2 \over \lambda\right)^{n \over 2}\sum_{p=2}^n v_n^p ( \cO_p-\la   \cO_p \ra_{\rm SYM})\,,
 \label{evenunmix}
\fe 
satisfy
\ie
\la   \cO_m^{\rm ren}  \bar{ {\cO}}_n^{\rm ren}\ra_{\rm SYM}
=
 \D_{mn}\,.
\fe
Similarly for $p,q\in 2\mZ+1$, we have
\ie
\la \cO_p \bar \cO_q\ra_{\rm SYM} =\frac{\left(\frac{\lambda }{4 \pi^2 }\right)^{\frac{p+q}{2}} 2 \Gamma \left(\frac{p+2}{2}\right)\Gamma \left(\frac{q+2}{2}\right)}{\pi  (p+q) \Gamma \left(\frac{p+1}{2}\right) \Gamma \left(\frac{q+1}{2}\right)}\,.
\fe
As before, we define the unmixing vector $ u^p_n$ for $p,n$ positive odd integers  
\ie
u^p_n=n   (2 i)^{p-n}
{\Gamma \left(\frac{n+p}{2}\right)  \over \Gamma (p+1) \Gamma \left(\frac{n-p+2}{2}\right)}\left(\frac{\lambda }{4 \pi^2 }\right)^{\frac{n-p}{2}}\,,
\fe  
then the re-normalized operators with odd $n$
\ie
  \cO_n^{\rm ren}\equiv  {1\over \sqrt{n}}\left ( 16\pi^2 \over \lambda\right)^{n \over 2}\sum_{p=1}^n u_n^p   \cO_p \,,
\label{unmixodd}\fe 
satisfy
\ie
\la   \cO_m^{\rm ren} \bar {  \cO}_n^{\rm ren}\ra_{\rm SYM}
=
\D_{mn}\,.
\fe

Putting together \eqref{1pforig} with \eqref{evenunmix}, the renormalized one-point-function in the presence of the D5-interface takes the simple form
\ie
\la    \cO_p^{\rm ren}\ra_{\rm dSYM}
=
 \frac{   i^p \sqrt{p}}{\pi (p^2-1)}\sqrt{\lambda }  \,.
 \label{1pfSYM}
\fe
for even $p$.

\subsubsection{Two-point functions of interface operators} 
Let's now consider correlation functions of BPS operators on the D5-brane interface. In this case, the interface CFT has a simple TQM sector 
  generated by a single $SU(N)$ singlet BPS operator
\ie
\cS=\tilde Q Q 
\fe
on the $S^1_{\rm TQM}$. 
Using the propagator in the TQM \eqref{QQprop}
\ie
\la  Q_i (\varphi_1)  \tilde Q^j  (\varphi_2)\ra 
=
-\D_i^j{s(\varphi_1-\varphi_2)+\tanh (\pi a_i)\over 4\pi } e^{-a_i (\varphi_1-\varphi_2)}
\fe 
 the Wick contraction gives, assuming $\varphi_1<\varphi_2$
\ie
\la \tilde Q Q(\varphi_1)\tilde Q Q(\varphi_2)\ra_a
=&
{1\over (4\pi)^2}
\left(
-\sum_{i=1}^N{1\over \cosh^2(\pi a_i)}
+ \sum_{i=1}^N \tanh^2(\pi a_i)
\right)
\fe
Then the two-point function of $\cS$ 

\ie
\la \cS (\varphi_1) \cS(\varphi_2)\ra =
{1\over Z_{\rm dSYM}}\int  [da]{
	e^{-\pi  {{\rm Im}\tau}  \sum_{i=1}^N a_i^2}  \over 
	\prod_{i=1}^N 2\cosh (\pi a_i)
}{1\over (4\pi)^2}
\left(
N-\sum_{i=1}^N{2\over \cosh^2(\pi a_i)}
\right)
\fe
which is related to the matrix model \eqref{mm} by
\ie
\la \cS (\varphi_1) \cS(\varphi_2)\ra = {1\over (2\pi)^4}\left.{\pa^2 \log Z_{\rm dSYM} \over \pa m^2} \right|_{m=0,t_{p}=0}\,.
\fe
Applying the large $N$ formulae, we have
\ie
\left.{\pa^2 \log Z_{\rm dSYM} \over \pa m^2}\right|_{m=0,t_p=0}
=&{N  \pi^2} \int_{-\m }^\m   dx \,\rho_0(x)  {1\over \cosh^2(\pi x)}
=2\pi N \int_{-1 }^1  dx     { \sqrt{1-x^2}\over \cosh^2(\m \pi  x)}\,.
\fe
To leading orders in the $1\over \lambda$ expansion, we may approximate
\ie
{1\over \cosh^2(  \m \pi x)} \sim {2\over  \m \pi}\D(x)\,.
\fe
Hence
\ie
\la \cS \cS \ra  ={1\over (2\pi)^4}{4 N\over \m}={N\over 2\pi^3 \sqrt{\lambda}}\,,
\fe
and we define renormalized defect operator by
\ie
  \cS^{\rm ren} = {\sqrt{2}\pi^{3\over 2}\lambda^{1\over 4}\over \sqrt{  N}} \cS\,.
\label{Snorm}
\fe

\subsubsection{Two-point functions of bulk and interface operators} 
  Next let's study the two-point functions of defect operator $\cS$ and bulk operator $\cO_p$. Performing the Wick contraction for the hypermultiplet scalars $(Q,\tilde Q)$, we obtain 
   \ie
 \la \cO_p \cS \ra= -\int  [da]   {
  	e^{-\pi  {{\rm Im}\tau}  \sum_{i=1}^N a_i^2}  \over 
  	\prod_{i=1}^N 2\cosh (\pi a_i)
  }{1\over 4\pi } \sum_{i=1}^N (a_i)^{p}
  \sum_{i=1}^N \tanh(\pi a_i)\,.
  \fe
  Clearly this is nonzero only if $p\in 2\mZ+1$. 
  This is related to the matrix model \eqref{mm} by 
  \ie
  \la \cO_p \cS\ra ={1\over (2\pi)^2 } \left.{\pa^2 \log Z_{\rm dSYM}\over \pa m \pa t_p}\right|_{t_{ p}=0,m=0}
  \fe
  where the matrix model has a potential $V(x)$ with $t_p=0$ (or $g_p=0$) for $p$ even except for $p=2$. In the rest of the section, we take $p$ to be an odd  positive integer.
  
 Clearly only the last term in the square bracket of \eqref{fulllogZ} is relevant for the $m$-derivative  and it gives
  \ie
  -{\pa  \log Z_{\rm dSYM}\over \pa m  } 
  =& {N  \over 2 } \int_{\m_-}^{\m_+}  dx \,M( x) \sqrt{(\m_+-x)(x-\m_-)} \tanh(\pi  x)
 \\
  =& {N  \over 2 } \left(
  \int_{0}^{\m_+}  dx \, M( x)   \sqrt{(\m_+-x)(x-\m_-)}
  -
  \int_{\m_-}^0  dx \, M( x)   \sqrt{(\m_+-x)(x-\m_-)}
  \right)
  \,,
  \fe
since to leading order in the ${1\over \lambda}$ expansion we can effectively replace $\tanh(\pi  x)$ by the step function $2\theta(x)-1$. 
 We then take derivatives with respect to $g_p$ and set $g_p=0$. Using the identity from \cite{Rodriguez-Gomez:2016ijh} 
 \ie
  g_2\pa_{g_p}  \m_+  = g_2\pa_{g_p} \m_-=-{1\over 2}\m ^{p-1}p \C_{p-1}\,,
  \fe
  valid for $g_{p\geq 3}=0$
  with 
  \ie
  \C_m=(1+(-1)^m){\Gamma\left( m+1\over 2 \right) \over 2\sqrt{\pi} \Gamma\left( m+2\over 2\right)}\,,
  \fe
  we get, 
  \ie
 &  \left.{\pa^2  \log Z_{\rm dSYM} \over \pa m \pa t_p}\right |_{m=0,t_{p}=0}=-{1\over N}\left.{\pa^2  \log Z_{\rm dSYM} \over \pa m \pa g_p}\right |_{m=0,g_{p\geq 3}=0}
  \\=&
  -{1 \over 2  }{ \m^{p-1} p\C_{p-1}} \left(
  \int_{0}^{\m}  dx \,  {x\over \sqrt{{\m}^2-x^2}}
  -
  \int_{-{\m}}^0  dx \,    {x\over \sqrt{{\m}^2-x^2}}
  \right)
  \\
  +&
  {1  \over 2 } \left(
  \int_{0}^{\m}  dx \, \pa_{g_p} M(x)  {  \sqrt{{\m}^2-x^2}}
  -
  \int_{-{\m}}^0  dx \, \pa_{g_p} M(x)    {  \sqrt{{\m}^2-x^2}}
  \right)
  \\
  =&
  -{ \m^{p-1} p\C_{p-1}} 
  \int_{0}^{\m}  dx \,   {x\over \sqrt{{\m}^2-x^2}}
  +
  {p {\m}^{p-3}\C_{p-3} }
  \int_{0}^{\m}  dx \,x {  \sqrt{{\m}^2-x^2}}\,,
   \fe
  where the first equality follows from
  \ie
 \left. g_2\pa_{g_p}\sqrt{(\m_+-x)(x-\m_-)}\right |_{g_{p\geq 3=0}} =-{x \m^{p-1} p\C_{p-1}\over2 \sqrt{\m^2-x^2}},\quad
  \left.M(x) \right |_{g_{p\geq 3=0}} = {16\pi^2\over\lambda}=2  g_2\,,
  \fe
  and in the second equality we have used \eqref{ck} and then made the replacement
  \ie
  \left. \pa_{g_p} M(x)\right |_{g_{p\geq 3=0}} =
  \sum_{k=0}^{p-2} x^k {p }\m^{p-k-2}(-1)^{p-k}\C_{p-k-2} 
  \to p x \m^{p-3} \C_{p-3} \,,
  \fe
  since we are interested in the leading effects in the $1\over \lambda$ expansion.
 
From the simple integrals
  \ie
  \int_{0}^\m  dx \,   {x\over \sqrt{\m^2-x^2}}=\m,\quad 
  \int_{0}^\m  dx \,   {x \sqrt{\m^2-x^2}}={\m^3\over 3}\,,
  \fe
we conclude
 \ie
&   \left.{\pa^2  \log Z_{\rm dSYM} \over \pa m \pa t_p}\right |_{m=0,t_{p }=0}
 =&
-{  {\m}^{p} p\C_{p-1}} 
+
{p {\m}^{p}\C_{p-3}\over 3}
=
-\frac{p (2 p-5)  \Gamma \left(\frac{p-2}{2}\right)}{6 \sqrt{\pi } \Gamma \left(\frac{p+1}{2}\right)}\mu ^p\,.
\fe 
  
  Thus
  \ie
  \la   \cO_p   \cS\ra_{\rm dSYM}=- {1\over (2\pi)^2}\frac{p (2 p-5)  \Gamma \left(\frac{p-2}{2}\right)}{6 \sqrt{\pi } \Gamma \left(\frac{p+1}{2}\right)}\mu ^p
  \fe
 Taking into unmixing with \eqref{unmixodd} and the normalization \eqref{Snorm}, we get for the renormalized bulk and defect operators,
  \ie
  \la     \cO_n^{\rm ren}   \cS^{\rm ren}\ra_{\rm dSYM}= & {1\over \sqrt{n}} \left( 16\pi^2 \over \lambda\right)^{n\over 2}{\sqrt{2}\pi^{3\over 2}\lambda^{1\over 4}\over \sqrt{  N}}  \sum_{p} u^p_n  \la     \cO_p    \cS\ra_{\rm dSYM}
  =
  - { \sqrt{2} \lambda^{1\over 4}\over 6\sqrt{  N \pi }}    \sqrt{n} \left( (-1)^{n-1\over 2} n+2\right)\,,
   \label{OS}
  \fe
  for $n$ odd.

 \subsubsection{Defect correlation functions from D5 branes in IIB string theory}
 In the large $N$ 't Hooft limit, the correlation functions of $\cN=4$ SYM are computed by Witten diagrams in the bulk supergravity on $AdS_5$ via the AdS/CFT correspondence. For defects in the SYM that are realized by boundary-anchored branes in the bulk, additional interactions are introduced  on the brane worldvolume in $AdS_5$, that couple the 5d supergravity fields as well as Kaluza-Klein (KK) modes from the $S^5$ reduction of IIB supergravity with the brane and its excitations. Consequently, correlation functions of bulk and defect local operators in the SYM receive contributions from brane vertices that are integrated over subspaces of $AdS_5$. 
 
For illustration, let us study the one-point function of the half-BPS operator $\cO_p$ in the presence of a D5-brane interface from the bulk perspective, to leading order in the $1\over \lambda$ expansion. 
This is computed by a simple Witten diagram that involves a bulk-to-boundary propagator, anchored at the boundary insertion, and integrated over the D5 world-volume in $AdS_5$.

Let us write the metric on $AdS_5\times S^5$ as
\ie
ds^2={ dz^2+dx_\m dx_\m  \over z^2}+(d\psi^2+\sin^2\psi d\Omega_2^2+ \cos^2\psi d\tilde\Omega_2^2)\,,
\fe
where $d\Omega_2^2$ and $d\tilde\Omega_2^2$ are line elements on the two unit $S^2$s. Compared to \eqref{AdS5S5m}, the $S^5$ angular coordinates above are related to the embedding coordinates $y^I$ with $|y^I|=1$ by
\ie
y^I =(
\cos\psi\cos\tilde\theta,\cos\psi\sin\tilde\theta\sin \tilde\phi,\cos\psi\sin\tilde\theta\cos \tilde\phi,\sin\psi\cos\theta,\sin\psi\sin\theta\sin \phi,\sin\psi\sin\theta\cos \phi
)\,.
\fe
The probe D5 brane extends along the $AdS_4 \subset AdS_5$ submanifold at $x_1=0$ and wraps the $S^2\subset S^5$ located at $\psi={\pi \over 2}$ with coordinates $(\theta,\phi)$. 

The ${1\over 2}$-BPS operator with general $SO(6)_R$ R-symmetry polarization can be represented as
\ie
\cO_{p}(Y_I,x_\m)_{\mR^4}\equiv {(8\pi^2)^{p\over 2}\over \lambda^{p/2}\sqrt{p}} \tr (Y^I \Phi_I)^p
\fe
where we have introduced a null polarization vector $Y^I$ and the coefficients are chosen to normalize the two-point functions on $\mR^4$ to $\D_{pq}$. Following the general AdS/CFT dictionary, in the large $N$ limit the one-point function with the D5 brane interface is
\ie
\la \cO_{p}(Y_I,x_\m)\ra_{\mR^4}=-\int {dzd \vec w\over z^4}  \underbrace{{\D S_{\rm D5}\over \D s(\vec w)} }_{\rm D5\, couplings }
\underbrace{{c_p z^p\over (z^2+x_1^2+(\vec x -\vec   w)^2)^p}}_{\rm bulk-boundary\,propagator}  \underbrace{{1\over 4\pi}\left.\int_{S^2} (Y^I y_I)^p\right|_{\psi={\pi\over 2}}}_{\rm internal\,wavefunction}
\label{1pfbulk}
\fe
with the normalization constant 
\ie
c_p={p+1\over 2^{2-p/2} N \sqrt{p}} 
\fe
chosen such that the two-point functions of $\cO_{p}(Y_I,x_\m)_{\mR^4}$ is normalized to $\D_{pq}$ \cite{Nagasaki:2012re}. Here $s(\vec w)$ denotes the source (spacetime part) for the operator $\cO_p$, and $S_{\rm D5}$ is the worldvolume action for the D5 brane\footnote{For our purpose, it suffices to turn off the NS 2-form $B$ and Ramond gauge fields $C_p$ except for $C_4$ since they are decoupled from the source $s(\vec w)$ \cite{Kim:1985ez}.}
\ie
S_{\rm D5}={T_5}\int d^6 \xi \sqrt{\det(G+ B+2\pi\A' F_{\rm D5})} +i T_5 \int e^{2\pi \A' F_{\rm D5}+B}\wedge \sum_p C_{p}
\fe
with D5 brane tension
\ie
T_5={1\over (2\pi)^5 \A'^3 g_s}= {2N\over (2\pi)^4}\sqrt{\lambda}\,.
\fe
Here $G$ is the induced metric on the brane, $B$ the NS two-form, $F_{\rm D5}$ the worldvolume field strength, and $C_p$ the RR gauge fields. To compute \eqref{1pfbulk}, we need the  first order term of $S_{\rm D5}$ in $\D s(\vec w )$. This analysis was carried out in \cite{Nagasaki:2012re} and the integral in \eqref{1pfbulk} reduces to\footnote{The holographic analysis in \cite{Nagasaki:2012re} is more general as they consider  D5 brane interfaces with nontrivial worldvolume flux $F_{\rm D5}$ which corresponds to, in the field theory, interfaces between SYMs of different ranks. In a subsequent publication, we perform the field theory computations for such interfaces and obtain matching with precise dependence on flux $F_{\rm D5}$ and 't Hooft coupling $\lambda$.}
 \ie
\la \cO_{p}(Y_I)\ra_{\mR^4}=& {1\over 4\pi}\left.\int_{S^2} (Y^I y_I)^p\right|_{\psi={\pi\over 2}} {\sqrt{\lambda}2^{p\over 2} \CC(p+{1\over 2})\over \pi^{3/2}\sqrt{p}\CC(p)} \int_0^\infty du {u^{p-2}\over (1+u^2)^{p+{1\over 2}}}\,,
\fe
where the operator is inserted at $x_\m=(1,0,0,0)$.

For our purpose, the operator $\cO_p$ is associated with the polarization vector
\ie
Y^I=(0,0,1,i,0,0)\,,
\fe
then  the $S^2$ integral of the internal wavefunction vanishes unless $p$ is even, in which case
\ie
 {1\over 4\pi }\left.\int_{S^2} (Y^I y_I)^p\right|_{\psi={\pi\over 2}}=\frac{i^p}{p+1}\,.
\fe
Hence we find from bulk computation, the one-point function of $\cO_p$ with a D5 brane interface is, for even $p$,
 \ie
 \la \cO_{p}\ra_{\mR^4}
 =
 \frac{   i^p \sqrt{p}}{2^{p\over 2}\pi (p^2-1)}\sqrt{\lambda }  \,.
 \fe
 Taking into account the $2^{p\over 2}$ factor between one-point functions on $\mR^4$ and those on $S^4$  (see \eqref{OS4} with $R={1\over2}$), we find  precise agreement with the field theory result \eqref{1pfSYM}.
 
A similar computation would determine the bulk-defect correlator $\la \cO_p \cS\ra$ but we will not pursue the details here. We simply note that the $N$ and $\lambda$ scalings of the D5 brane worldvolume couplings imply that  
\ie
\la \cO_p \cS\ra\sim \lambda^{1\over 4}N^{-{1\over 2}}
\fe
for normalized bulk operator $\cO_p$ and interface operator $\cS$ \cite{DeWolfe:2001pq}. This is in agreement with what we find from the field theory \eqref{OS}.
As noted in \cite{DeWolfe:2001pq}, the fractional power dependence on $\lambda$ is the signature of  strong-coupling effects of the $\cN=4$ SYM in the 't Hooft limit.
 
 \section{Conclusion and Discussion}
 \label{sec:conclusion}
 In this paper, we have initiated  the study of general defect observables in the 4d $\cN=4$ super-Yang-Mills preserving a single supercharge $\cQ$. We extend the previous work \cite{Drukker:2007yx,Drukker:2007qr,Pestun:2009nn} by classifying general defect observables involving interfaces (boundaries), surface operators, line operators and local operators in the SYM that form ${1\over 16}$-BPS defect networks preserving $\cQ$ (see Figure~\ref{fig:network}). For interface (boundary) defects, we carry out the localization computation with respect to $\cQ$, building upon previous works of \cite{Pestun:2009nn} and \cite{Dedushenko:2016jxl}. As a result, we have uncovered an effective 2d/1d theory that controls the dynamics of the $\cQ$-preserving observables in the SYM,  described by 2d Yang-Mills coupled to 1d topological quantum mechanics, which we call the defect-Yang-Mills (dYM). We note that our derivation is not completely rigorous since we do not explicit evaluate the one-loop determinant associated to the operator $\cQ^2$ as in \cite{Pestun:2009nn} though we provide nontrivial consistency checks with known results in the literature that suggest the determinant is indeed one. We then provide explicit dictionary between $\cQ$-cohomology observables in the SYM and those in the dYM.  Applied to the D5-brane type interface in $U(N)$ SYM in the large $N$ 't Hooft limit, we extract the one-point functions of bulk half-BPS operators, and two-point functions between bulk and defect local operators in the strong coupling limit. The results are in perfect agreement with a Witten diagram computation  in the bulk IIB supergravity on $AdS_5\times S^5$ with a probe D5-brane that corresponds to the interface in the boundary SCFT.

There are a number of interesting future directions. Here we have mostly focused on boundary conditions (and corresponding interfaces from the unfolding trick) of the Dirichlet and Neumann types. The $\cN=4$ SYM is known to admit a much larger family of boundary conditions \cite{Gaiotto:2008sa} that form orbits under the $SL(2,\mZ)$ duality of the bulk theory \cite{Gaiotto:2008ak}. In particular, the Nahm pole boundary condition, which is relevant for general D5 type defect (e.g. with world-volume flux) is S-dual to the Neumann boundary condition. In an upcoming work \cite{nahmp}, we study such general D5 defects in the large $N$ SYM and obtain exact results in the 't Hooft coupling $\lambda$, which agrees and interpolates between the integrability results at weak coupling \cite{deLeeuw:2015hxa, Buhl-Mortensen:2016jqo,Buhl-Mortensen:2017ind} and the IIB supergravity results at strong coupling \cite{Nagasaki:2012re}. 

Another class of important interface defects in $\cN=4$ SYM are the Janus domain walls  \cite{Bak:2003jk,DHoker:2006qeo,DHoker:2007hhe,DHoker:2007zhm,Gaiotto:2008sd,Bobev:2020fon} and related duality interfaces \cite{Gaiotto:2008sd,Gaiotto:2008ak}. The Janus domain wall interpolates between different values of the exactly marginal parameter $\tau$ of $\cN=4$ SYM and was first introduced in the context of AdS/CFT correspondence with IIB string theory \cite{Bak:2003jk} where they correspond to solutions of IIB supergravity with a nontrivial profile for the axion-dilaton $\tau$. With specific couplings on the interface \cite{DHoker:2006qeo,Gaiotto:2008sd}, the Janus domain wall is half-BPS and lives in the $\cQ$-cohomology. A closely related defect is the duality interface in $\cN=4$ SYM \cite{Gaiotto:2008sd,Gaiotto:2008ak}. We start with a superconformal Janus domain wall at $x_1=0$ that interpolates between $G$ SYM with gauge couplings related by an S-transform, and then preform an S-duality transformation on the half-space $x_1\geq 0$. The resulting interface now carries nontrivial local excitations  described by a  3d $\cN=4$ SCFT known as the $T[G]$ theory with global symmetry $G\times G^\vee$ that are gauged by the 4d vector multiplets \cite{Gaiotto:2008ak}.  
Because it couples SYM with Langlands-dual gauge groups (or equivalently it effects an S-transform  of $\tau$), this interface is known as the duality interface (or S-duality wall). This construction also generalizes by considering Janus domain walls that relates values of $\tau$ by general $SL(2,\mZ)$ group elements \cite{Terashima:2011qi}.\footnote{See \cite{Inverso:2016eet,Assel:2018vtq,Guarino:2019oct,Guarino:2020gfe} for corresponding solutions in IIB supergravity.}
It would be interesting to study the Janus domain walls and duality defects using the dYM, which can shed light on the $SL(2,\mZ)$ properties of the SYM.

Finally it would be interesting to feed  the ${1\over 16}$-BPS OPE data captured by the dYM into the defect/boundary bootstrap program to determine the non-BPS spectrum and their OPE. The simplest case is to consider the two-point function of the half-BPS operators $\cO_p$, which now admit a single nontrivial cross ratio. Depending on whether we take the bulk OPE limit or the bulk-boundary OPE limit, we arrive at two decompositions that involve exchanging bulk and boundary operators respectively. The bulk-to-boundary crossing relations constrain the spectrum and OPE of such intermediate operators (see \cite{Liendo:2016ymz} for such crossing relation in the SYM). Combined with the dynamical input from the dYM, one can hope to determine the non-BPS data either by numerical methods (an extension of \cite{Beem:2013qxa,Beem:2016wfs,Chang:2019dzt})
or recent  analytic functional bootstrap methods  \cite{Kaviraj:2018tfd,Mazac:2018biw}.

\section*{Acknowledgements}
The author thanks Ofer Aharony, Simone Giombi, Michael Gutperle, Shota Komatsu,  Bruno Le Floch, and Yuji Tachikawa for interesting discussions and correspondences. The author is also grateful to Michael Gutperle and Shota Komatsu for useful comments on the draft.
 This work  is supported in part by the US NSF under Grant No. PHY-1620059 and by the Simons Foundation Grant No. 488653. 
The author also thanks the Weizmann Institute for hospitality  where this work was initiated.

\appendix

\section{4d $\cN=4$ superconformal algebra}
\label{app:4dsca}
The 4d $\cN=4$ superconformal algebra $\mf{psu}(2,2|4)$ is generated by 16 Poincar\'e supercharges $(Q_{\A a\dot a},\bar Q_{\dot \A a\dot a})$ and 16 superconformal charges $(S_{\B b\dot b},\bar S_{\dot \B b\dot b})$. Here $\A,\B=1,2$ and $\dot{\A},\dot \B$ are 4d chiral and anti-chiral spinor indices respectively. They are raised and lowered by the epsilon tensor as\footnote{Our convention for the epsilon tensor in this paper is $\ep_{21}=\ep^{12}=1$.}
\ie
u^\A  \equiv \epsilon^{\A\B} u_\B,~u_\A=\ep_{\A\B} u^\B \,.
\fe
 We introduce the relevant gamma matrices in the (anti)chiral basis
\ie
\sigma^\m_{\A \dot \A}=(\vec\sigma, -i),\quad \bar \sigma^{\m \dot\B\B}=(\vec\sigma, i) \,,
\fe
which satisfy
\ie
\sigma^\m_{\A \dot \A} \ep^{\A\B} \ep^{\dot \A\B}= -\bar \sigma^{\m \dot \B\B},\quad \sigma_{\{\m}\bar \sigma_{\n\}}=\bar\sigma_{\{\m} \sigma_{\n\}}= \D_{\m\n}\,.
\fe

The spacetime rotation generators $J_{\A \B}$ and $J_{\dot \A\dot \B}$, are related to the $M_{\m\n}$ which we have used in the main text by
\ie
J_{\A \B}= {1\over 2}(\sigma_{\m\n})_{\A\B} M^{\m \n},\quad J_{\dot \A\dot \B}= {1\over 2}(\bar\sigma_{\m\n})_{\dot \A\dot \B} M^{\m \n}
\,,
\fe 
where 
\ie
\sigma_{\m\n}=-{1\over 2} \sigma_{[\m}\bar \sigma_{\n]},\quad \bar\sigma_{\m\n}=-{1\over 2} \bar\sigma_{[\m} \sigma_{\n]}\,.
\fe
These rank-two gamma matrices are self-dual and anti-self-dual respectively,
\ie
{1\over 2}\epsilon_{\m\n\rho\lambda} \sigma^{\rho\lambda}=\sigma_{\m\n},\quad {1\over 2}\epsilon_{\m\n\rho\lambda} \bar\sigma^{\rho\lambda}=-\bar \sigma_{\m\n}\,,
\fe 
and satisfy
\ie
{}[\sigma_{\m\rho},\sigma_{\n\lambda}]=&\D_{\m\n} \sigma_{\rho\lambda}+\D_{\rho\lambda} \sigma_{\m\n}
-\D_{\m\lambda}\sigma_{\rho\n}-\D_{\rho\n} \sigma_{\m\lambda}\,,
\\
{}[\bar\sigma_{\m\rho},\bar\sigma_{\n\lambda}]=&\D_{\m\n} \bar\sigma_{\rho\lambda}+\D_{\rho\lambda} \bar\sigma_{\m\n}
-\D_{\m\lambda}\bar\sigma_{\rho\n}-\D_{\rho\n} \bar\sigma_{\m\lambda}\,.
\fe

The bosonic conformal subalgebra of $\mf{psu}(2,2|4)$ is
\ie
{}[J_{\A\B},J_{\C\D}]=&\ep_{\B\C}J_{\A\D}+\ep_{\A\D}J_{\C\B},\quad
[J_{\dot \A\dot \B},J_{\dot \C\dot \D}]=\ep_{\dot \B\dot \C}J_{\dot \A\dot \D}+\ep_{\dot \A\dot \D}J_{\dot \C\dot \B} \,,
\\
[K_{\A\dot \A}, P_{\B\dot \B}]=&
\ep_{\A\B}\ep_{\dot \A\dot \B} D+\ep_{\A\B} J_{\dot \A\dot \B}+J_{\A\B} \ep_{\dot \A\dot \B}\,,
\\
[J_{\A\B},P_{\C\dot \C}]=&-\ep_{\C(\A}P_{\B)\dot \C} ,\quad[J_{\dot \A\dot \B},P_{\C\dot \C}]= -\ep_{\dot \C(\dot \A}P_{\C \dot \B)}\,,
\\
[J_{\A\B},K_{\C\dot \C}]=&-\ep_{\C(\A}K_{\B)\dot \C} ,\quad[J_{\dot \A\dot \B},K_{\C\dot \C}]= -\ep_{\dot \C(\dot \A}K_{\C \dot \B)} \,,
\fe
where the translation and special conformal transformations generators are denoted here as
\ie
P_{\A\dot \A}\equiv P_\m \sigma^\m_{\A \dot \A },\quad 
K_{\dot \A \A }\equiv K_\m \bar \sigma^\m_{ \dot \A \A}\,.
\fe
They act on the supercharges by
\ie 
{}[ K_{\A\dot \A},  Q_{\B a\dot a} ]=&\ep_{ \A \B} \bar S_{\dot \A a \dot a} ,\quad 
[K_{\A\dot \A}, \bar  Q_{\dot \B a\dot a} ]=\ep_{ \dot \A \dot \B}  S_{ \A a \dot a} \,,
\\
[   P_{\A\dot \A},  S_{\B a\dot a} ]=&\ep_{ \A \B} \bar Q_{\dot \A a \dot a} ,\quad 
[P_{\A\dot \A}, \bar  S_{\dot \B a\dot a} ]=\ep_{ \dot \A\dot  \B}  Q_{ \A a \dot a} \,,
\\
[J_{\A\B}, Q_{  \C a \dot a}]=&  -\ep_{  \C ( \A}   Q_{ \B)a\dot a},\quad [J_{\A\B},  S_{ \C a \dot a}]=  -\ep_{  \C (  \A}  S_{  \B)a\dot a}\,,
\\
[J_{\dot\A\dot \B}, \bar Q_{\dot \C a \dot a}]=&  -\ep_{\dot \C (\dot \A} \bar Q_{\dot \B)a\dot a},\quad [J_{\dot \A\dot \B}, \bar S_{\dot \C a \dot a}]=  -\ep_{\dot \C (\dot \A} \bar S_{\dot \B)a\dot a}\,.
\fe

The R-symmetry indices $a,b=1,2$ and $\dot a,\dot b=1,2$ transform as doublets under the maximal subgroup $SU(2)_H\times SU(2)_C \subset SO(6)_R$. The $SU(2)_H\times SU(2)_C$ is generated by $T^H_A$ and $T^C_{\dot A}$ respectively with $A,\dot A=1,2,3$ while the remaining generators are denoted by $M_{A\dot A}$. They are related to $R_{IJ}$ which we have used in the main text by
\ie
T_A^C=-i(R_{67},R_{75},R_{56}),~T_A^H=-i(R_{90},R_{08},R_{89}),~M_{A\dot A}=-iR_{A+4,\dot A+7}\,.
\fe
They satisfy the commutation relations
\ie
{}[T^H_{A}, M_{B \dot A}]=& i\ve_{ABC} M_{C\dot A},\quad [T^C_{\dot A}, M_{A \dot B}]= i\ve_{\dot A\dot B\dot C} M_{A\dot C}\,,
\\
[M_{A\dot A},M_{B\dot B}]=&i(\D_{\dot A\dot B}\epsilon_{ABC} T^H_{ C}+\D_{AB}\epsilon_{\dot A \dot B\dot C} T^C_{\dot C})
\,,\\
[T^H_A,T^H_B]=&i\ve_{ABC}T^H_C,\quad [T^C_{\dot A},T^C_{\dot B}]=i\ve_{\dot A\dot B\dot C}T^C_{\dot C}\,,
\fe
and act on the supercharges by
\ie	       
{}[T^H_A, Q_{\A a\dot a}]  =&{1\over 2}(\tau_A)^b{}_a Q_{\A b\dot a},\quad 
[T^H_A, S_{\A a\dot a}]  ={1\over 2}(\tau_A)^b{}_a Q_{\A b\dot a}\,, 
\\
[T^C_{\dot A}, Q_{\A a\dot a}]  =&{1\over 2}(\tau_{\dot A})^{\dot b}{}_{\dot a} Q_{\A a\dot b},\quad 
[T^C_{\dot A}, S_{\A a\dot a}]  ={1\over 2}(\tau_{\dot A})^{\dot b}{}_{\dot a} S_{\A a\dot b}\,, 
\\
[T_A^H, \bar Q_{\dot \A a\dot a}]  =&{1\over 2}(\tau_A)^b{}_a  \bar Q_{\dot \A b\dot a},\quad 
[T_A^H,  \bar S_{\dot \A a\dot a}]  ={1\over 2}(\tau_A)^b{}_a \bar  Q_{\dot \A b\dot a}\,, 
\\
[T_{\dot A}^C, \bar  Q_{\dot \A a\dot a}]  =&{1\over 2}(\tau_{\dot A})^{\dot b}{}_{\dot a} \bar  Q_{\dot \A a\dot b},\quad 
[T_{\dot A}^C, \bar  S_{\dot \A a\dot a}]  ={1\over 2}(\tau_{\dot A})^{\dot b}{}_{\dot a}  \bar S_{\dot \A a\dot b}\,, 
\\
{}[M_{A {\dot A}}, Q_{\A a\dot a}]  =&-{1\over 2}(\tau_A)^b{}_a (\tau_{\dot A})^{\dot b}{}_{\dot a} Q_{\A b\dot b},\quad 
[M_{A {\dot A}}, S_{\A a\dot a}]  ={1\over 2}(\tau_A)^b{}_a (\tau_{\dot A})^{\dot b}{}_{\dot a} S_{\A b\dot b}\,, 
\\
[M_{A {\dot A}}, \bar Q_{\A a\dot a}]  =&{1\over 2}(\tau_A)^b{}_a (\tau_{\dot A})^{\dot b}{}_{\dot a}  \bar Q_{\A b\dot b},\quad 
[M_{A {\dot A}}, \bar  S_{\A a\dot a}]  =-{1\over 2}(\tau_A)^b{}_a (\tau_{\dot A})^{\dot b}{}_{\dot a}  \bar  S_{\A b\dot b}\,, 
\fe
where $(\tau_A)^a{}_b$ and $(\tau_{\dot A})^a{}_b$ are given by usual Pauli matrices and
\ie
(\tau_A)_{ab}\equiv \ep_{ac}(\tau_A)^c{}_b,\quad(\tau_A)^{ab}\equiv (\tau_A)^a{}_c  \ep^{cb}  
\fe 
similarly for $(\tau_{\dot A})^a{}_b$.

Finally the anti-commutators of the supercharges are
\ie
\{ Q_{\A  a\dot a}, S_{\B b \dot b}\}=&\ep_{\dot a \dot b}\ep_{ab} J_{\A\B}-{1\over 2}\ep_{\A\B}
\left(
\epsilon_{ab} T_{\dot a\dot b}+ \epsilon_{ \dot a \dot b} T_{  a  b} + M_{ab \dot a\dot b}+ \epsilon_{ab} \epsilon_{\dot a\dot b}D
\right)\,,
\\
\{ \bar Q_{\dot \A  a\dot a}, \bar  S_{\dot \B b \dot b}\}=&\ep_{\dot a \dot b}\ep_{ab} J_{\dot \A\dot \B}-{1\over 2}\ep_{\dot \A\dot \B}
\left(
\epsilon_{ab} T_{\dot a\dot b}+ \epsilon_{ \dot a \dot b} T_{  a  b}-M_{ab \dot a\dot b}+ \epsilon_{ab} \epsilon_{\dot a\dot b}D
\right) \,,
 \\
\{  Q_{\A a\dot a},\bar Q_{\dot\A  b\dot b}\}=&\ep_{\dot a \dot b}\ep_{ab} \sigma^\m_{\A \dot \A} P_\m
,\quad
\{ S_{\B a\dot a}, \bar  S_{\dot\B b \dot b}\}=\ep_{\dot a \dot b}\ep_{ab} \sigma^\m_{\B \dot \B} K_\m\,,
\fe
where we have introduced for convenience
\ie
T_{ab}\equiv (\tau_A)_{ab} T^H_A,\quad T_{\dot a\dot b}\equiv (\tau_{\dot A})_{\dot a\dot b} T^C_{\dot A},
\quad M_{ab\dot a\dot b}\equiv(\tau_{ A})_{ a b} (\tau_{\dot A})_{\dot a\dot b} M_{A\dot A}\,.
\fe

\section{3d $\cN=4$ superconformal algebra}
\label{app:3dsca}
Here we present the half-BPS subalgebra $\mf{osp}(4|4) \subset \mf{psu}(2,2|4)$ induced by the involution $\iota$ with $\cP_\iota=\CC_{1890}$ (see Section~\ref{sec:halfbpsalg}). This involution 
 fixes the hyperplane $x_1=0$ and the corresponding 3d supercharges preserved by $\cP_\iota$ are given by
\ie
\cQ_{\A a \dot a}=Q_{\A a \dot a}+ i(\sigma_3)^{\dot \A}_\A\bar Q_{\dot \A a \dot a},
\quad
\cS_{\A a \dot a}=S_{\A a \dot a}+i (\sigma_3)^{\dot \A}_\A\bar S_{\dot \A a \dot a}\,.
\fe
We define the 3d gamma matrices in relation to the 4d ones
\ie
\C^i_{\A\B}= (\sigma_2, \sigma_3,-i 1_2)\cdot (\sigma_3)=( i\sigma_1, 1,- i \sigma_3) 
\fe
with $i=2,3,4$ denoting the 3d spacetime directions.

The 3d bosonic conformal generators are related to those in 4d by
\ie
\cJ_{\A\B} =J_{\A\B}-(\sigma_3)_\A^{\dot \A} (\sigma_3)_\B^{\dot \B}  J_{\dot \A\dot \B},\quad 
P_{\A\B} = i P_i \C^i_{\A \B},\quad  K_{\A\B} =-i K_i  \C^i_{\A \B}\,,
\fe
which generate the bosonic conformal subalgebra 
\ie
{}[\cJ_{\A\B},\cJ_{\C\D}]=&\ep_{\B\C}\cJ_{\A\D}+\ep_{\A\D}\cJ_{\C\B}\,,
\\
[K_{\A\B}, P_{\C \D}]=&
4\ep_{\C (\A}\ep_{\B) \D} D-2\ep_{\A\C} \cJ_{\B\D}-2\cJ_{\A\C} \ep_{\B\D}\,.
\fe
The R-symmetry now is simply $SU(2)_C\times SU(2)_H$ whose generators are defined by 
\ie
T_{ab}\equiv (\tau_A)_{ab} T^H_A,\quad T_{\dot a\dot b}\equiv (\tau_{\dot A})_{\dot a\dot b} T^C_{\dot A}.
\fe
The (anti)commutators involving the supercharges are
\ie
\{ \cQ_{\A  a\dot a}, \cS_{\B b \dot b}\}=& \ep_{\dot a \dot b}\ep_{ab} \cJ_{\A\B}-\ep_{\A\B}
\left(
\epsilon_{ab} T_{\dot a\dot b}+ \epsilon_{ \dot a \dot b} T_{  a  b} + \epsilon_{ab} \epsilon_{\dot a\dot b}D
\right)\,,
\\
\{  \cQ_{\A a\dot a}, \cQ_{\B  b\dot b}\}=&\ep_{\dot a \dot b}\ep_{ab} \sigma^i_{\A\B} P_i
,\quad
\{ \cS_{\A a\dot a}, \bar \cS_{ \B b \dot b}\}=-\ep_{\dot a \dot b}\ep_{ab} \sigma^i_{\A \B} K_i\,,
\\
[ K_{\A\B},  \cQ_{\C a\dot a} ]=& 2\ep_{  \C(\A}  \cS_{\B) a \dot a}  
,\quad 
[   P_{\A\B},  \cS_{\B a\dot a} ]=-2\ep_{ \C (\A }  Q_{ \B) a \dot a} \,,
\\
[\cJ_{\A\B}, \cQ_{  \C a \dot a}]=&  -\ep_{  \C ( \A}   \cQ_{ \B)a\dot a},\quad [\cJ_{\A\B},  \cS_{ \C a \dot a}]=  -\ep_{  \C (  \A}  \cS_{  \B)a\dot a}\,.
\fe

\section{Coordinates on $S^4$}
\label{app:coords}

We start with the embedding coordinates for $S^4$ in Euclidean $\mR^5$ 
\ie
X_1^2+X_2^2+X_3^2+X_4^2+X_5^2=R^2\,.
\fe
The stereographic coordinates $x_\m$ are given as
\ie
X_\m={R x_\m \over 1+ {x^2\over 4R^2}},\quad X_5=R{1- {x^2\over 4R^2} \over 1+ {x^2\over 4R^2}}\,,
\label{embtostereo}
\fe
and the metric takes the form
\ie
ds^2=e^{2\Omega} dx^2,\quad e^\Omega ={1 \over 1+ {x^2\over 4R^2}}\,.
\fe
The angular coordinates with metric 
\ie
ds^2= R^2(d\zeta^2 +\cos^2\zeta d\tau^2 +\sin^2\zeta (d\theta^2+ \sin^2 \theta d\phi^2))\,,
\fe
are related to the embedding coordinates by
\ie
(X_\m,X_5)=&R \left  (\sin \zeta \sin \theta \sin \phi ,\sin \zeta \sin \theta \cos \phi , \sin \zeta \cos \theta  ,
\cos \zeta  \sin \tau  ,\cos \zeta  \cos \tau \right) \,,
\fe
and to the stereographic coordinates by
\ie
x_\m=R\left({2 \sin \zeta \sin \theta \sin \phi  \over 1+\cos \zeta \cos \tau },~
{2 \sin \zeta \sin \theta \cos \phi  \over 1+\cos \zeta \cos \tau },~
{2 \sin \zeta   \cos \phi  \over 1+\cos \zeta \cos \tau },~
{2 \cos \zeta  \sin \tau   \over 1+\cos \zeta \cos \tau }\right)\,.
\fe
Finally we have the hybrid coordinates $(\tau, \tilde x_i)$ with $i=1,2,3$, in which case the $S^4$ metric takes the following form
\ie
ds^2={d\tilde x^2\over  \left( 1+ {\tilde x^2\over 4R^2} \right)^2}+R^2  \left( {1- {\tilde x^2\over 4R^2}\over  1+ {\tilde x^2\over 4R^2}} \right)^2d\tau^2 \,,
\fe
related to the embedding coordinates by
\ie
X_i={R \tilde x_i \over 1+ {\tilde x^2\over 4R^2}},\quad X_4=R{1- {\tilde x^2\over 4R^2} \over 1+ {\tilde x^2\over 4R^2}}\sin \tau ,\quad X_5=R{1- {\tilde x^2\over 4R^2} \over 1+ {\tilde x^2\over 4R^2}}\cos \tau \,,
\fe
and related to the stereographic coordinates by
\ie
x_i={\tilde x_i } { 1+\cos \zeta\over 1+\cos \zeta \cos \tau },\quad 
x_4=R {2\cos \zeta \sin \tau \over 1+\cos \zeta \cos \tau },\quad \cos \zeta={1- {\tilde x^2\over 4R^2} \over 1+ {\tilde x^2\over 4R^2}}\,.
\fe

\bibliographystyle{JHEP}
\bibliography{defREF,SYMdefect}

\end{document}